\documentclass[final,5p,times,twocolumn]{elsarticle}
\usepackage{url}
\usepackage{rotating}

\usepackage{graphicx}

\def\be{\begin{equation}}
\def\ee{\end{equation}}
\def\bea{\begin{eqnarray}}
\def\eea{\end{eqnarray}}

\journal{Astroparticle Physics}

\begin{document}
\begin{frontmatter}

\title{Active Galactic Nuclei under the scrutiny of CTA}

\author[ad1]{H. Sol}
\author[ad1]{A. Zech}
\author[ad1]{C. Boisson}
\author[ad2]{U.~Barres~de~Almeida}
\author[ad3]{J. Biteau}
\author[ad4]{J.-L. Contreras}
\author[ad3]{B.Giebels}
\author[ad4]{T.~Hassan}
\author[ad5]{Y.~Inoue}
\author[ad6]{K.~Katarzy{\'n}ski}
\author[ad7]{H. Krawczynski}
\author[ad4]{N.~Mirabal}
\author[ad8]{J.~Poutanen}
\author[ad9,ad10]{F.~Rieger}
\author[ad11]{T.~Totani}
\author[ad12]{W.~Benbow}
\author[ad1]{M.~Cerruti}
\author[ad13]{M.~Errando}
\author[ad14]{L.~Fallon}
\author[ad15]{E. de Gouveia Dal Pino}
\author[ad16]{J.A.~Hinton}
\author[ad17]{S.~Inoue}
\author[ad18]{J.-P.~Lenain}
\author[ad19]{A.~Neronov}
\author[ad17]{K.~Takahashi}
\author[ad20]{H.~Takami}
\author[ad16]{R.~White}
\author {on behalf of the CTA collaboration}

\address[ad1]{
LUTH, Observatoire de Paris, CNRS, Universit\'e Paris Diderot, 5 Place Jules Janssen, F-92190
 Meudon, France
}

\address[ad2]{
Max-Planck-Institut fur Physik, Foehringer Ring 6, 80801 - Muenchen, Deutschland
}

\address[ad3]{
Laboratoire Leprince-Ringuet, Ecole Polytechnique, CNRS/IN2P3, F-91128 Palaiseau, France
}

\address[ad4]{
Dpto. de Fisica Atomica, Molecular y Nuclear, Universidad Complutense de Madrid, Spain
}

\address[ad5]{
Kavli Institute for Particle Astrophysics and Cosmology, Department of Physics, and
SLAC National Accelerator Laboratory, Stanford University, Stanford, California 94305, USA
}

\address[ad6]{
Toru{\'n} Centre for Astronomy, Nicolaus Copernicus University, ul. Gagarina 11, 87-100 Toru{\'n}, Poland
}

\address[ad7]{
Washington University in St Louis, Physics Department, and McDonnell Center for the Space Sciences, 1 Brookings Drive, CB 1105, St Louis, MO 63130, USA
}

\address[ad8]{
Department of Physics, Astronomy Division, University of Oulu, PO Box 3000, FI-90014, Finland
}

\address[ad9]{
Max-Planck-Institut f\"ur Kernphysik, P.O. Box 103980, D 69029 Heidelberg, Germany
}

\address[ad10]{
European Associated Laboratory for Gamma-Ray Astronomy, jointly supported by CNRS and MPG
}

\address[ad11]{
Department of Astronomy, School of Science, Kyoto University, Sakyo-ku, Kyoto 606-8502, Japan
}

\address[ad12]{
Harvard-Smithsonian Center for Astrophysics, 60 Garden St, Cambridge, MA 02138, USA
}

\address[ad13]{
Department of Physics and Astronomy, Barnard College, Columbia University, NY 10027, USA
}

\address[ad14]{
Dublin Institute for Advanced Studies, 31 Fitzwilliam Place, Dublin 2, Ireland
}

\address[ad15]{
Instituto de Astronomia, Geofisica e Ciencias Atmosfericas, Universidade de Sao Paulo, R. do Matao, 1226, Sao Paulo, SP 05508-090, Brazil
}

\address[ad16]{
Department of Physics and Astronomy, The University of Leicester, University Road, Leicester,
 LE17RH, UK
}

\address[ad17]{
Graduate school of science and technology, Kumamoto University, 2-39-1 Kurokami, Kumamoto 860-8555, Japan
}

\address[ad18]{
Landessternwarte, Universit\"at Heidelberg, K\"onigstuhl, D 69117 Heidelberg, Germany
}

\address[ad19]{
ISDC Data Center for Astrophysics, Geneva Observatory, Chemin d'Ecogia 16, 1290 Versoix, Switzerland
}

\address[ad20]{
Max Planck Institute for Physics, Fohringer Ring 6, 80805, Munich, Germany
}

\begin{abstract}
Active Galactic Nuclei (hereafter AGN) produce powerful outflows which
offer excellent conditions for efficient particle acceleration in internal and external
shocks, turbulence, and magnetic reconnection events.
The jets as well as particle accelerating regions close to the supermassive 
black holes (hereafter SMBH) at the intersection of plasma inflows and outflows, 
can produce readily detectable very high energy gamma-ray
emission. As of now, more than 45 AGN including 41 blazars and 4 radiogalaxies 
have been detected by the present ground-based gamma-ray telescopes, which 
represents more than one third of the cosmic sources detected so far in the
VHE gamma-ray regime.
The future Cherenkov Telescope Array (CTA) should boost the sample of AGN detected in
the VHE range by about one order of magnitude, shedding new light on AGN population 
studies, and AGN classification and unification schemes. 
CTA will be a unique tool to scrutinize the extreme high-energy tail of accelerated particles 
in SMBH environments, to revisit the central engines and their associated 
relativistic jets, and to study the particle acceleration and emission mechanisms,
particularly exploring the missing link between accretion physics, SMBH magnetospheres and jet formation. 
Monitoring of distant AGN will be an extremely rewarding observing program which 
will inform us about the inner workings and evolution of AGN. Furthermore these 
AGN are bright beacons of gamma-rays which will allow us to constrain 
the extragalactic infrared and optical backgrounds as well as the intergalactic 
magnetic field, and will enable tests of quantum gravity and other "exotic" phenomena.
\end{abstract}


\end{frontmatter}

\section{Introduction}

During the last decade, a new branch of astrophysics has emerged at the high-energy
end of the electromagnetic spectrum as our cosmos is revealing its amazing richness 
in the TeV gamma-ray band. In the extragalactic domain, more than 45 AGN have been 
identified in the very high energy (hereafter VHE) band above 30 GeV. The majority of these sources belongs to the blazar class, 
a peculiar type of AGN with a strong Doppler boosted continuum emission emitted by
a relativistic jet closely aligned with the line of sight.
Monitoring of bright blazars led to the discovery of flux variability down to minute time scales, 
never so well resolved before in any AGN at longer wavelengths, and brought new constraints on the physical 
mechanisms at work \cite{Al501,AhBigFlare}. 
Another type of active extragalactic sources has recently emerged as TeV emitters, with the radiogalaxies M 87 \cite{AhM87,Ac,AlM87} 
and Cen A \cite{CenAdisco} for which we now have detailed TeV gamma-ray studies. 
Radiogalaxies have a much higher space density than blazars and their detection could be less
affected by their alignment to the line of sight. VHE studies of radiogalaxies are thus likely
to contribute significantly to a better understanding of AGN unification schemes. 

The VHE\ observations already challenge current theories of particle acceleration, 
to explain how particles are accelerated to $>$TeV energies
in regions relatively small compared to the fiducial scale of the black hole 
event horizon \cite{Al501,AhM87,Ah1713,Aharonian}. 
Emission models are poorly constrained, with both leptonic and hadronic models 
able to fit most of the available spectral data \cite{Ah2155,Ah1713first} but have difficulties 
to explain fast variability. The energy spectra of distant AGN raise specific questions. 
Their observed shape depend on the intrinsic emitted VHE energy spectra, complex absorption and/or cascading processes in the AGN host galaxy
and in extragalactic space, and even on cosmological 
expansion and the star formation history of the universe \cite{Blanch}. 
Indirect constraints on the infrared diffuse background radiation and the stars and 
galaxies which produce it, can therefore be deduced from the observed VHE 
spectra \cite{AhEBL1,AhEBL2,Al3C279}. The observations taken so far show already
that our universe is apparently more transparent to VHE gamma-rays than previously thought. 
Disentangling the different effects definitively requires high quality spectra 
with better sensitivity and higher spectral resolution at energies above TeV energies  
for large samples of AGN at different redshifts. 

CTA will provide a unique opportunity to address such issues. 
With an order-of-magnitude improved flux sensitivity compared to existing instruments, 
it will offer a large dynamic range of more than $10^4$ for studies of bright VHE flaring epochs. 
Temporal resolution down to the sub-minute time scales will become possible, 
making CTA a perfect tool for gamma-ray timing analysis and for studying AGN micro-variability. 
About four orders in magnitude will be covered in energy, from typically 20 GeV to beyond 100 TeV, 
with an energy resolution of typically 10\% to 15\%, ideally suited for the search for 
spectral features from extragalactic absorption processes.
For the better defined events with multiple triggered Cherenkov telescopes, one  
arc-minute angular resolution can be achieved, and astrometric positions can be
determined with a precision better than 10 arcsec, which will ensure the reliable identification 
of AGN, and possibly also the detection of VHE\ emission from compact and extended AGN components. 

CTA will be operated as an open observatory, offering a
multi-functional tool with several configurations and observation
modes. The flexibility of CTA will be particularly useful for the
study of AGN. The whole array may be used simultaneously for deep
observation of specific fields and targets, or some sub-arrays may
work independently for monitoring, alert, or survey purposes,
increasing the global capabilities of the infrastructure.  Two CTA
sites are foreseen to allow a survey of the whole sky, one in the
southern hemisphere and one in the northern one. The latter will
largely be devoted to observe AGN and extragalactic sources
\cite{CTA-DS}. 
 
In this paper we examine how CTA will contribute
to answer open questions on AGN physics such as the nature of the black
hole magnetosphere, the formation of jets and the acceleration of
particles, the total energy budget, and the origins of
variability, and how it sheds new light on AGN classification
and unification schemes. After a short presentation of the current
knowledge on AGN at VHE in section 2, we analyze the performances of
CTA in terms of AGN population studies in section 3. Section 4
illustrates how the high quality spectra obtained with CTA can test
emission and absorption models and constrain various parameters. The
capability of CTA to probe AGN variability is specifically addressed
in section 5. Finally we exemplify in section 6 that observing AGN
with CTA will provide original clues on intergalactic media and
diffuse backgrounds.

\section{Current status of Active Galactic Nuclei at VHE}

\begin{figure}[h!]
\begin{center}
\includegraphics[width=0.5\textwidth]{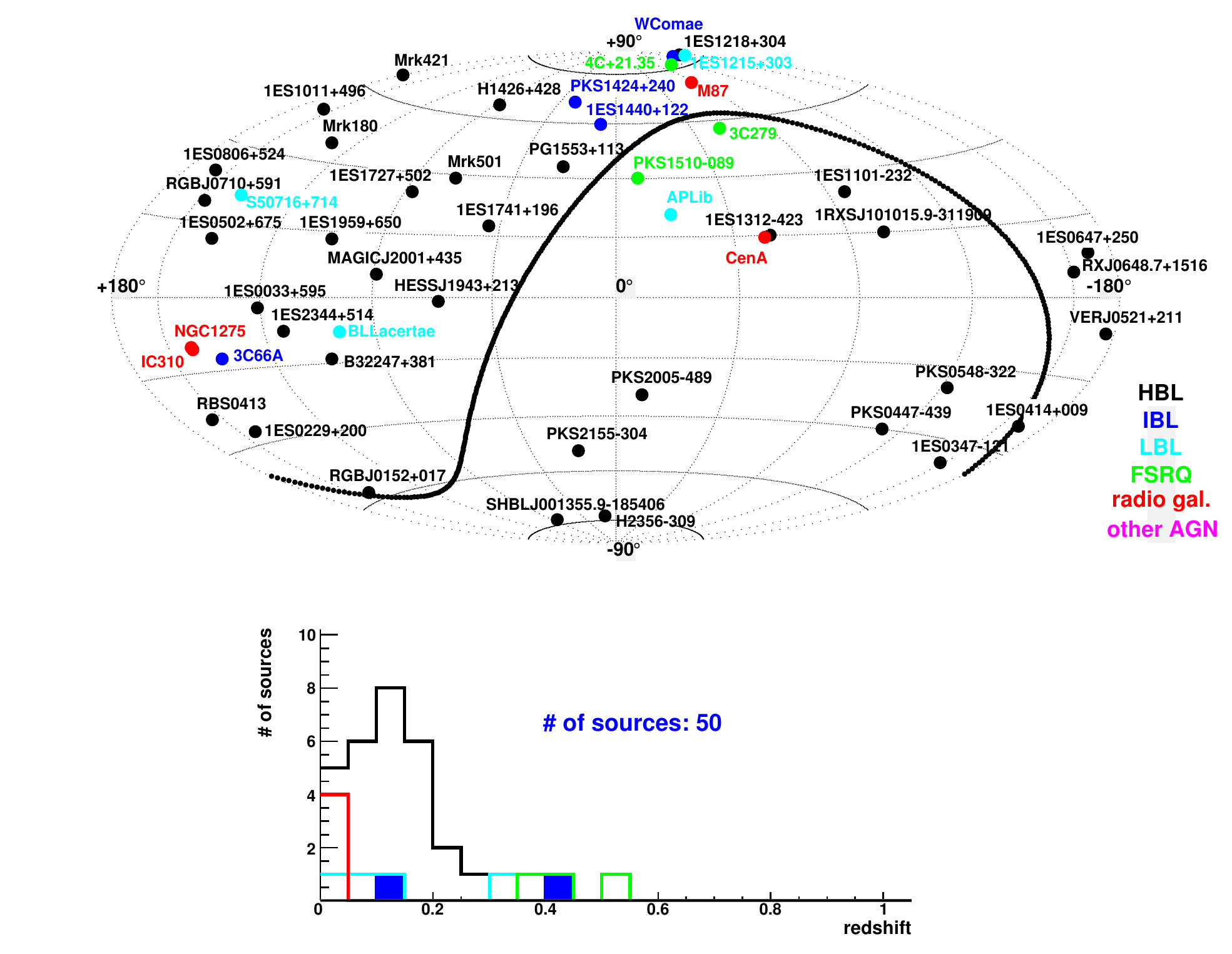}
\caption{Distribution in galactic coordinates of the VHE active galactic nuclei currently detected by the present generation of Cherenkov experiments H.E.S.S., MAGIC and VERITAS, as listed in the TeVCat catalog in October 2012). The detection of some fifty sources in the TeV domain is now firmly confirmed. Various types of AGN are shown, as well as their distribution in redshift. 
\label{fig:tevagn}}
\end{center}
\end{figure}

AGN are currently believed to harbor a central massive black hole surrounded by an accretion disk 
of matter spiraling towards the black hole. In about 10\% of the cases (50\% for the most energetic sources), 
energetic particle beams are emitted along the rotation axis of the black hole, giving rise to the so-called 
``radio-loud'' AGN with well-collimated radio jets. 
There is a large variety of AGN classes, with different observational characteristics \cite{Anton,Urry}. 
Standard AGN unification schemes attempt to classify the sources according to their viewing angle $\theta$. 
The blazar family is the least numerous class, as the blazars are believed to be
radio loud AGN with their jets aligned to within a few degrees of the line of sight to the observer.
This leads to a strong relativistic boosting effect which amplifies the observed luminosity 
by a factor of $\sim \delta^4$ and shortens the observed flare time scales by a factor of $\delta$, 
where the Doppler factor of the relativistic bulk motion is 
\begin{equation}
\delta = \frac{1}{\gamma_{\rm bulk} (1- \beta \cos{\theta})}
\end{equation}
where $\beta = V_{\rm bulk}/c$, and the bulk relativistic Lorentz factor
$\gamma_{\rm bulk} = 1 / (1 �- \beta^2)^{1/2}$.  This relativistic
beaming also contributes to the reduction of internal absorption of
$\gamma$-rays by reducing the intrinsic luminosity of the source and
allowing intrinsically larger emitting zones.  Indeed, most of the VHE
bright AGN belong to the blazar family (Fig.\ref{fig:tevagn}).  Their
TeV radiation comes from extremely relativistic jets, and strong
Doppler boosting favors their detection. Blazars include various types
of AGN such as HBL (high-frequency peaked BL Lac), IBL and LBL
(Intermediate and low-frequency peaked BL Lac), and FSRQ (Flat
Spectrum Radio Quasars).  Another class of AGN, the radio galaxies,
has also been detected in the TeV range at low redshifts.  It is not yet clear how
gamma-ray properties of these sources, believed to have relativistic
jets oriented at larger viewing angles, compare to the ones of TeV
blazars.

\begin{figure}[h!]
\includegraphics[width=0.45\textwidth]{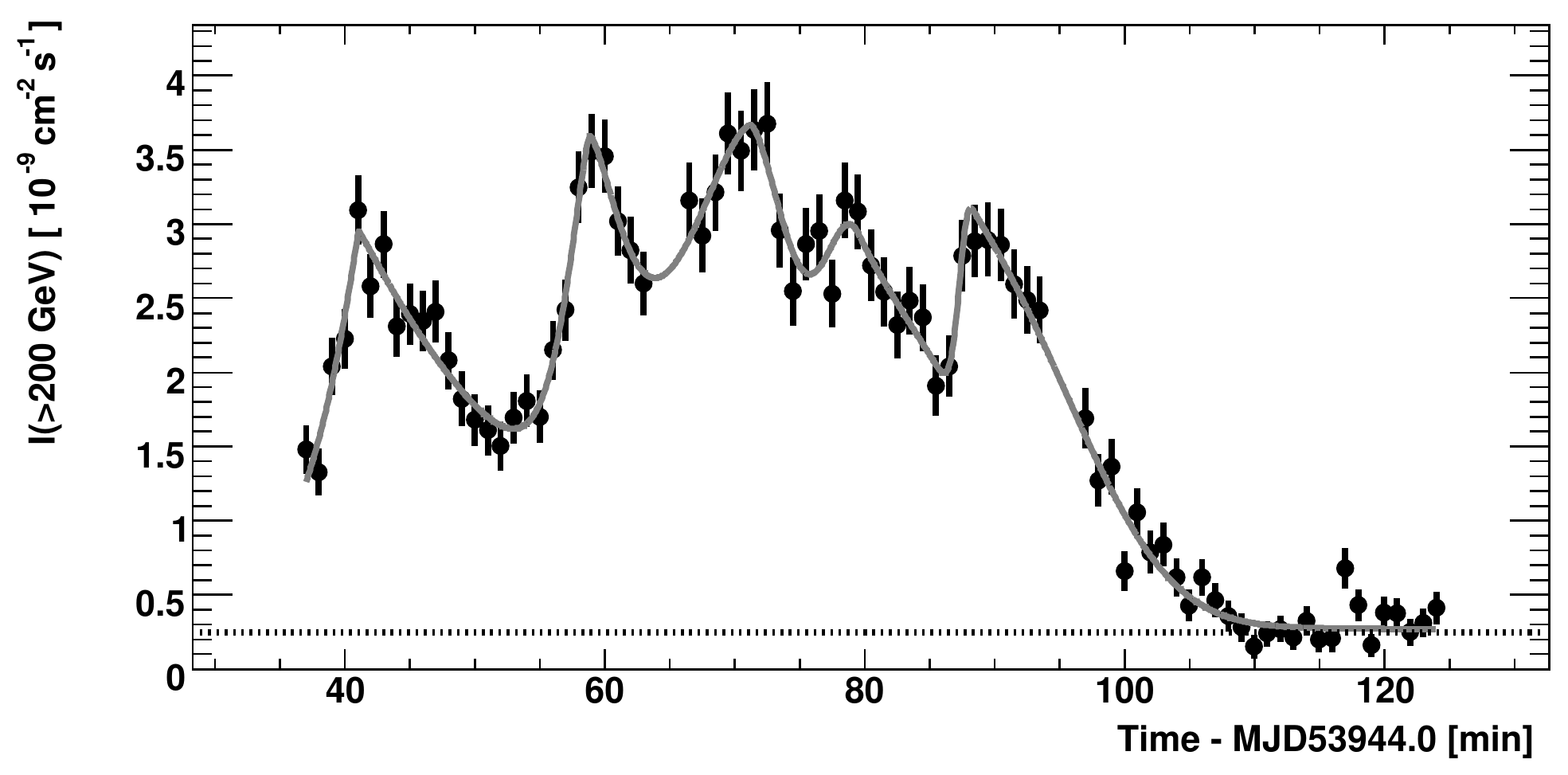}
\caption{The big flare of PKS 2155-304 observed in 2006 by H.E.S.S., with variability down to a few minute scale \cite{AhBigFlare}. Fig.\ref{fig:lcext} of part 5 shows how CTA could have seen it.}
\label{fig:BigFlare}
\end{figure}

The sample of AGN detected at VHE currently includes 49 published sources, spread in redshifts from z = 0.0018 to z = 0.536, with 45 blazars (33 HBL, 4 IBL, 4 LBL, 3 FSRQ, and one blazar of unknown type), four Fanaroff-Riley type 1 radio galaxies, and a few other cases to be confirmed. The additional case of the Galactic Center, considered as a weak AGN, will be further discussed in section 3.4. Sources can show quiescent, low, and highly active VHE flaring states. The AGN fluxes at TeV energies range from 0.003 to $\sim 20$ Crab units, from low states to the brightest events.  The VHE energy spectra of most sources can be described by simple power-laws, with observed photon index $\Gamma_{\rm obs}$ between 1.9 and 4.6. Variability is frequently observed, namely in about 20 sources out of the 45, despite sparse time coverage for many sources. Flux variability has been detected on all time scales from years, months, and days, down to the minute scale for three flaring sources (Fig.\ref{fig:BigFlare}).

Highly variable VHE events of AGN likely originate at small distances (less than 1 {\rm pc}) from the central engine since rapid variability ($t_{\rm var}$) on time scales from a few days \cite{Fossati08} down to a few minutes \cite{Aharonian07a} limits the  comoving size of the emitting region to 
\begin{equation}
R' \le \frac{c t_{\rm var} \delta}{1+z},
\end{equation}
where $\delta$, the Doppler factor, is in the range from a few up to a few tens, gives $R' \sim 10^{14}$ to $ \sim 10^{17}$ {\rm cm}. VLBI radio monitoring of the closest objects confirms this argument by showing growing evidence of correlation between VHE activity and VLBI core evolution \cite{Charlot,M87_campaign,Piner10,Lister}. The existence of additional extended VHE emission remains an interesting possibility \cite{Lukasz1,Lukasz2,Hard11}.

\par

The number of detected TeV sources per class of AGN appears extremely peculiar since the blazars are usually the smallest population among all types of AGN seen across the electromagnetic spectrum, yet they by far dominate the TeV sample. The current TeV HBL sample is not flux-limited and is highly biased. The dynamical range of present-day IACT remains usually below 5000, while active states can easily amplify the fluxes by factors of 10 to 200, and Doppler boosting by factors of $10^4$ to $10^6$, or even more. Comparatively, a factor 7 in redshift (from Mrk 421 to 1ES 1011+496 for instance) would decrease the flux typically by a factor 50 only. So the VHE sample appears to be largely incomplete and suffer from strong observational biases towards large Doppler boosting and active states, due to the present sensitivity limits and the strategy of observations, often done under VHE and multiwavelength alerts. Luminosity functions at VHE are still impossible to derive due to poor statistics and to insufficient time coverage to firmly distinguish the quiescent stationary states from bright flares.

\begin{figure}[h!]
\begin{center}
\includegraphics[width=0.5\textwidth]{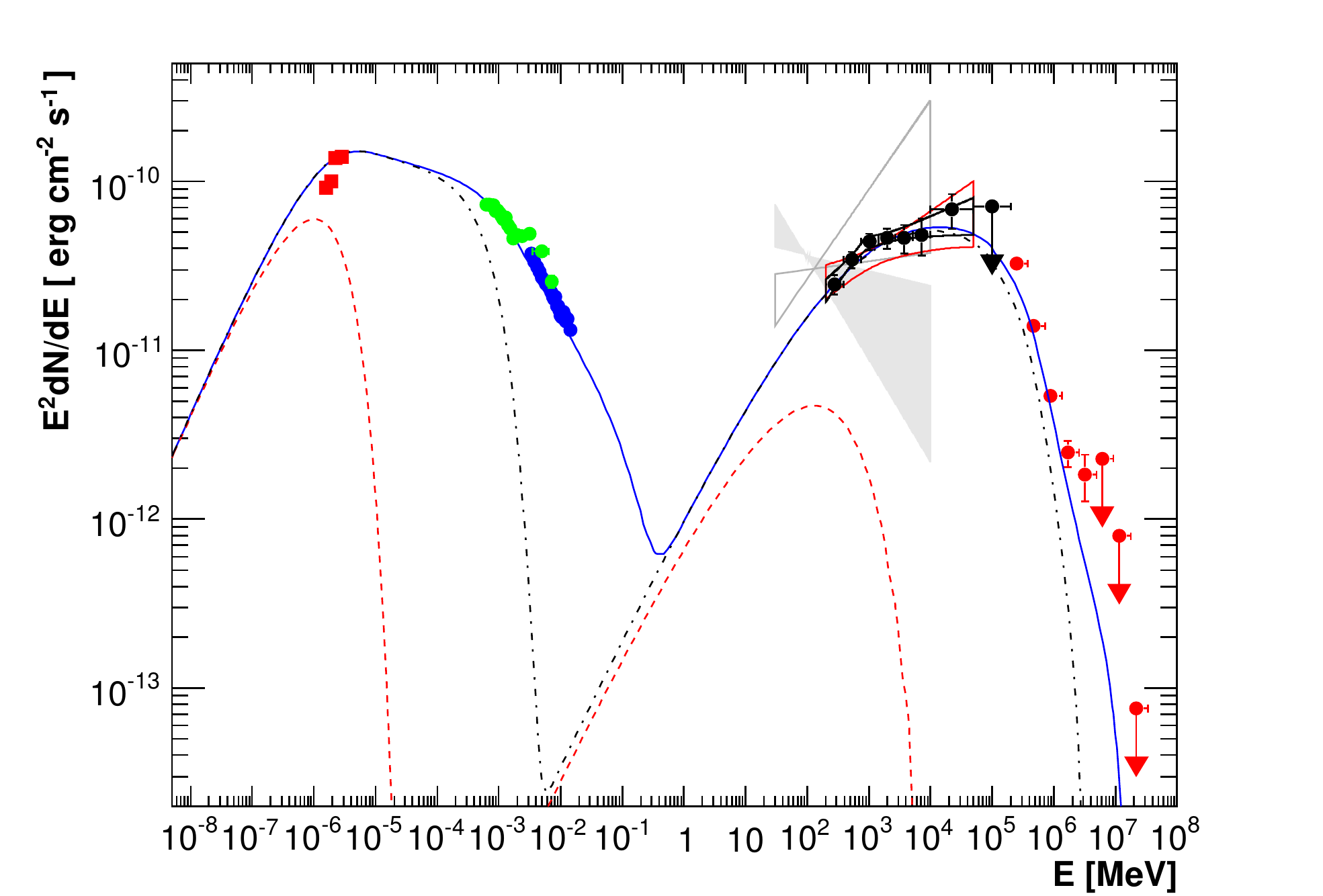}
\caption{Example of the double-peaked SED of TeV AGN as seen with current instruments, reproduced by a basic SSC scenario. The figure shows the quiescent high energy state of PKS 2155-304 observed during a multiwavelength campaign in 2008 with simultaneous data from ATOM, RXTE, Swift, Fermi and H.E.S.S. \cite{PKS2155mwl_09}.    
\label{fig:SSC}}
\end{center}
\end{figure}

The spectral energy distribution of TeV AGN appears double-peaked, with a first bump in X-rays and a second one in gamma-rays (Fig.\ref{fig:SSC}). Additional emission at lower energies comes mainly from the stellar population of the host galaxy in the optical range, and from extended jets, hot spots and lobes in the radio range. Both leptonic and hadronic scenarios are widely invoked to describe the radiation processes in the VHE domain. The low energy bump that is visible in the SED of blazars in the optical to X-ray range is dominated in both leptonic and hadronic models by the synchrotron emission from a relativistic electron population. In leptonic models one generally assumes synchrotron self-Compton processes to explain the high energy bump in high-frequency peaked BL Lacs (HBLs), with additional contributions from inverse Compton scattering on external photon fields to account for the SEDs of BL Lacs peaking at lower energies (IBLs, LBLs) and flat-spectrum radio quasars (FSRQs). The inverse-Compton interaction $(e + \gamma_0 \rightarrow e + \gamma)$ of the X-ray synchrotron emitting electrons with ambient photons produces the gamma-ray bump at energies $h\nu \sim min [\gamma_{\rm e}^2 h\nu_0, \gamma_{\rm e} m_{\rm e} c^2]$, where $\gamma_{\rm e}$ is the electron Lorentz factor making the peak synchrotron photons. This scenario is called "synchrotron-self-Compton" (SSC) if the ambient photons are the synchrotron emission of the same electron population. If the photons come from another radiation field, namely from the accretion disk, clouds, or dusty torus of the AGN, or from other parts of the jet (e.g. a slow envelope, or upstream or downstream plasma) the scenario is called "external inverse-Compton" (EIC). In hadronic models, very high energy protons are at the origin of the gamma-ray emission, either directly by their synchrotron (or curvature) emission, or through their interaction with local gas and radiation background, creation of pions and subsequent decay into VHE photons (mainly $\pi^0 \rightarrow 2 \gamma)$  with typical energies $E_{\rm \gamma} \sim  E_{\rm \pi} /2 \sim  10\% E_{\rm p}$. Decay of pions into muons also produces neutrinos, and secondary electrons which can radiate in the X-ray range. A detailed presentation of the above scenarios can be found for instance in \cite{Weekes,FAAharonian}. 

An additional effect modifies the shape of the spectra at high energy gamma-rays, due to the extragalactic background light (EBL) absorption which becomes quite important for high redshift sources. While traveling through the extragalactic space, VHE gamma-rays interact with the EBL photons and suffer strong absorption due to pair creation, 
\begin{equation}
\gamma_{\rm EBL} \gamma_{\rm VHE} \rightarrow e^+ e^-, 
\end{equation}
with $h\nu_{\rm EBL}\times h\nu_{\rm VHE} > (m_{\rm e} c^2)^2$. The absorption of VHE photons is mainly due to the infrared background generated by the integrated light of stars, galaxies and dust (see the EBL article in this issue for more details). The observed photon spectrum $\Phi_{\rm obs}(E, z)$ of a source at redshift z is attenuated compared to the intrinsic spectrum $\Phi_{\rm em}(E')$ emitted at the source, so 
\begin{equation}
\Phi_{\rm obs}(E, z)  = e^{-{\tau_{\rm \gamma}(E, z)}} \Phi_{\rm em}, 
\end{equation}
where $\tau_{\rm \gamma}(E, z)$ is the optical depth computed along the line of sight, and $E' = E(1+z)$. In the energy range 0.2 to 2 TeV, the emitted and observed photon indexes $\Gamma$ are approximately related by   
\begin{equation}
\Gamma_{\rm obs}(z) \sim \Gamma_{\rm em} + \Delta \Gamma(z)
\end{equation}
where $\Delta \Gamma(z)$ is an increasing function of z \cite{Stecker2006}. Applying various reasonable models of the EBL, one can deduce intrinsic photon indices in the range $\sim 1.7$ to $\sim 3$ from the observed spectra, compatible with data at low redshifts. Future data  especially  from interplanetary space missions should be able to significantly improve the direct measurements of the EBL which will be mandatory for a better access to the intrinsic spectra of AGN. 

\begin{figure}[h!]
\begin{center}
\includegraphics[width=0.45\textwidth]{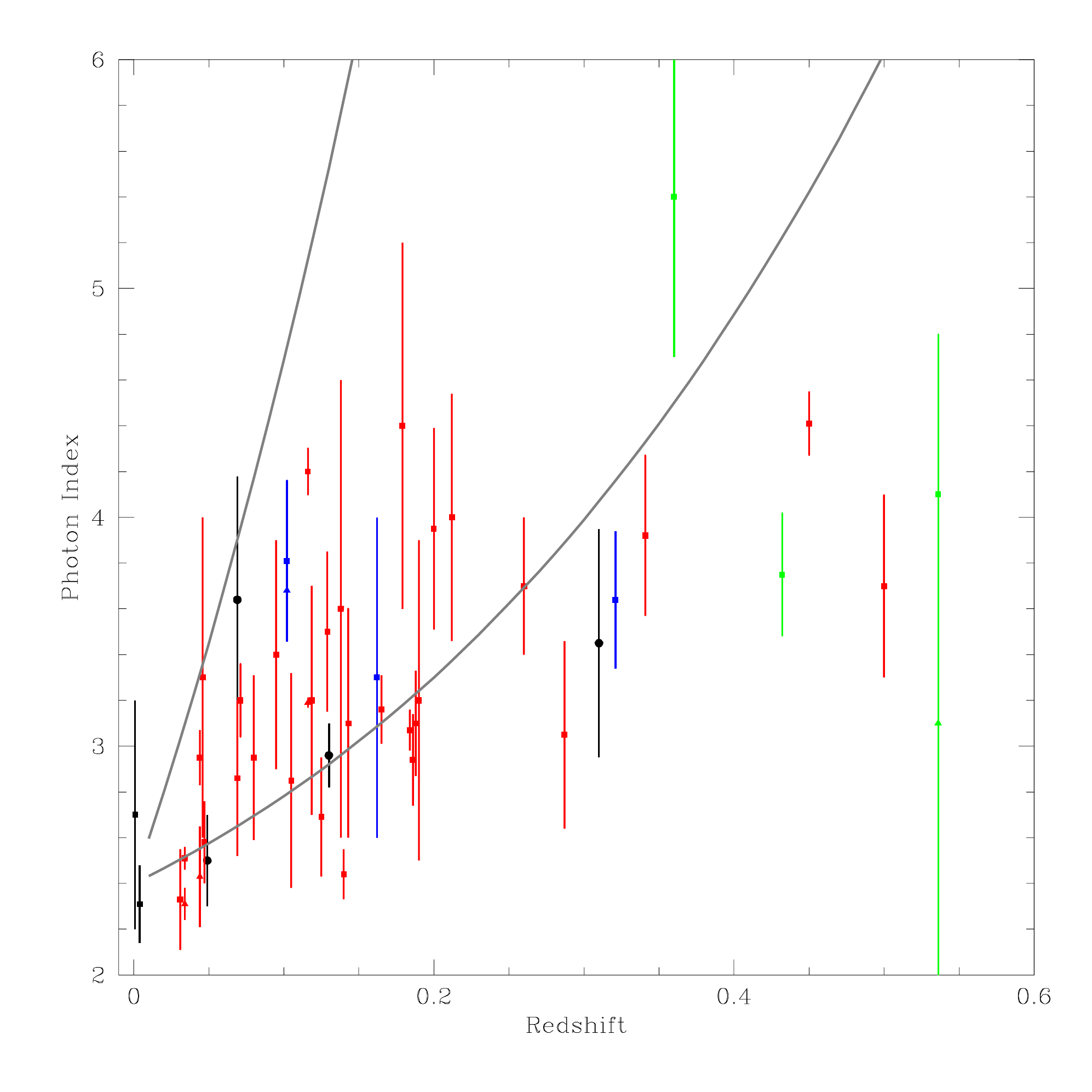}
\caption{The present $\Gamma_{\rm obs}$-z distribution. This figure includes only published sources with confirmed redshift (plus two lower limits at $z \sim 0.32$ and $0.5$) and good spectral index determination. The error bars correspond to the statistical uncertainties. The observed spectral indices are mostly derived in the 0.2-2 TeV range, the relevant EBL range in that case being 0.25-2.5 eV. The spectral energy distribution of the EBL departs from a power-law in this range, showing an emission bump. Bracketing the spectral photon number density versus energy by power laws, De Angelis et al (2009) \cite{Angelis} derived an approximate analytic expression for the optical depth within the Franceschini et al (2008) model \cite{Franceschini:2008} in the local Universe. The two grey curves in the figure  correspond to the expected lower and upper limit after taking into account the cosmological effects, assuming that $\Gamma \sim 2.35$ at $z = 0$. Apart from two radiogalaxies at very low redshift, various types of blazars are shown in color (red = HBL, black = LBL, blue = IBL, green = FSRQ). Triangles indicate confirmed flaring states. A few sources are shown in both flaring and non-flaring states (Mrk 501, PKS 2155 and 1ES2344), and two are shown only in flaring states (W Comae and 3C 279).
\label{fig:Gamma-z}}
\end{center}
\end{figure}

If the intrinsic AGN spectra were all similar, then the distribution of observed photon index versus redshift should show an apparent increase of $|\Gamma_{\rm obs}|$ with redshift \cite{Persic}, which is not really obvious at the moment, even when considering only one specific sub-class of blazars (Fig.\ref{fig:Gamma-z}). This is currently a matter of debate, although the departure from standard power laws and potential observational biases, such as the current difficulty to measure large values of $\Gamma_{\rm obs}$, need to be further investigated. 
Non-standard explanations have been proposed, including various evolutionary effects in the blazar population \cite{Stecker_Scully,Aharonian_2008a,Boettcher_2008}, ultra-high energy cosmic rays (UHECR) effects \cite{Dermer,Essey2012} or the existence of axion-like particles which could  significantly reduce the EBL dimming \cite{Angelis,DeAngelis}. More conservative solutions exist, such as a hardening of intrinsic spectra during active states \cite{Lefa}, together with the fact that high redshift objects might be  preferentially detected during flaring states (which is indeed the case of 3C 279). However none of them is fully satisfactory yet. 
Increasing the statistics on AGN at all redshifts and gathering high quality spectra with CTA should clarify this intriguing situation.

\section {AGN population studies with CTA}

The jump of sensitivity with CTA will offer large samples of VHE sources and open the way towards statistical studies of the blazar and AGN populations.  Large and homogeneous samples of sources are necessary for unbiased statistical analyzes and to construct useful luminosity functions. To be able to gather them, a significant step forwards will be to firmly detect and study quiescent stationary VHE states for comparison with bright flares of blazars and AGN. Currently, this appears possible only for a few bright sources such as the BL Lac PKS 2155-304. Dealing with real stationary fluxes (if any) will at last allow a good statistical approach to the gamma-ray AGN samples, by comparing objects in the same activity state. This will become possible with CTA. It will allow us to explore the relation between HBL, IBL, LBL and FSRQ, and to clarify the validity of the so-called �blazar sequence� and its extended versions \cite{Fossati,Padovani1,Cacc,Padovani2} as well as the recent "blazar envelope" view \cite{Georgano_envelope} or other tentative unifying schemes which try to explain the trends by leptonic scenarios with a decreasing relative importance of the SSC to the EC emission from HBL to FSRQ, or to reproduce all observed SED as a function of black hole mass and accretion rate \cite{Ghisel1,Ghisel2}. Furthermore, comparing properties such as variability characteristics between quiescent and flaring states will contribute to elaborate a global scenario for VHE phenomena in AGN. Is it the same emission mechanism which dominates both quiescent and active states, or are there quite different origins for the VHE emission, with several mechanisms at work depending on the activity level of the source? CTA will answer such questions, which is important for a better understanding of the AGN physics and also to elucidate the observational biases known to affect their VHE detection. 

\begin{figure}[h!]
\begin{center}
\includegraphics[width=0.45\textwidth]{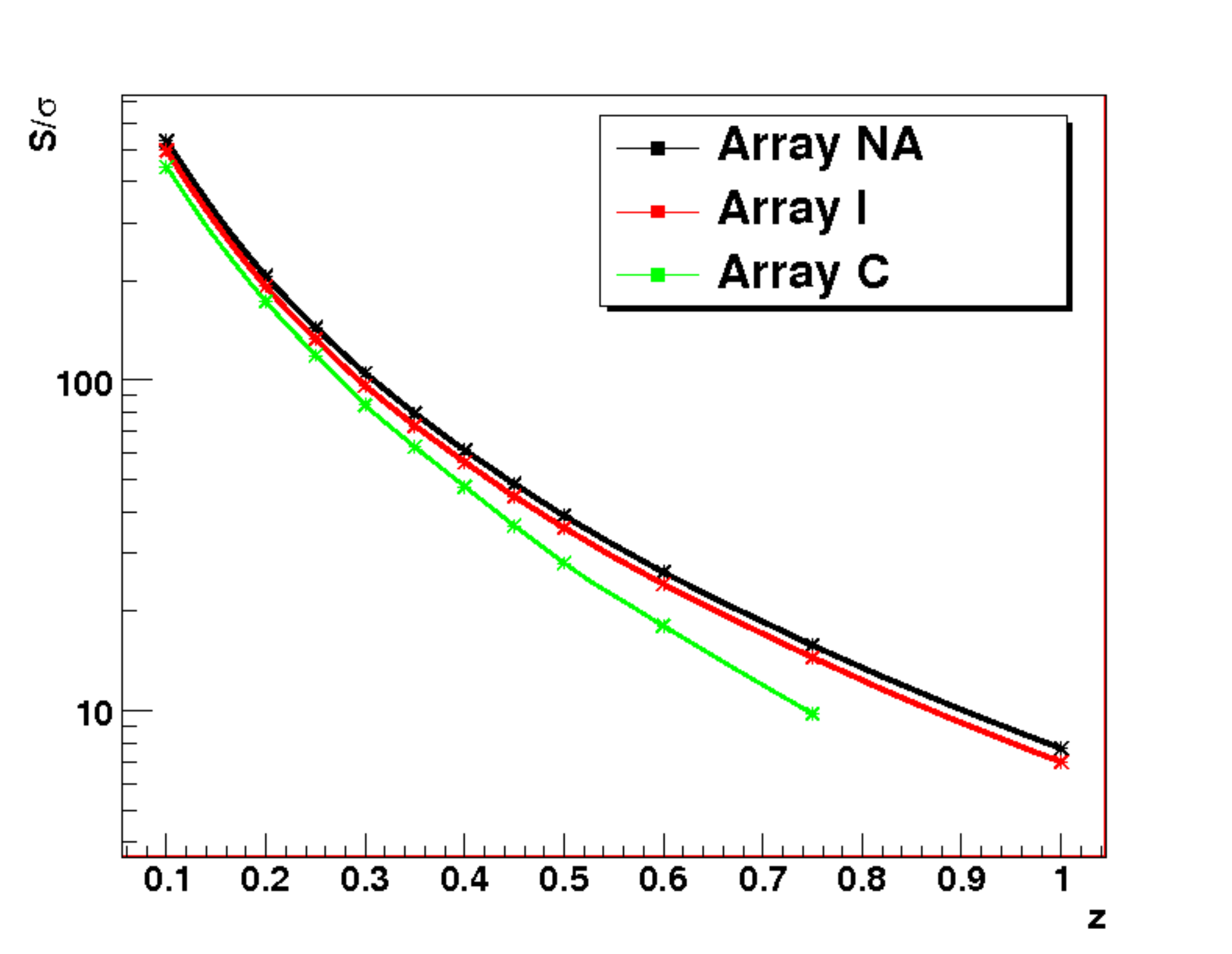}
\caption{Significance of detection in units of the standard deviation $\sigma$ versus redshift for HBL flares of the type of PKS 2155-304, assuming a typical flare with an integrated flux of 10 Crab  at $z = 0.1$, a spectral index of 2.5, and 3 hours of observing time, for three different CTA arrays. 
\label{fig:Flare-z}}
\end{center}
\end{figure}

\begin{figure}[h!]
\begin{center}
\includegraphics[width=0.45\textwidth]{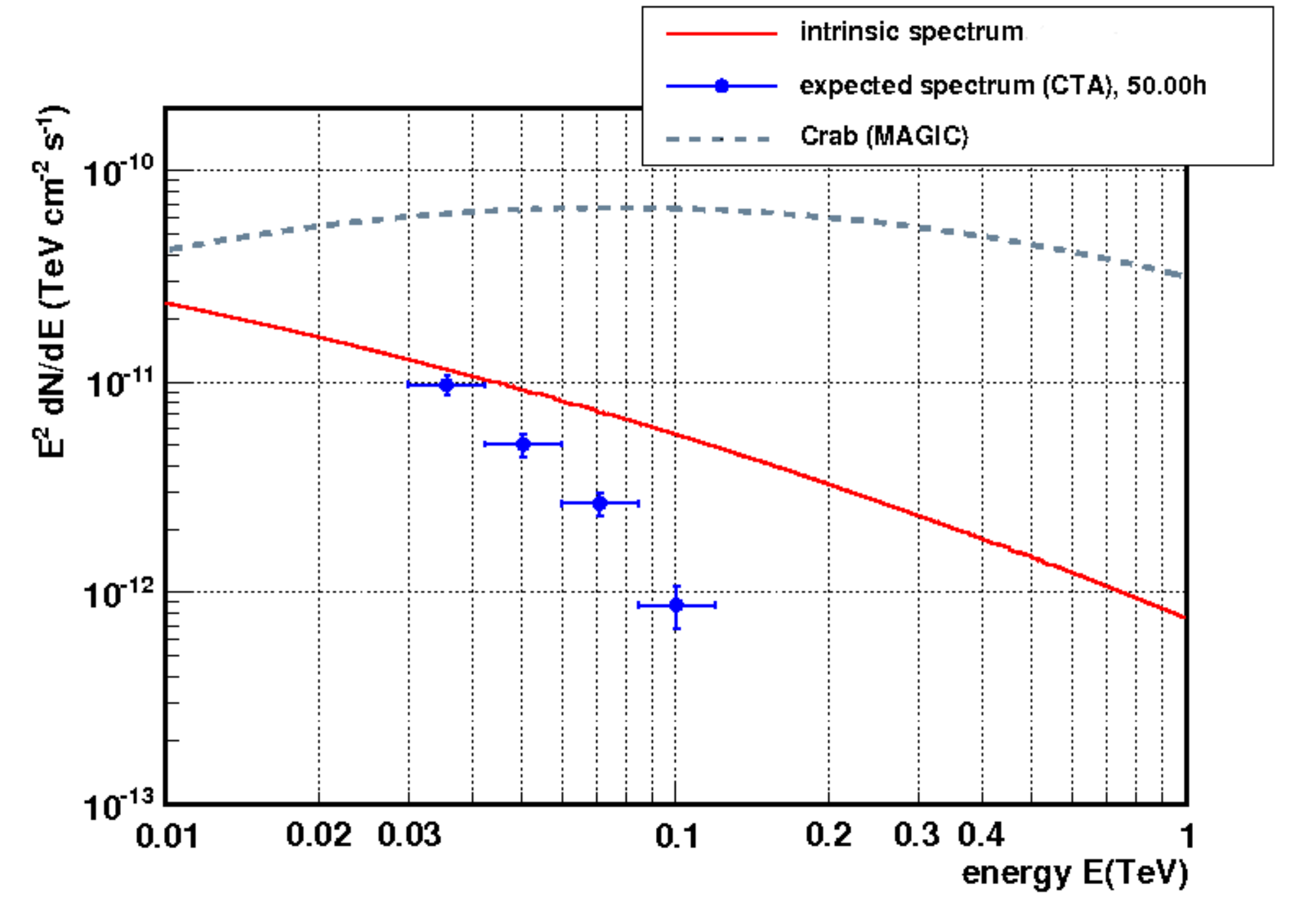}
\caption{CTA detection at highest redshifts: the Fermi blazar 2FGLJ1504.3+1029 at redshift $z = 1.839$ as possibly seen by CTA. Here the VHE spectrum of the blazar during a quiescent state has been extrapolated from the Fermi data assuming a log-parabolic spectral shape, and including EBL absorption effects deduced from \cite{Franceschini:2008}, which are clearly visible in this figure. Detection by CTA can be achieved in 50 hours at $13.6 \sigma$ (array configuration B, zenith angle of 20 degrees). A small sample of blazars should be reachable by CTA at redshifts $\sim 2$ during quiescent states. Bright and long flaring states may allow to obtain useful spectra from sources at even higher redshifts, such as the distant Fermi blazar 1FGL J1344.2-1723, at redshift $z \sim 2.49$.
\label{fig:J1344}}
\end{center}
\end{figure}

The distribution in redshift of known BL Lac sources peaks around $z \sim 0.3$, with a large majority of the population within $z < 1$ \cite{BZCAT,Giommi2012}, a domain of redshift easily reachable with CTA in flaring and even non-flaring states for the brightest sources (see Fig.\ref{fig:Flare-z} and Fig.\ref{fig:FermiAGN}). Conversely, bright FSRQ should be detectable beyond redshift $z\sim 2$ with long exposure time of 50 hours (see Fig.\ref{fig:J1344} and Fig.\ref{fig:FermiAGN_z}). CTA will therefore clarify the actual distribution of observed photon index versus redshift (Fig.\ref{fig:Gamma-z}) and should offer the possibility to study evolutionary effects at VHE, at least for the blazar class.

\begin{figure}[h!]
\begin{center}
\includegraphics[width=0.45\textwidth]{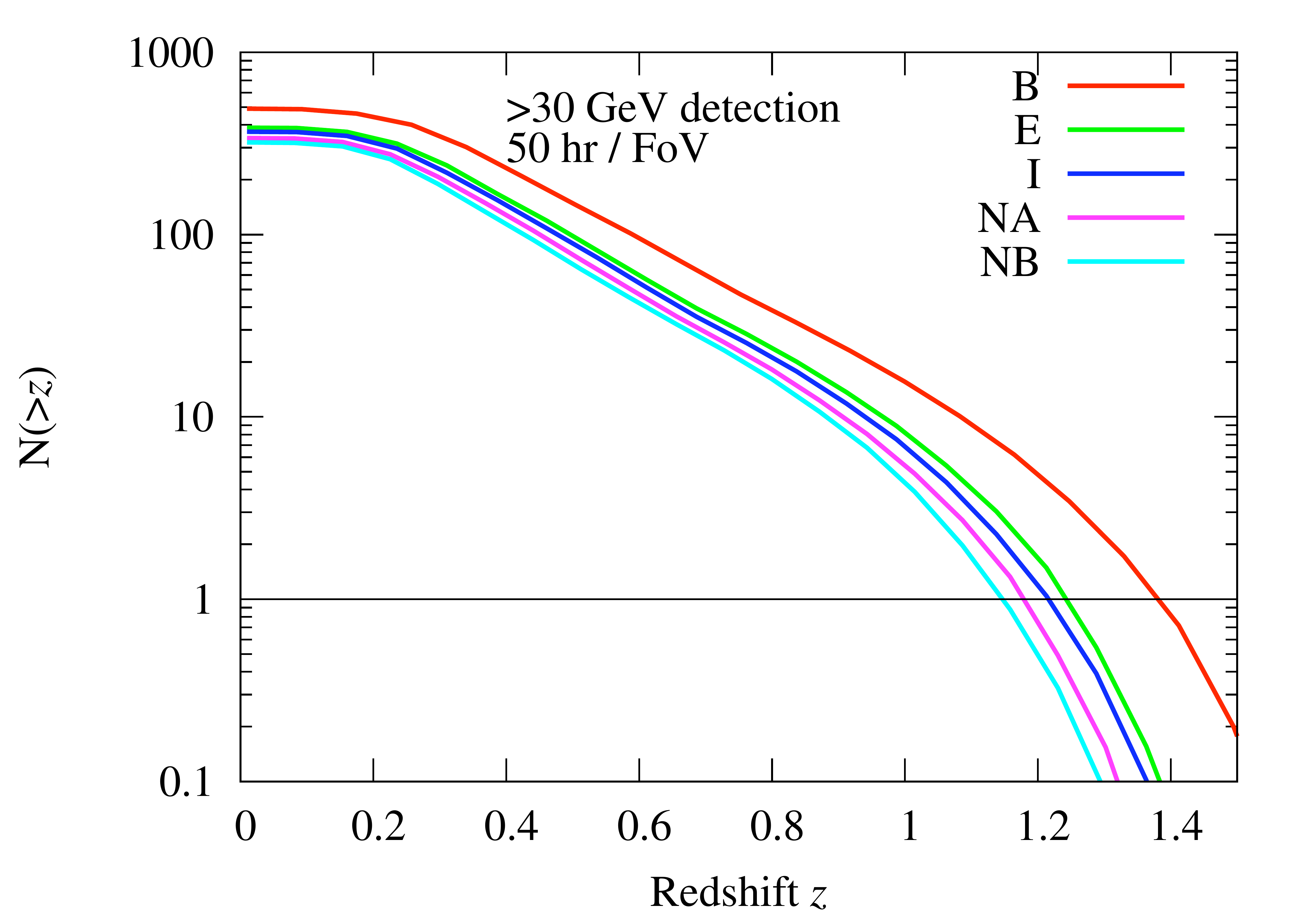}
\caption{Cumulative redshift distribution of blazars above 30 GeV for various CTA array configurations (50 hours observing time for each field, 20 degrees zenith angle observations). Here SED have been simulated following the standard blazar sequence scenario, which seems to underestimate the importance of VHE emission, at least at high redshifts (see Fig.\ref{fig:J1344} and Fig.\ref{fig:FermiAGN_z} for comparison), so these distributions should be considered as lower limits. 
\label{fig:BLLac-z}}
\end{center}
\end{figure}

\subsection{The population of blazars}

What will be the best strategy to gather complete samples of sources? A blind imaging survey is the most unbiased way to observe in a given waveband, and potentially includes the possibility of discovering unexpected objects, such as isolated black holes expelled from host star clusters \cite{Sijacki,Lousto} or final evaporation of primordial black holes \cite{Linton,Schroedter}. Given our present ignorance of AGN properties and statistics at VHE, it is not yet certain how many AGN can be detected by deep surveys for a realistic observing time with CTA. 

Estimates have been done for the blazar class, including BL Lac sources and FSRQs, based on a blazar gamma-ray luminosity function model consistent with recent Fermi results \cite{Yoshi1,Yoshi2} and under the conservative assumption of the standard blazar sequence to simulate the SED. The number of blazars detectable by CTA in blank fields has been estimated for various array configuration (see also the survey article in this issue). The best case leads to about 0.36 blazars detected above 30 GeV in 40 $deg^2$, for 50 hours of exposure time. This results into 370 blazars potentially detectable by CTA, for an all-sky survey with 50 hours/FoV. This requires a very long observing time, and more than 30 years \footnote{Throughout this section, we assume a CTA field of view (FoV) of 7 degrees and 1500 hours of observing time per year. One can  anticipate that typically one third of the whole CTA observing time, including the two sites, could be devoted to AGN programs}, comparable with the expected lifetime of CTA. However, relatively bright objects should readily come out in a full-sky survey at $\sim 1\%$ Crab level in about one year. The expected cumulative redshift distribution in the entire sky is shown in Fig.\ref{fig:BLLac-z}. For a conservative view of blazar statistics, which tends to minimize VHE emission, CTA has the potential to find at least 20 blazars at redshift larger than 1 with its configuration B \cite{Yoshi3,Yoshi4}. Should the standard blazar sequence be revised with more high redshift TeV blazars, still higher detection rates would be expected. 

Blank field sky surveys are mandatory because they provide samples of sources with minimal observational biases, especially  from other waveband observations or from detections triggered by alerts, which is essential for any reliable statistical analysis of the blazar population. As deep surveys are very time consuming, they will be extended over many years, in the fields of specific targets. One can anticipate that for a FoV of 7 degrees, more than 100 serendipitous discoveries of blazars should be obtained in the field of other extragalactic sources in about ten years.

\subsection{Extrapolating from Fermi sources}

Follow-up observations of sources already detected at lower frequencies offer another promising approach --- biased but faster --- in the search for TeV AGN. Here we consider the case of the {\it Fermi} satellite, which has already provided detections of a large sample of AGN in the high energy range. The detection of blazars in the {\it Fermi} LAT energy band ($\sim$0.1 to 100 GeV), i.e. at energies below or partially overlapping with  the ones accessible with CTA, provides important information about the intrinsic spectral properties of extragalactic sources, which are mostly unaffected by  $\gamma$-ray absorption on the EBL in the {\it Fermi} LAT band. The combination of same-epoch {\it Fermi} and CTA measurements of AGN spectra will be of great importance since it will provide a broad-band measurement over $\sim 5$ decades in energy. 

In order to make predictions for CTA, we extrapolated Fermi/LAT AGN spectra into VHE frequencies taking EBL attenuation into account \cite{Franceschini:2008}. Two independent analyses, based on slightly different selections of sources and assumptions on their detection, lead to compatible estimates and predict a large sample of AGN within the reach of CTA. 

For this work, the effective areas and cosmic ray backgrounds simulated for different array configurations have been used to estimate the  instrumental response of CTA (cf. the Monte Carlo contribution in this issue). For simplicity, we consider here full CTA arrays in north and south, although a full configuration with arrays like E, I and C is not planned at the moment for the northern site. With 24 months of accumulated data, the 2FGL catalog contains 1873 point sources characterized in the 100 MeV to 100 GeV energy range \cite{fermi}. Many of these sources have not yet been discovered in VHE due to the lack of sensitivity of existing instruments, but they will be accessible to CTA. In fact, about 85\% of the VHE active galactic nuclei detected by ground-based Cherenkov observatories are found in the 2FGL catalog and an extrapolation of  the 2FGL data to higher energies seems a sensible procedure for the compilation of a mock catalog of CTA sources. 

All associated/affiliated extragalactic sources were selected from the 2FGL catalog \cite{fermi,1FGL}. Out of these 1098 extragalactic sources, only those with a measured redshift can be used for our studies, as it is needed for applying the EBL absorption. For a first study, aimed at providing a conservative estimate of source counts, AGN were only selected for further processing if no analysis problems were flagged in the 2FGL catalog. In a second study, designed to include a maximum number of Fermi sources,  no cut was applied to the 2FGL flags. Technical details on the source selection and the significance estimates can be found in ~\ref{sec:app_fermi}.

\begin{figure}[h!]
\begin{center}
\includegraphics[width=7cm]{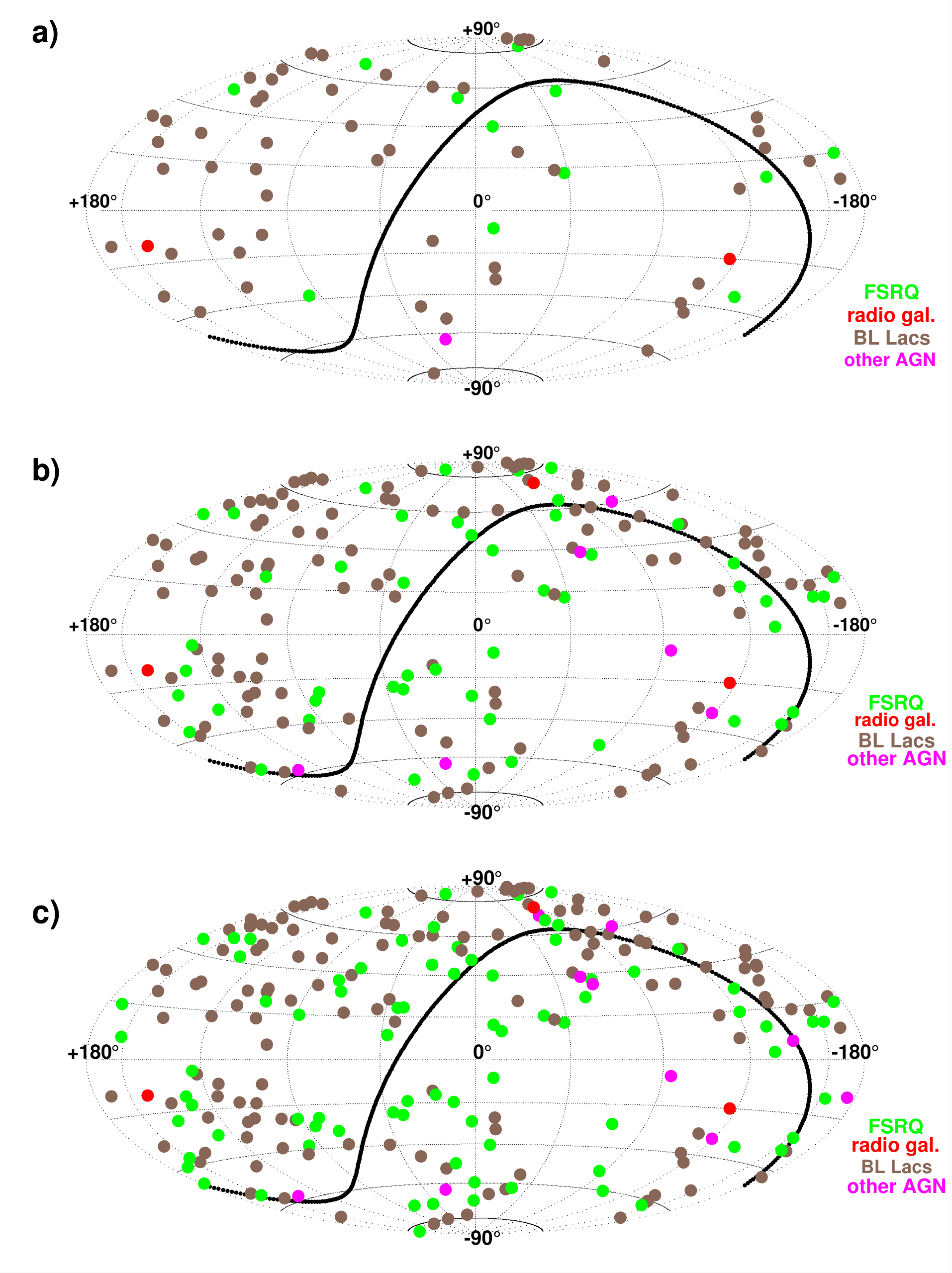}
\caption{Skymaps of AGN with confirmed redshift from the second Fermi  catalog (2FGL) that are potentially detectable by CTA in 5 hours (panel a), 50 hours (panel b) and 150 hours (panel c) maximum exposure time per FoV (assuming the array configuration B and a 20 degree zenith angle over the whole sky). It should be noted that several extragalactic sources that are already known to be TeV emitters do not appear in this figure, due to the tight selection criteria applied here and the existence of non-Fermi VHE sources. Such skymaps should be obtained in less than two months (a), in about three years (b) and in less than 10 years (c) with CTA, assuming 1500 hours observing time per year.} 
\label{fig:FermiAGN}
\end{center}
\end{figure}

The first study focuses on the array configuration B with the best sensitivity at low energy. It yields about 170 detectable AGN within 50 hours of maximum exposure time per source (Fig.~\ref{fig:FermiAGN}, panel b). Out of these, 27 sources (resp. 70) should already be detectable after half an hour (resp. 5 hours) of exposure time with the full array (Fig.~\ref{fig:FermiAGN}, panel a). With a maximum exposure time of 150 hours per source, which is not unusual for observations of very faint sources with the current IACT, about 230 sources should be detectable (Fig.~\ref{fig:FermiAGN}, panel c) in less than 10 years. 

\begin{figure}[h!]
\begin{center}
\includegraphics[width=7cm]{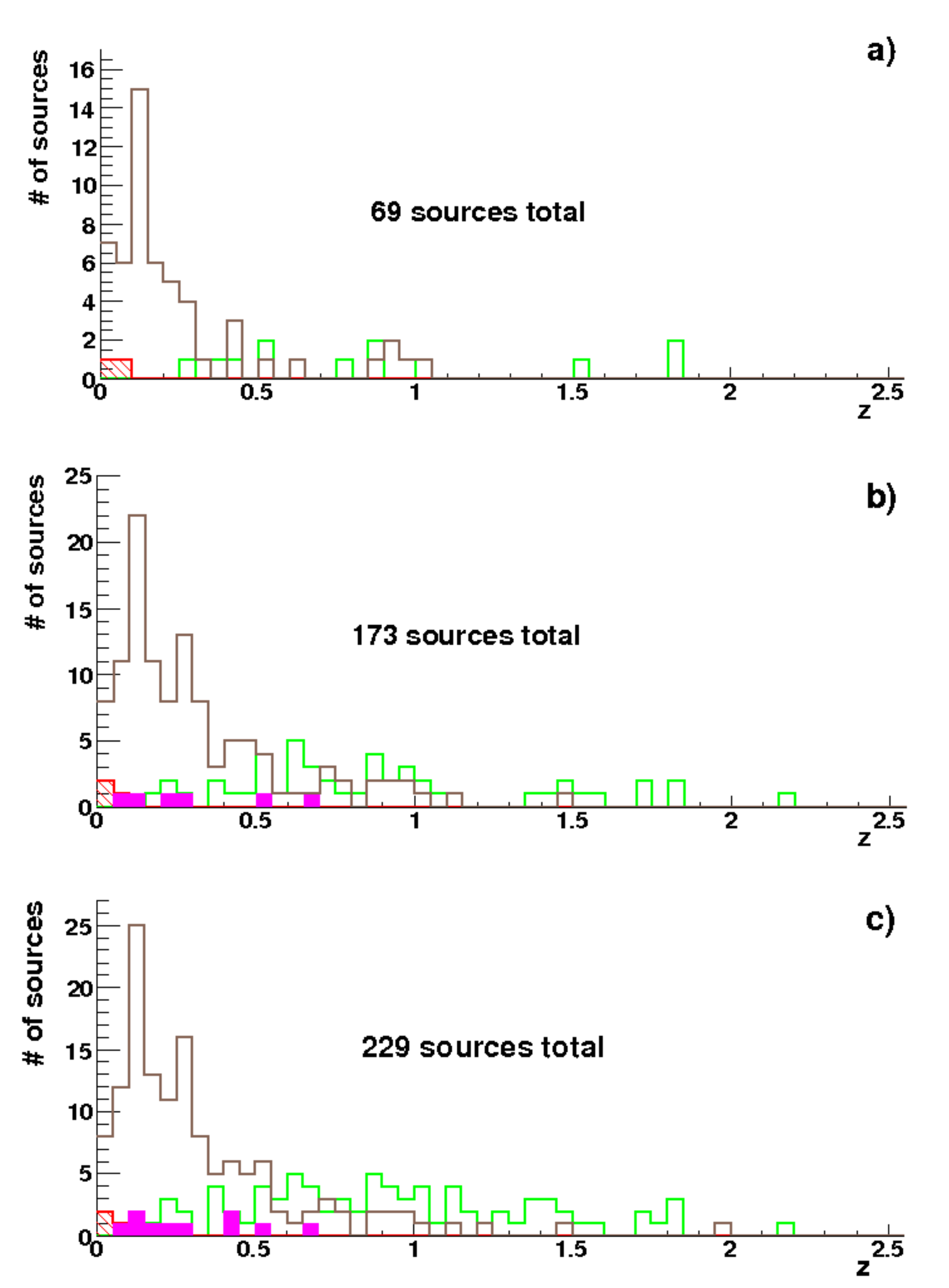}
\caption{Redshift distributions of the AGN with confirmed redshift from the second Fermi catalog (2FGL) that are potentially detectable by CTA in 5 hours (panel a), 50 hours (panel b) and 150 hours (panel c) maximum exposure time per FoV (assuming the array configuration B and a 20 degree zenith angle over the whole sky). These distributions correspond to the skymaps shown in Fig~\ref{fig:FermiAGN}. The same colour code is used: brown for BL Lacs, green for FSRQs, red for radio galaxies and magenta for other types of AGN.} 
\label{fig:FermiAGN_z}
\end{center}
\end{figure}

Apart from augmenting the total number of extragalactic TeV sources, CTA should also increase dramatically the number of objects visible at high redshifts. Fig.~\ref{fig:FermiAGN_z} shows the distributions of AGN as a function of redshift. The most distant quiescent AGN predicted here is at $ z = 2.2$ and under certain conditions this limit might be raised to even higher redshifts. In particular, our estimations indicate that certain flaring FSRQs with a gamma-ray flux increase of a factor of 10 and moderate spectral hardening $\Delta \Gamma = 0.3$ could produce further detections at $z > 2$.

The second study explores the performances of different array configurations (see Table~\ref{tab:agnforecast}). We obtain typically more than $\geq 140$ extragalactic detections, for 50 hours of maximum exposure time per source. In particular, we note that CTA will be most efficient for hard sources ($\Gamma<2$) and shall reveal all the complexities of the extragalactic population extending beyond the reach of current detections. For softer sources that are currently only accessible during flares, CTA will provide unique access to quiescent states.

\begin{table*}
\vspace{0.5 cm}
\caption{Number of detectable Fermi AGN with redshift for different array configurations (50 hours of maximum exposure time). AGN with unknown type are classified as "other AGN".}
\centering
\vspace{0.5 cm}
\begin{tabular*}{0.6\textwidth}{lccccccc}
\hline
Array & FSRQs & BL Lacs & other AGN & SBGs & RGs & Seyferts & Total \\
\hline
\hline
B               &                46     &       117       &      19       &       3       &       6       &       1       &     192 \\
C               &                17     &        84       &      17       &       3       &       6       &       1       &     128 \\   
E               &                32     &       111       &      18       &       3       &       6       &       1       &     171 \\
NA              &                33     &       109       &      18       &       3       &       6       &       1       &     170 \\
NB              &                27     &       103       &      17       &       3       &       6       &       1       &     157 \\
\hline
\end{tabular*}

\label{tab:agnforecast}
\end{table*}

The array configurations that do not include large-size telescopes (LSTs), such as array ``C'', yield the poorest performance. They give access to significantly fewer FSRQs, which usually have very steep spectra, but the same is also true for BL Lacs. Array B, with the best coverage at low energies, yields the best results in terms of source statistics. The ``compromise'' solutions, such as configuration E, and the northern array NA, remain a good option.

The results from these two studies illustrate the remarkable capabilities of CTA compared to current IACT. The detection of at least 140 extragalactic sources is expected in less than two years with a maximum of 50 hours per source. Actually, the total number of Fermi AGN detectable in 50 hours would reach about 370 if the 2FGL BL Lacs without known redshifts were detected with the same proportion (namely one third for our two studies) than the ones with confirmed redshift \footnote{However, the percentage of detection could be smaller for BL Lacs without redshift, since they have a higher probability to be remote objects}. In fact, the main difficulty in elucidating the full BL Lac population lies in obtaining direct redshift measurements from their mostly featureless optical/UV spectra. A possible workaround might come from a direct measurement of the shape of the EBL, which would allow us to set an upper limit on unknown redshifts (see EBL contribution in this issue). The artificial break introduced here at 100 GeV for hard sources might also lead to an underestimation of the number of detections, but on the other hand, some of the sources might have intrinsic spectral breaks or cutoffs above the {\it Fermi} LAT energy range, which would reduce their signal in the VHE band. Only CTA will inform us on the actual spectra beyond 1 TeV for most of these sources. 
 
Discoveries such as extreme AGN or extragalactic sources not adapted to the {\it Fermi} LAT band and with emission peaking in the CTA energy range should further increase the sample of AGN seen by CTA. Given that currently 6 out of 45 TeV AGN have not yet been detected with the {\it Fermi} LAT \cite{Lott}, one can anticipate a fraction of about 15\% of missing CTA sources when making predictions based on the 2FGL catalog. All these estimates are in agreement with the lower limit predictions made from the blazar population analysis presented in part 3.1 (which includes Fermi and non-Fermi blazars, with or without known redshift). They prove that within its lifetime, CTA should provide significant samples of AGN of various types, suitable for statistical analyses of the VHE population.

\subsection{Radiogalaxies and extended sources}

Detection and monitoring of four radio-galaxies at TeV energies definitely proved that AGN other than blazars actually radiate at VHE. This is a key information which came out rather recently. Indeed, although quite rewarding by itself, detecting only blazars at VHE energies provides biased information about AGN jets, always seen at small viewing angles. Studying AGN with moderate or negligible Doppler boosting with CTA will provide access to a 2D view of AGN jets at VHE and will shed new light on our current understanding of radio-loud AGN. In particular, the performances of CTA should allow a detailed analysis of the relation between VHE and non-thermal radio emission and will contribute to the long-standing question of the origin of radio-loudness in AGN. 

The four radiogalaxies detected so far at VHE, namely M 87 \cite{AhM87,HegraM87}, Cen A \cite{CenAdisco}, IC 310 \cite{IC310_MAGIC,IC310_Neronov}, and NGC 1275 \cite{NGC1275_MAGIC}, have all been tentatively classified in the literature as Fanaroff-Riley type 1 radiosources, each with some "peculiarities". Apart from that, these active galaxies and their nucleus and jets show very different properties which, based on knowledge already gathered at lower energies, do not suggest any specific prominent common features, except being TeV sources. These VHE galaxies form an emerging class of AGN, which should find its place within AGN grand unification. This simple fact illustrates that observing at VHE probes new aspects of AGN not yet explored at any other energies, and offers a fully independent tool of investigation. One possibility would be that the VHE band directly catches the emission from the base of an inner beam or jet when it has a right orientation, independently of any other properties and classification of the radiosource at others frequencies. 

This emphasizes the strong interest of studying such types of sources with CTA, but precludes at the moment any convincing prediction of a sample to be detected at VHE in the future. In particular, in the current small sample of VHE radiogalaxies, the average detected TeV fluxes do not seem to be related to the average non-thermal radio and X-ray fluxes of the sources. One common trend is that the four sources are located in rich environments, and show sign of galaxy interaction or mergers. Both M 87 and NGC 1275 are dominant cluster galaxies with a very massive central black hole. Cen A is presumably a recent merger, in a group of galaxies, and IC 310 is located in the Perseus cluster where radio jets can interact with the intracluster gas. Other common properties could be to have an intermediate viewing angle (about 20 degrees for M 87, 40 for Cen A, between 20 and 50 for NGC1275 and $\le 38$ for IC 310), weak or moderate Doppler beaming, and no direct alignment between their radio compact VLBI core and their extended radio structures (grossly misaligned  by $\sim 70$ degrees in M 87 and $\sim 45$ in Cen A), possibly enhanced by projection effects and inhomogeneous external medium, or related to specific properties of their central engines. Moreover, the detection of some transient BL Lac-type phenomena has been reported or discussed in the literature for the four  radiogalaxies discovered up to now at VHE. Such properties should help to identify promising candidates for further observation with CTA. 

Among this new class of AGN, M 87, the first non-blazar detected in the VHE range, has been the most studied in the literature. It questions our global understanding of TeV emission scenarios for AGN, and to some extent our general view of AGN classification. Even assuming a high Doppler factor, the TeV variability of M 87 requires very small emitting zones, of the order of a few Schwarzschild radii $R_{\rm s}$ of its $3 \times 10^9$ M$_{\rm \odot}$ black hole, under causality argument \cite{AhM87,Ac,AlM87,M87_campaign}. This raises the critical question of particle acceleration mechanisms in such small regions, and excludes the Virgo cluster, the radio lobes, the host galaxy, the large scale jet and its brightest knot A as dominant TeV emission zones. Three different main emitting zones have been considered for M 87, (i) the peculiar knot HST-1 located at about 65 {\rm pc} from the nucleus, (ii) the inner VLBI jet, and (iii) the central core itself, namely the accretion disk or the inner black hole magnetosphere. 

The coordinated campaign in 2005 found a possible correlation between a VHE flare and an X-ray outburst of HST-1. However, another multiwavelength campaign organized in 2008 concluded that the X-ray light curve of HST-1 obtained by Chandra does not follow the VHE one. Conversely, the radio and X-ray emissions from the core are correlated with the VHE flux. In radio, regular monitoring of M 87 by the VLBA at 43 GHz also allows one to explore the sub-{\rm mas} scale in order to probe the jet formation and collimation zone at about $100 R_{\rm s}$ from the black hole, with an outstanding angular resolution of 0.21 {\rm mas} x 0.43 {\rm mas} ($0.5 {\rm mas} \sim 0.04 {\rm pc} \sim 140 R_{\rm s}$ at the distance of M 87). The VLBA detected a significant rise of the flux of the radio core at the time of the 2008 VHE activity, together with enhanced emission along the VLBI jet.  These results favour scenarios where most of the VHE emission comes from the inner VLBI jet (multi-zone models inspired from standard blazar scenarios \cite{Lenain08,Tav08,Giannios10}) or from the central core (particle acceleration in the black hole magnetosphere \cite{Neronov07,Rieger08,Levinson11}). However the situation still remains unclear, and suggests the existence of different types of VHE flares, as discussed in a recent work reporting on the 2010 large joint monitoring campaign \cite{AbraM87_2011}. 

Current magnetospheric models for M 87 \cite{Neronov07,Rieger08,Levinson11} imply that minimum variability timescale should always be larger than a few Schwarzschild light crossing times $R_{\rm s}/c \sim (0.2 - 0.4)$ days (in the absence of strong Doppler effect) and that the TeV spectra should exhibit a clear break, due to internal $\gamma \gamma$ absorption and maximum energy constraints, well below 50 TeV. Confirming or rejecting such magnetospheric scenarios with "particle acceleration close to the SMBH" by future high-sensitivity observations with CTA would boost our understanding of AGN central engines: in the former case, new jet physics beyond current developments would be demanded, in the latter case, a strong link between jet formation, particle acceleration and disk physics would be established. 

\begin{figure}[h!]
\begin{center}
\includegraphics[width=0.45\textwidth]{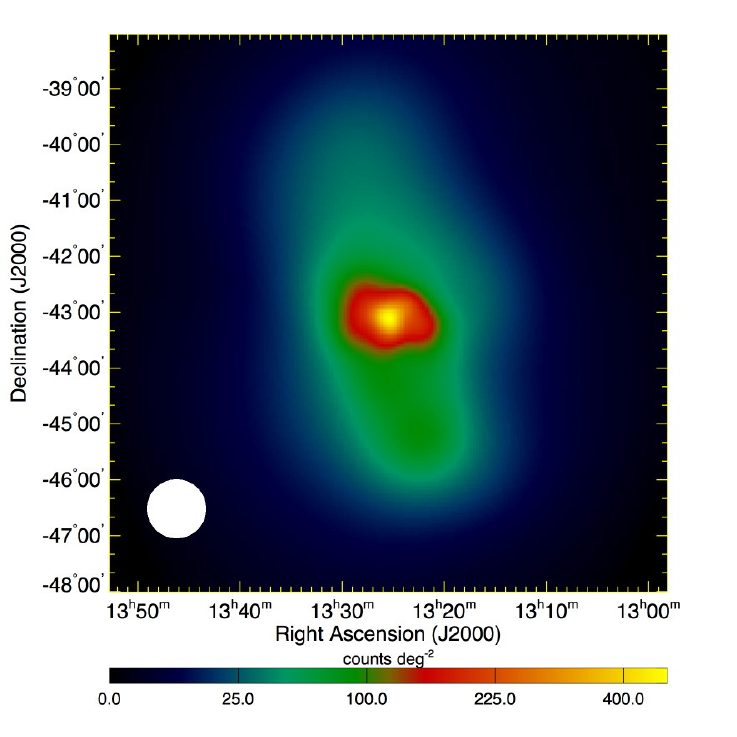}
\caption{The diffuse gamma-ray emission detected by Fermi from the giant lobes of the radiogalaxy Cen A \cite{CenAdiffuse}. Depending on the actual VHE spatial distribution, and the sensitivity and angular resolution performance of CTA, the structure of the extended VHE emission on the {\rm kpc} scale in the central part of the galaxy can be probed (here the white circle corresponds to the LAT PSF of 1 degree). 
\label{fig:CenA_Fermi}}
\end{center}
\end{figure}

\begin{figure}[h!]
\begin{center}
\includegraphics[width=0.45\textwidth]{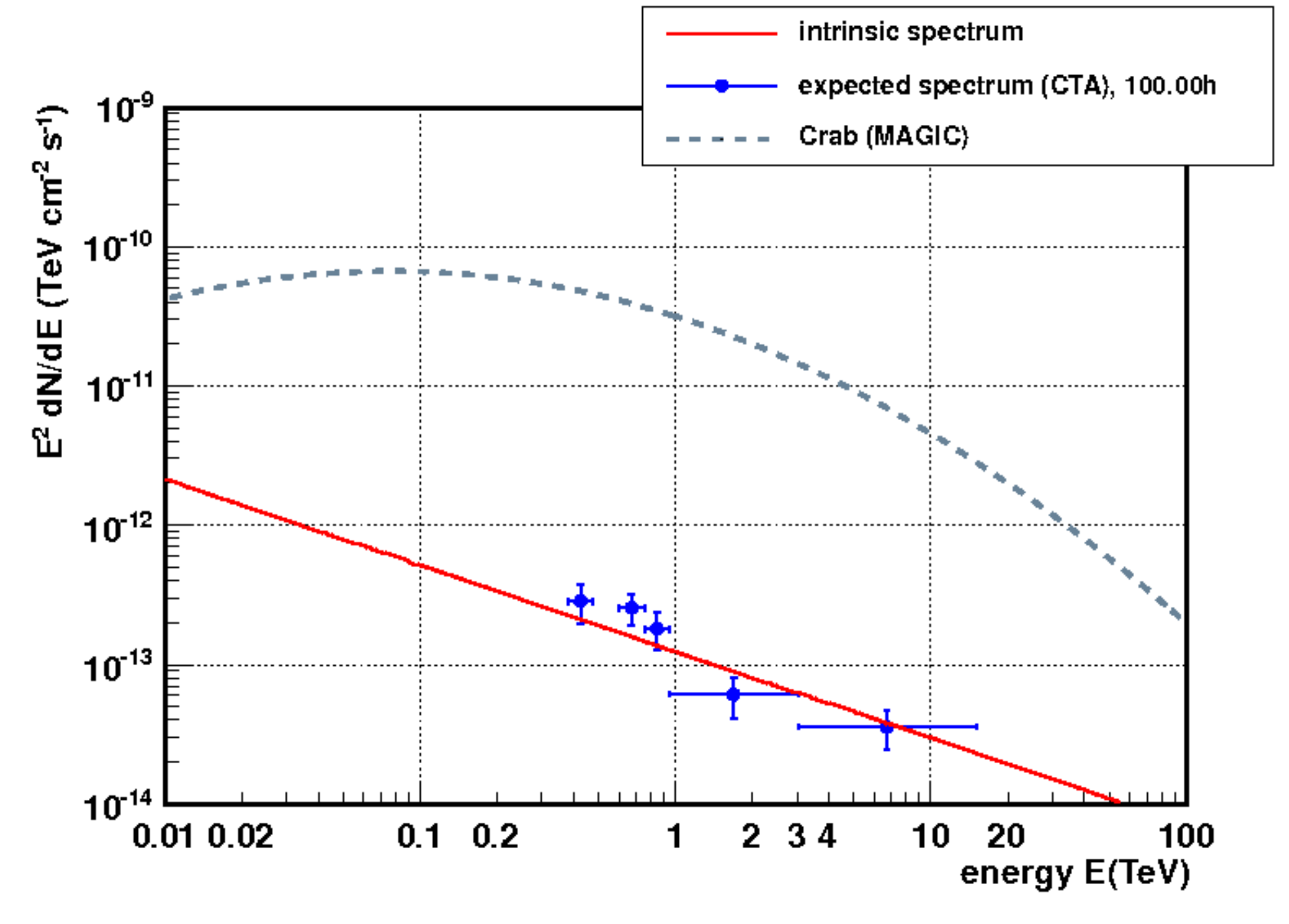}
\caption{CTA detection of extended extragalactic VHE emission: under certain conditions, a halo of extended TeV emission around M 87 could be within reach of CTA. The figure shows the expected spectrum from the CTA array configuration I, for 100 hours of observing time, assuming an extended emission of 0.2 degrees with a flux of 50\% of the total flux detected during low states of M 87 (see text). 
\label{fig:halo_M87}}
\end{center}
\end{figure}

In the second radiogalaxy discovered at VHE, the nearby source Cen A, the origin of the dominant VHE signal is even less clear than in M 87, as both the radio core and the {\rm kpc} jets of Cen A are within error bars of its position on the sky. This results in many possible emitting zones such as the black hole magnetosphere, the base of the jet, the large scale jets and inner lobes, or even a pair halo in the host galaxy. A better accuracy on the absolute astrometry at VHE expected with CTA and obtaining high quality light curves with good temporal coverage  should clarify this decisive issue, and at least distinguish between a dominant core emission or a dominant extended component related to the {\rm kpc} jets. Indeed, very extended diffuse gamma-ray emission has been recently found by Fermi at lower energies \cite{CenAdiffuse} (see Fig.\ref{fig:CenA_Fermi}). The gamma-ray emission above 100 MeV coming from the giant lobes can be described by EIC models on the cosmic microwave background and the EBL. This emission is quite important, with a total flux slightly higher than the one from the core \cite{CenAcoreFermi} and a power comparable to the kinetic power required in the jets. The high sensitivity and improved angular resolution of CTA should allow one to look for and possibly map any extended structure at higher energies, thereby opening a completely new view on VHE particle acceleration, transfer and radiative losses. A simple extrapolation of the Fermi halo to the VHE band shows that its VHE counterpart remains out of reach by CTA, assuming the same spatial extension of about 2 degrees. However, VHE detection can be expected from regions of enhanced gamma-ray emission in shocks, knots, or hot spots, which could be identified with the angular resolution of CTA. Conversely, the detection of an extended VHE halo around M 87 would be very challenging but could be possible with CTA (see Fig.\ref{fig:halo_M87}), assuming, as seen by Fermi in Cen A, a flux in the lobes of 50\% the flux detected in the TeV range during a low state of the source \cite{AhM87} for an extension of 0.2 degrees corresponding to the extension of radio maps.

In contrast to the three other VHE radiogalaxies, IC 310 was initially not recognized as having a specifically remarkable non-thermal activity. Its serendipitous discovery at VHE in the field of NGC 1275 emphasized our current poor knowledge on VHE populations. Only recently it appeared in the Fermi catalog and was then identified as a potential TeV source. Two emission zones can be considered, the central engine and the inner jet as commonly described for TeV BL Lacs, or the bow shock created by interaction of the fast moving host galaxy with the intracluster gas \cite{IC310_Neronov}. Astrometric and angular resolution capabilities of CTA should distinguish between them. However, the first option appears favored because of the detection of few-dayscale variability. Indeed, IC 310 was already mentioned in the literature as a FR I source which may have a non-thermal activity related to the BL Lac phenomena, but at weaker levels than characterized by the standard definition of BL Lac objects \cite{Owen96}. Moreover recent VLBI data show a blazar-like one-sided core-jet structure at intermediate angle to the line of sight \cite{Kadler}. VHE instruments are therefore possibly on the way to solve the long standing problem of the "missing BL Lac" and to firmly identify the still elusive transition population between beamed BL Lacs and unbeamed FR I galaxies \cite{Rector99}, a difficulty of the standard unification scheme which proposes that BL Lac are FR I radiogalaxies seen along their jet axis. Surveys at VHE could have the capability to recognize a population of low luminosity or misdirected BL Lacs, difficult to identify at lower energies, and thus "bridge the gap" between genuine BL Lacs and FRI radiogalaxies. It will be interesting to further investigate such a view in the context of recent blazar classification scenarios \cite{Giommi2012}. 

Generally speaking, nearby radio galaxies offer the opportunity of unique studies of extreme acceleration processes in relativistic jets and in the vicinity of supermassive black holes. Given the proximity of the sources and the larger jet angle to the line of sight compared to BL Lac objects, the outer and inner {\rm kpc} jet structures are potentially resolvable by CTA, enabling us to look for possible VHE radiation from large scale jets and hot spots besides the central core and VLBI jet, and to spatially pin down the main site of the emission. With the help of simultaneous multiwavelength observations and temporal correlation studies, different sections of the jet and the core can be probed, down to the smallest {\rm pc} (milliarcsecond) scale, only accessible to VLBI radio observations or timing analysis. Further studies of variability with CTA will strengthen the limits on the size of the emission region and clarify the correlations with other wavelengths. Long-term monitoring and the search for intra-night variability would be two major goals to constrain the physics and start characterizing this new population of sources. Remembering the basic classification of extragalactic radio sources, one could consider highly variable VHE radiogalaxies as likely core-dominated gamma-ray sources, and poorly variable ones as possibly lobe-dominated gamma-ray sources. This VHE population is still lacking a standard unifying model and deserves further analysis. 

Recent and fast developments on VHE radiogalaxies show that present VHE instruments start to provide an original view of non-thermal activity from  central AGN engines and inner jets, with the capability to directly probe a very specific region, still not fully identified and unreachable by other means, in the close vicinity of SMBH, such as for M 87. The next generation of IACT will explore this still missing link between SMBH magnetospheres and the physics of jets and extended radiosources. One can also anticipate that it could provide decisive constraints on the fundamental question of the total energy budget of some non-thermal sources where the contribution of the extended gamma-ray emission appears quite significant, such as for Cen~A.

\subsection{Seyfert galaxies}

There is a growing evidence that relativistic jets are not only seen in blazars and radio galaxies but in several types of Seyfert galaxies as well. About 5\% of narrow-line Seyfert 1 (NLS1) galaxies are radio-loud (RL) \cite{komossa}, and show flat spectra together with variability in the radio band, suggesting the presence of relativistic jets. This hypothesis has recently been confirmed by the detection of a small number of RL-NLS1s with {\em Fermi}-LAT \cite{fermi-rl-nls1}. The measured GeV spectra are typically steep, with $\Gamma = 2.5-2.8$, which makes the detection of RL-NLS1s with CTA challenging. However, at least two RL-NLS1s (PMN~J0948+022 and SBS~0846+513) have shown significant variability \cite{0948-flare,donato-perkins}, with gamma-ray luminosities that would make them detectable with CTA during gamma-ray outbursts. 

The broadband emission of RL-NLS1 galaxies appears similar to that of blazars, being explained by SSC and EIC scenarios. Thermal emission from the disk, broad-line region (BLR) and the infrared torus dominate the IR/optical output, and provide seed photons for EIC scattering \cite{fermi-rl-nls1}. Even though the radiative properties of the jets in RL-NLS1 galaxies look similar to that of blazars, they originate in different host galaxies. Blazars are hosted in elliptical galaxies, while RL-NLS1s are likely in spirals. In addition, gamma-ray emitting RL-NLS1s have inferred SMBH masses typically 1--2 orders of magnitude smaller than blazars, while their accretion rate reaches extreme values, up to 80\% of the Eddington rate, which have never been found in gamma-ray loud AGNs but are usual for NLS1s \cite{Foschini}. Detecting RL-NLS1s with CTA would significantly increase the range of parameters of the accretion-ejection process explored at VHE. 

Two Seyfert 2 galaxies have also been detected by {\em Fermi}-LAT, namely NGC~1068 and NGC~4945 \cite{lenain-s2}. Both galaxies exhibit AGN and starburst activity \cite{lester,iwasawa}, and a simple extrapolation of their GeV spectra yields VHE flux levels detectable with CTA. Given their similarity with the known gamma-ray emitting starburst galaxies, it seems likely that their GeV emission originates in the interstellar medium of the galaxy. However, \cite{lenain-s2} find evidence that the AGN component could be dominating in NGC~1068. A detection of a significant spectral steepening or flux variability in the VHE range would point towards AGN emission, disentangling it from the steady starburst component. 

Classical radio-quiet Seyfert galaxies could also emit gamma-rays originating from starburst or AGN activity. Classical Seyferts also show jet-like structures, but with flows that are typically slow, weak, and poorly collimated when compared to the relativistic jets of radio galaxies and blazars. So far, only two possible associations between GeV sources and radio-quiet Seyfert galaxies have been found (ESO~323--G077 and NGC~6814) but chance spatial coincidences with these objects cannot be ruled out \cite{fermi-sy}.

A detailed characterization of the gamma-ray emission from the different classes of Seyfert galaxies will test our knowledge of jet launching mechanisms in AGNs with SMBH masses and accretion rates very different from the well-studied gamma-ray loud blazars. This will shed light on fundamental questions in AGN physics like formation and propagation of jets, the physical cause of radio-loud/radio-quiet distinction, the fundamental parameters governing the central engine, and how the host galaxy influences the active nucleus.

\subsection{Low-luminosity AGN, supermassive black holes, and the Galactic Center}

Applying standard SSC scenarios to other low luminosity AGN (LLAGN) such as NGC 4278 shows that they may be detected by CTA \cite{Takami11}. Furthermore, scenarios developed for M 87 show that magnetospheres of rotating SMBH can easily generate VHE particles and radiation, as long as the accretion disk remains under-luminous to avoid strong internal absorption by the ambient radiation. This raises a very general question. Is VHE emission a generic feature of AGN and SMBH?  Indeed, many normal galaxies harbor SMBH in their nuclei and the search of some VHE signal from them will be a challenging topic for CTA which could shed new light on the crucial question of missing SMBH, AGN evolution and feedback between AGN and host-galaxies. Spheroidal systems such as elliptical or lenticular galaxies, and bulges of early-type spiral galaxies, are believed to host SMBH with masses between $10^6$ and $10^9$ solar masses. During the early stages of galaxy evolution, the SMBH accrete matter at high rates and are observed as bright QSOs. The radiative output in the optical decays from redshift $z > 3$ to $z = 0$ by almost 2 orders of magnitude. The majority of SMBH in the local universe are hosted in those evolved systems with low accretion rate, not embedded in dense radiation fields. This enables high energy and VHE $\gamma$-rays, if generated, to escape from the nuclear region without suffering from strong absorption via photon-photon pair absorption. A good candidate for such detection could be, for instance, NGC 1399, as discussed in detail in \cite{Rieger11,Peda11}. Any positive signal of this kind would be a real breakthrough for AGN and SMBH physics as well as for co-evolution schemes of SMBH and galaxies. It could also raise important issues related to the origin of extragalactic cosmic rays, and to the still debated results from the Pierre Auger Observatory (PAO) \cite{AUGER}, and offer an explanation for the correlation claimed by the PAO collaboration between the arrival directions of the most energetic cosmic rays and the spatial distribution of low redshift AGN, which mainly involves apparently weak AGN. Further statistical analysis and monitoring of AGN samples at very high energies will tackle these still open questions. 

One remarkable case of weak AGN is the Galactic Center. Indeed, TeV emission has been detected from the central region of our Galaxy, although it is not yet known whether it comes from the SMBH Sgr A* itself or from the pulsar wind nebula G359.95-0.04 at 8.7 arcseconds, still inside the spatial error bars of present IACT, but likely resolvable with CTA. Moreover, an interesting strategy to fix the position of the VHE emission at the milliarcsecond scale is to look for pair-production VHE eclipses, when stars orbiting the central SMBH approach the line of sight. Such phenomena can be very well described since trajectories are precisely known for several stars. Light curves can be predicted and appear within reach of CTA if the VHE emission zone is really compact \cite{Abra10}. Any positive detection of such new-type of eclipse in the next decades would allow to reconstruct for the first time the position of a VHE source with a tremendous milliarcsecond accuracy. 

If CTA concludes that the VHE source in the Galactic Center region is actually associated with the central SMBH Sgr A*, one could then try to extrapolate its characteristics to the galactic nucleus of other nearby galaxies, such as M31 (Andromeda). Although its distance (770 {\rm kpc}) is about 100 times larger than that of Sgr A*, its nuclear X-ray luminosity ($L_{\rm X} \sim 10^{36}$ erg/s \cite{Garcia05}) is about 1000 times higher ($L_{\rm X} \sim 10^{33}$ erg/s for Sgr A* \cite{Baga03}). Therefore, crudely assuming $L_{\rm VHE} \propto L_{\rm X}$, one can expect a VHE flux from M31 of about 10\% of the one of Sgr A*, which would be within reach by CTA. This is an additional clue for future possibilities of studying weak SMBH in nearby galaxies. 

Transient phenomena may also offer new signatures of SMBH, like the recent event (Swift J164449.3+573451) discovered by Swift in X-rays which may reveal a Tidal Disruption Event \cite{Bloom2011,Burrows2011}. Indeed, the tidal disruption of a star by the otherwise "dormant" SMBH of the compact galaxy hosting this burst may have activated the formation of a beamed jet. Depending on the synchrotron or Inverse-Compton origin of the X-ray emission, high energy gamma-rays could be produced \cite{Aliu11}. VHE instruments should therefore contribute to constraining such phenomena that are still basically unexplored, which may bring to light the basic building block of the accretion-ejection cycle around SMBH.

\section{Particle acceleration and emission models}

Non uniform velocity fields in collisionless plasmas allow a variety of particle acceleration processes where the density in high energy particles can grow until it affects the flow itself and the acceleration mechanism. Indeed, most of the VHE sources detected so far harbor powerful flows, especially AGN which appear as perfect laboratories to study relativistic plasmas under extreme conditions inaccessible to experiments. Fermi acceleration processes of the 1st and 2nd order in shocks and turbulence, respectively, have been extensively explored in the literature and are clearly expected in AGN \cite{Melrose2009,Kotera2011,Spit2008,Sironi2009,Sironi2011,Summer2011,Nishi2005,Nishi2009}, as well as magnetic reconnection phenomena \cite{Roma1992,Zweibel2009}, and possibly direct electric fields in gaps or centrifugal forces around rotating SMBH \cite{Rieger11,BZ77,Beskin2000,Rieger2000,Rieger2008,Istomin2009}. There is a growing interest especially in relativistic reconnection in recent years since intermittent reconnection events in sheets or small magnetic islands in jets or in the SMBH vicinity could be recognized as efficient particles accelerators and nicely match AGN non-thermal emission properties, especially fast VHE variability \cite{Bete,Drake2006,Lyub2008,Giannos1,Giannos2,Lyub2010,Nale2010,Elisabete2010,Cerutti2011,Kowal2011,Elisabete2011,Elisabete2012}.  Together with cosmic ray and neutrino experiments, CTA is among the most promising projects for detailed studies of these extreme cosmic accelerators.   

VHE observations provide an uncluttered view of the physics by pointing out only the most extreme phenomena at the highest energies and by  disentangling non-thermal effects from thermal ones radiating at longer wavelengths. They should help identifying dominant processes in the global energetics of the source. Powerful phenomena inducing VHE emission are potentially the sources of several other secondary events observed at longer wavelengths, the explanation of which possibly lies in the primary VHE data. CTA will therefore offer an invaluable tool for exploring the violent mechanisms at work in SMBH environment, especially on the jet physics, including formation, collimation, and propagation as well as the development of shocks and turbulence along the jet. Depending on the precise origin of the VHE radiation, which is still to be settled, CTA data could also constrain properties of the SMBH and the accretion regime. 

As presented in part 2, several scenarios have been proposed to explain the TeV emission of blazars but none of them is yet fully self-consistent. In the absence of a definitely convincing global picture, a first goal for CTA will be to constrain model-dependent parameters in a given scenario. For instance, basic stationary one-zone SSC scenarios have typically eight main free parameters, three macroscopic ones for the size, the magnetic field, and the Doppler factor of the emitting zone, and five others to describe the microphysics of the emitting particle distribution (density factor, slopes of a broken power-law, maximal individual Lorentz factor, and Lorentz factor at the break). High quality VHE spectra and detailed simultaneous SED are needed during quiescent as well as active states to get a similar number of observational constraints (as frequencies and fluxes at the synchrotron and the inverse-Compton peaks, break frequencies, and spectral indices). This will be achievable thanks to the large spectral range, high sensitivity and high spectral resolution of CTA, together with the coordination of multiwavelength campaigns. In such a way, the physics of basic radiation models will be well constrained by CTA, characterizing the properties of the particle distribution from its emission at VHE, which should help to identify the acceleration mechanisms at work, and decide which part of the models should be corroborated or ruled out. 

The second most difficult goal will be to distinguish between the different remaining options and to firmly identify the dominant acceleration and radiation mechanisms. Detection of any specific spectral features, break, cut-off, absorption or additional components which could be radiative signatures of specific acceleration processes, would be mandatory in this regard. Observations above several TeV will be an important test for the most popular SSC scenarios which may have difficulties to generate hard enough spectra because of the limitation due to the Klein-Nishina regime. This could rule out simple one-zone models by requiring alternate photon sources from, for example, a structured jet. Such intense research will benefit from improved performances of current particle-in-cell (PIC) codes. Probing AGN variability at all time scales, down to the shortest ones at the few-seconds scale with CTA, will significantly constrain acceleration and cooling times, instability growth rates, shocks and turbulence time evolution. Deep analysis of the light curves should explore all timescales from several years down to several seconds, look for periodic and quasi-periodic behaviour, and characterize the additive or multiplicative processes trying to distinguish macrophysics from microphysics effects at the origin of the variability. Periodicities of about 10 to 50 minutes, related to the last stable orbit around SMBH of $10^7$ solar masses, should be reachable with a good temporal coverage with CTA even during low activity states (see section 5). The role of CTA as a time explorer will be decisive for constraining both radiative phenomena, and global geometry and dynamics of the AGN engine.

\subsection{Testing and constraining leptonic scenarios with CTA}

In this section, we illustrate some aspects of leptonic scenarios that the unique capabilities of CTA should be able to characterize, confirm or invalidate. Leptonic models assume the emission of relativistic electrons with energy $E=\gamma m_{\rm e} c^2$ and individual Lorentz factors up to about $10^6$. First order Fermi acceleration processes in shocks can for instance convert about $10 \%$ of the bulk jet kinetic energy into random energy of fast  particles and provide a power-law particle energy distribution $N(\gamma) \propto \gamma^{-n}$ required to explain the observed power-law spectra of blazars, with a synchrotron flux $F_{\rm syn} \propto \nu^{-\alpha}$, with spectral index $\alpha=(n-1)/2$. These electrons can up-scatter their own synchrotron photons (SSC) or other externally produced low-frequency photons (EIC), increasing their energy by several orders of magnitude and inducing strong VHE radiation. 

\subsubsection{Peak frequency and emission level correlation in SSC models}

A noticeable characteristic of SSC scenarios resides in the correlation which should be found in the evolution of the high energy IC peak. Indeed, such types of correlation have been already found at lower energies for the synchrotron bump, as shown for instance by a detailed analysis of the UV and X-ray spectra of the well-known BL Lac Mrk 421 \cite{Tanihata04} which display evidence for a clear correlation between the position of the synchrotron peak 
($E_{\rm peak}$) and the emission level at the peak $E_{\rm peak} \propto 
\nu F^{0.77 \pm 0.02}_{\rm peak}$. However, such effects remain out of reach for many sources at VHE  with current IACT 
because their study requires precise monitoring of the spectrum
over a wide range of energies. This is especially difficult in the case of high 
energy peaked sources where the IC peak is located close to the high energy threshold
of orbital observatories (e.g. CGRO, Fermi) and low energy threshold of 
ground based Cherenkov instruments (e.g. H.E.S.S., MAGIC, Veritas). 

CTA with its improved low 
energy threshold and high sensitivity will offer the unique opportunity for detailed 
investigations of the IC peak evolution of bright VHE blazars such as Mrk 421, Mrk 501 and PKS 2155-304. Detecting the peak of the VHE bumps and monitoring its temporal evolution on short timescales for a small sample of blazars will be a real benchmark to constrain emission models, which can be achieved by CTA. Improved CTA performances, or the hypothetical observation of an exceptionally bright event, could extend such analysis to the family of radiogalaxies, which would be a significant advance.   

\begin{figure}[h!]
\begin{center}
\includegraphics[width=0.45\textwidth]{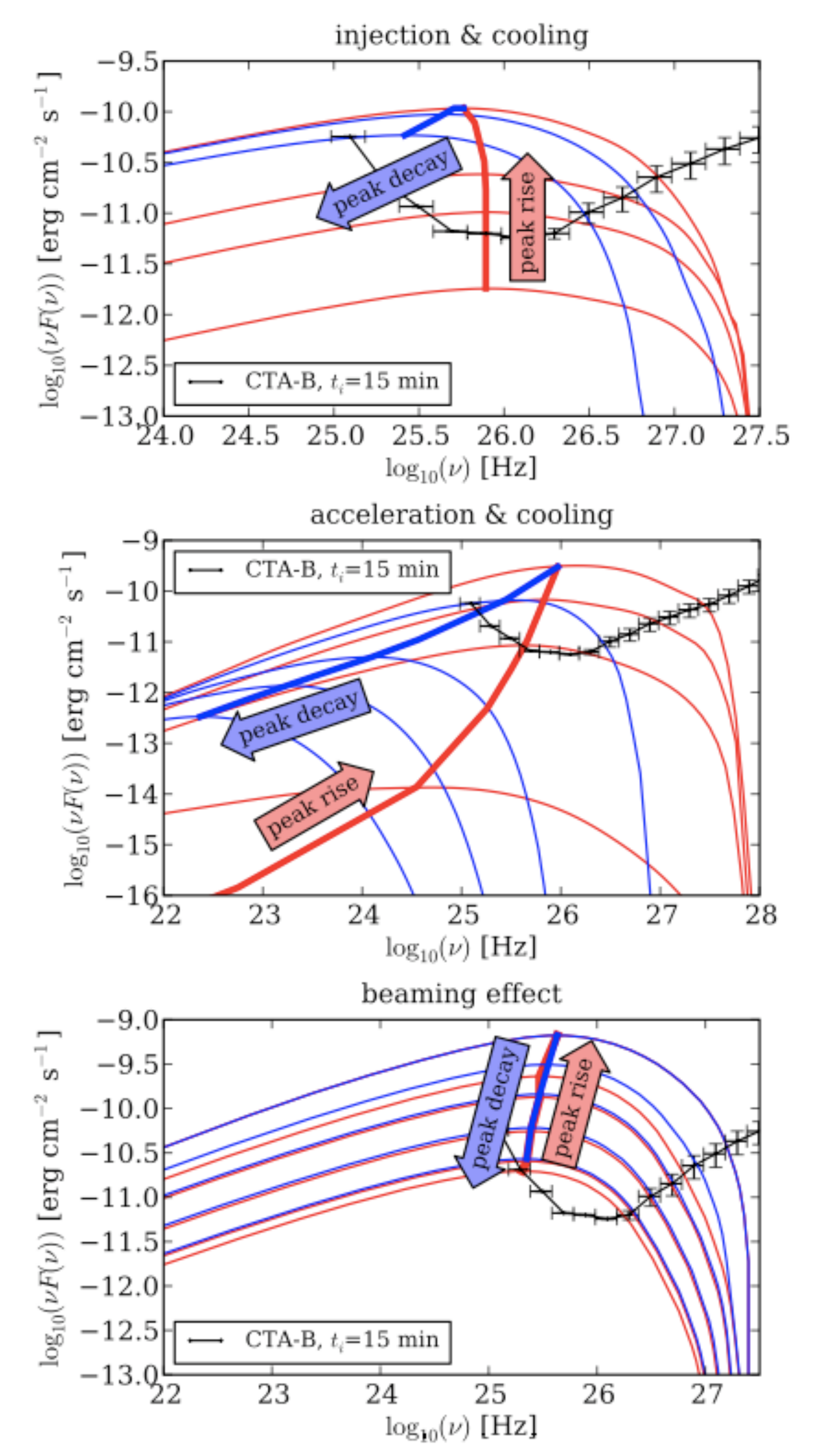}
\caption{Examples of the IC bump and peak evolution for various time-dependent SSC scenarios. Spectra during rise (resp., decay) time are shown at  different moments in red (resp., blue) thin lines. The IC peaks of the spectra are connected with red and blue bold lines which show a shape characteristic of the different scenarios, namely a `"1" like shape characteristic of an "injection \& cooling" scenario (top panel), a "wing" like shape line drawn by the IC peaks for an "acceleration \& cooling" scenario (middle panel) and a "single line" for a scenario with a simple change of the Doppler factor (bottom panel). The three examples given here represent typical bright VHE flares expected from nearby TeV BL Lac sources. The middle panel specifically reproduces the rise and decay of a Mrk 501 flare \cite{Djannati99} while the bottom one describes an active event of Mrk 421 over a few hours \cite{Fossati08}. Bold black lines show the CTA sensitivity curve for 15 minutes of integration time with the B array. Only the high sensitivity of CTA provides temporal resolution good enough to follow such fast sub-hour evolution. 
\label{fig:CTAPaper_all}}
\end{center}
\end{figure}

Indeed, high quality data on the evolution of the synchrotron and IC peaks can drastically constrain SSC modelling. As shown in Fig.\ref{fig:CTAPaper_all}, the short timescale evolution of the IC peak during flares is directly related to the physical origin of the variable event and can characterize it. In an "injection \& cooling" scenario, the evolution of the particle energy spectrum ($N(\gamma, t)$) can be described by the kinetic equation
\begin{equation}
\frac{\partial N(\gamma,t)}{\partial t} - \frac{\partial }{\partial \gamma} 
\left[ \left\{ \dot{\gamma}_{\rm syn} + \dot{\gamma}_{\rm IC}\right\} N(\gamma,t) \right] = Q(\gamma, t)
\end{equation}
where $ \dot{\gamma}_{\rm syn}$ and $\dot{\gamma}_{\rm IC}$ are the synchrotron and IC cooling rates respectively. Here non-radiative losses and escape are neglected. The particle acceleration is a free term, just described by a given injection rate $Q(\gamma,t)$. In such a case, the density of high energy particles significantly increases inside the source during the injection phase, with a quick increase of the observed emission. When the acceleration becomes inefficient and the injection stops, the radiative cooling takes control over the particle energy evolution and reduces the number of high energy particles. The two processes, "injection \& cooling",  can explain observed flares giving characteristic ``1'' like shape for the peak evolution, as shown in Fig. \ref{fig:CTAPaper_all} (top panel). This model predicts that the correlation should be different for the rise and decay of a flare. This can characterize some AGN events, but not all of them, remembering the case of Mrk 421 where the correlation measured for the synchrotron bump \cite{Tanihata04} suggests that they are similar.

\begin{figure}[h!]
\begin{center}
\includegraphics[width=0.45\textwidth]{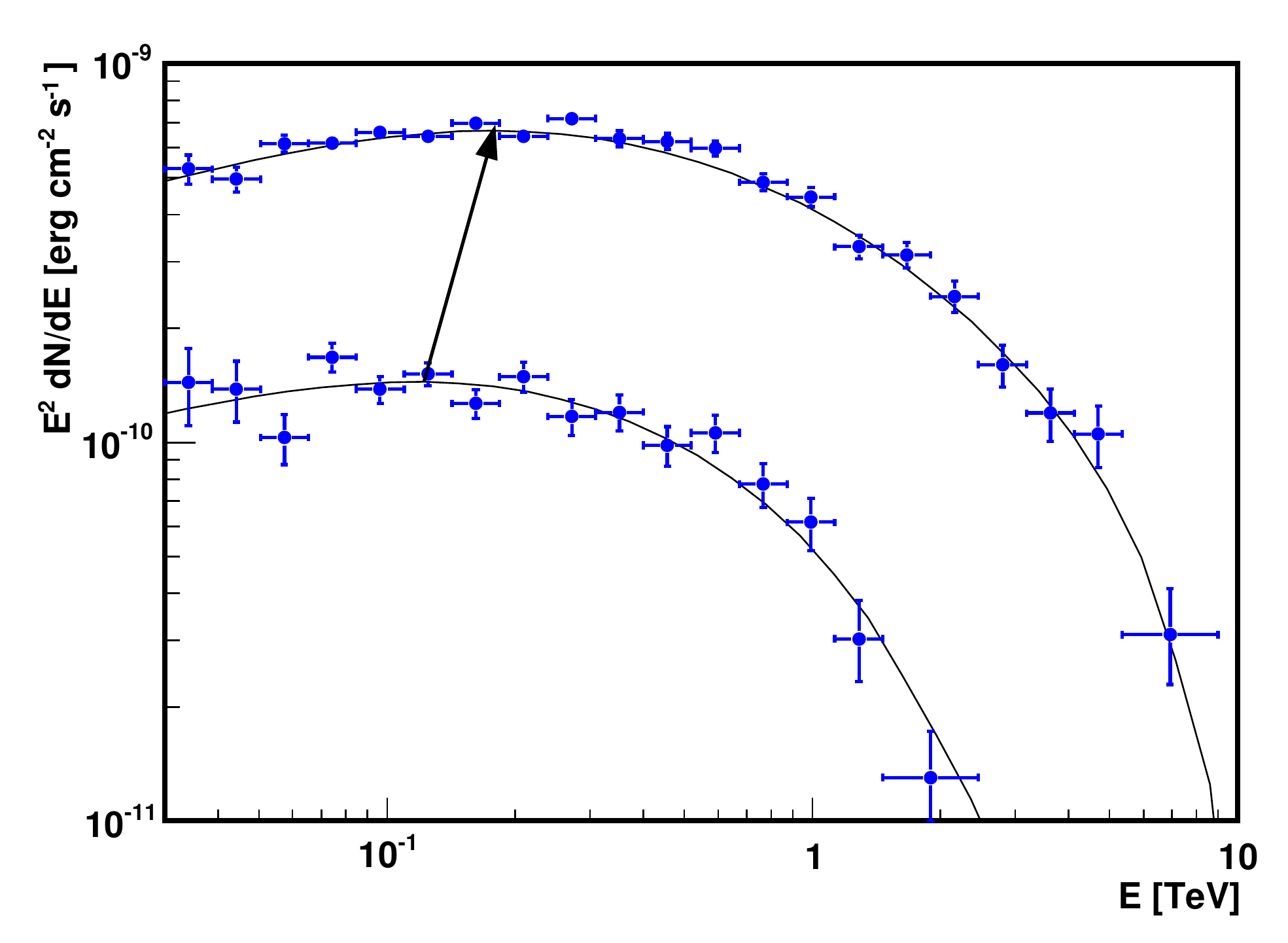}
\caption{Simulations of spectra obtained in 15 minutes by CTA (B array) during the rise of the flare for the case of the simple change of Doppler factor illustrated in Fig.\ref{fig:CTAPaper_all} (bottom panel). This illustrates the decisive importance of having good sensitivity and spectral resolution over a large range in frequencies, especially at low energies, in order to allow a good determination of the peak motion during the flare. 
\label{fig:SSC_simul}}
\end{center}
\end{figure}

In an "acceleration \& cooling" scenario, one can assume that the particles are systematically 
accelerated inside the source, following a kinetic equation of the form
\begin{eqnarray}
\frac{\partial N(\gamma,t)}{\partial t} + \frac{\partial }{\partial  
\gamma}
\left[\left\{ A(\gamma, t) - (\dot{\gamma}_{\rm syn} +  
\dot{\gamma}_{\rm IC})\right\} N(\gamma,t)  \right.& & \nonumber \\
\left. + D(\gamma, t) \frac{\partial N(\gamma,t)}{\partial \gamma} \right]  &  
= &  0,
\end{eqnarray}
where two additional terms describe acceleration $(A(\gamma,t))$ and stochastic diffusion $(D(\gamma,t))$ in the energy domain \cite{Katar_2006}. The maximum energy particles can reach depends on the efficiency of acceleration versus radiative cooling and on the duration of the acceleration process. When acceleration becomes inefficient, cooling starts reducing the particle's energy. The evolution of the particle energy spectrum then draws a "wing" like shape for the IC peak path, as shown in Fig. \ref{fig:CTAPaper_all} (middle panel). In this case as well the correlation should be different for the rise and the decay, except for a fine tuning of the parameters.  

In the alternative case where the variable event is simply due to a change of the source Doppler factor, and not to the evolution of the particle energy distribution, the same peak correlation occurs for the rise and decay of a flare, as shown in Fig. \ref{fig:CTAPaper_all} (bottom panel). Such scenario can be relevant when the emitting zone travels on slightly curved trajectory and its apparent Doppler factor $\delta$ significantly varies for different values of the viewing angle $\theta$. Since the observed flux is proportional to a high power of the Doppler factor a small change of $\theta$ (e.g., from 5 to 3 degrees) induces a significant increase of the observed emission \cite{Katarzynski10}.  Current available data do not constrain well the evolution of the VHE bump since the IC peak cannot be monitored yet, being below the threshold of current IACT and difficult to measure on short time scale by Fermi. As clearly shown in Fig. \ref{fig:SSC_simul}, the performances aimed with CTA should allow for the first time to follow these behaviors.

\subsubsection{TeV and X-ray emission correlation}

One strong piece of evidence in favour of SSC models is the correlation often found between the evolution of X-ray and the VHE flares. This is not necessarily the case for EIC scenarios where there may be a competition between the two types of radiative losses, synchrotron and external Inverse-Compton, which can even result in an anticorrelated behavior. However some hadronic models, such as the synchro-proton scenario where protons and electrons are simultaneously accelerated, can also explain such a correlation, though this deserves a more detailed investigation. 

This correlation is not universal, as definite orphan VHE flares have been observed, with no X-ray counterpart (and conversely). Light curves in X-rays and gamma-rays can sometimes look very different, as shown for instance by Fermi for 3C279 \cite{Hayashida2012}. In this source, comptonization of external radiation from the dusty torus and the broad line region should be taken into account for the generation of gamma-rays, and X-rays seem to have a different origin than in HBL, the steady X-ray emission being possibly an inverse Compton component corresponding to a synchrotron bump peaking in the mm/sub-mm range. Intensive and long term coordinated monitoring in VHE and lower frequencies should throw more light on these questions, studying the variation of the SED as a function of source activity \cite{Mankuz1}. Pursuing X-ray campaigns will be essential in this regard, and will contribute to the characterization of the emitting particle distribution.

\subsubsection{Hard intrinsic spectra}

Despite the present uncertainties on the EBL values, some AGN observations suggest that intrinsic TeV spectra could be unexpectedly hard, with intrinsic energy spectral index $\alpha_{\rm em}$ equal to 0.5 or even smaller for mid or high levels of the EBL \cite{Aharonian06}.

For instance, the observations of 1ES 1101-232 made by H.E.S.S. show a TeV spectrum that can be approximated by a power-law 
function with spectral index $\alpha_{\rm obs} = 1.88 \pm0.17$. The emission of 
this distant ($z=0.186$) source is absorbed by the EBL and current uncertainties on the EBL flux level and spectrum \cite{Kneiske04,Kneiske10,Primack_2011} introduce some difficulties in defining how hard the intrinsic spectrum is. However, in the case of 1101-232, it tends to be very hard, with intrinsic spectral index $\alpha \sim 0.5$ when corrected for low EBL levels, and higher than 0.5 if the true EBL was significantly higher than its strict lower limit derived from galaxy count \cite{Aharonian06,Aharonian1101}. Such hard intrinsic spectra could be also expected in a number of extreme blazars like 1ES 0229+200 and 1ES 0414+009 \cite{Aharonian07b,Abramowski0414}. 

At a first glance, it seems difficult to generate such hard spectra within simple SSC models and standard particle acceleration models. Then, the IC scattering that gives the TeV emission usually  occurs in the Klein-Nishina regime and 
the efficiency of the scattering decreases very quickly with increasing  
electron and photon energies, resulting in a soft intrinsic spectrum 
with $\alpha_{\rm em} \le 1$). 

A simple solution to this problem is to assume a sharp low energy cut-off in the 
particle energy spectrum \cite{Katarzynski06}. If the minimum 
energy of the particles is high enough ($E_{\rm min} =\gamma_{\rm min} m_{\rm e}
c^2, \gamma_{\rm min} 
\gtrsim 10^5$) then the intrinsic spectrum can be hard with spectral index 
$\alpha_{\rm em} \geq -1/3$. This limiting value of the spectral index comes 
from the fact that the particles are scattering `tail' photons of the synchrotron 
emission. In such approach the value of $\alpha_{\rm em}$ depends on the
$\gamma_{\rm min}$ value and can extend from -1/3 up to 1. Moreover,
the scenario predicts an abrupt break in the spectrum at energies of several
TeVs. Therefore, it should be easy to test it, using CTA to obtain large band spectra 
from the GeV range up to a few tens of TeVs. Various ways to generate such specific particle distributions 
and hard VHE spectra are presented in \cite{Lefa}. Such investigation aims to 
find a simple and robust description 
for the VHE emission of the TeV blazars, to be able to predict their intrinsic
spectra, to relate them to the AGN characteristics, and to better constrain intervening media. 

\subsubsection{Time delay in TeV light curves}

Detailed observations of the rapid activity of Mrk 501 conducted by the 
MAGIC telescope on July 9th, 2005 \cite{Albert07} show 
delays between the light curves obtained in different energy ranges. 
In particular a several minutes delay was seen between the light curve 
obtained in the range 150-250 GeV and the one 
in the range 1.2-10 TeV. A natural explanation for such
delay is an on-going particle acceleration process just caught at the time of the observation, as proposed by \cite{Mastichiadis08,Bednarek}. Quantum nature of the gravity is an alternative possibility \cite{Albert08}. Such type of delays deserves 
precise investigation \cite{Ulisses09,Anita,Bolmont} and more advanced data, especially the spectral shape evolution is 
crucial to confirm or reject the acceleration scenario. This will be possible 
only with CTA, with sensitivity good enough to provide high quality spectra 
of such rapid flaring events.

\subsection{Hadronic and lepto-hadronic models}

Observational evidence and theoretical considerations on the energetics of AGN \cite{Sol89,Sik2000,Cel2008} suggest that AGN jets can contain a non-negligible fraction of hadrons which can constitute an important fraction of the particle content. The characteristics of these hadrons, e.g., their energy spectrum and number density, are however not known and it is not yet clear if they contribute significantly to the radiative emission. 

Hadronic emission models explore scenarios where radiative processes involving relativistic hadrons contribute to the high energy emission from AGN. These scenarios provide a direct connection between the detectable gamma-ray emission and the (as of yet undetected) emission of neutrinos and ultra-high energy cosmic rays (UHECRs) from AGN. Despite difficulties of establishing a firm link with the arrival direction of UHECRs, AGN remain one of the few candidate sources for these particles, and the nearby radio-galaxy Centaurus A is still considered a potential source for several UHECR detected with the PAO \cite{Abr2007}. Current and future neutrino telescopes, such as IceCube and KM3NeT, might detect very energetic neutrinos from certain classes of AGN, which would provide firm evidence that hadronic processes take place in these objects. The higher temporal and energy resolution of CTA and its improved sensitivity will help investigate such scenarios.

\subsubsection{Hadronic emission scenarios}
Different hadronic models have been developed to describe the SEDs of different types of blazars and radiogalaxies. They usually require magnetic fields of the order of 10 to 100 G, much higher than in the leptonic description. The Synchrotron Proton Blazar (SPB) model \cite{Aharonian_2000,Mue2001,Mue2003,Boe2008} ascribes the VHE emission in HBLs mainly to synchrotron emission from ultra-relativistic protons, pions and muons, while the X-ray emission is dominated by electron synchrotron photons. 
The ``Proton Induced Cascade'' models \cite{Man1992,Man1993}, in which photo-meson (mainly pion) production by ultra-relativistic protons with the internal radiation field in the  source dominates the VHE peak, are better adapted to describe the SEDs of LBLs than that of HBLs.
In FSRQs, photo-meson production on external photons from the disk, the BLR and the dust torus adds to the other hadronic emission processes \cite{Atoyan2001}.

A general problem for hadronic models is that short-term variability appears more difficult to account for in hadronic than in leptonic scenarios given the anticipated longer time scale for the interaction processes. However, if the variability is originated by changes in the viewing angle of the jet, the effect due to the variation of the bulk Doppler factor is the same for any process taking place in the jet's frame \cite{Romero1995,RomeroReynoso}. They need also some elaborated solutions to explain the TeV and X-ray emission correlation, which is not expected in the simple case of co-acceleration of electrons and protons in the same region. Indeed, alternative scenarios have been proposed that might be able to account for rapid variability and correlated flares \cite{Dar1997,Bea1999,Rac2000,Poh2000,Mue2001,Mue2003,Araudo,Rey2010}. One recent example \cite{Bar2010} in which AGN jets interact with gas envelopes from red giant stars and emit proton synchrotron radiation can even explain the very rapid variability seen during the big flare of PKS~2155-304 in 2006.

\subsubsection{Testing hadronic scenarios with CTA}

\begin{figure}[h!]
  \includegraphics[width=0.5\textwidth]{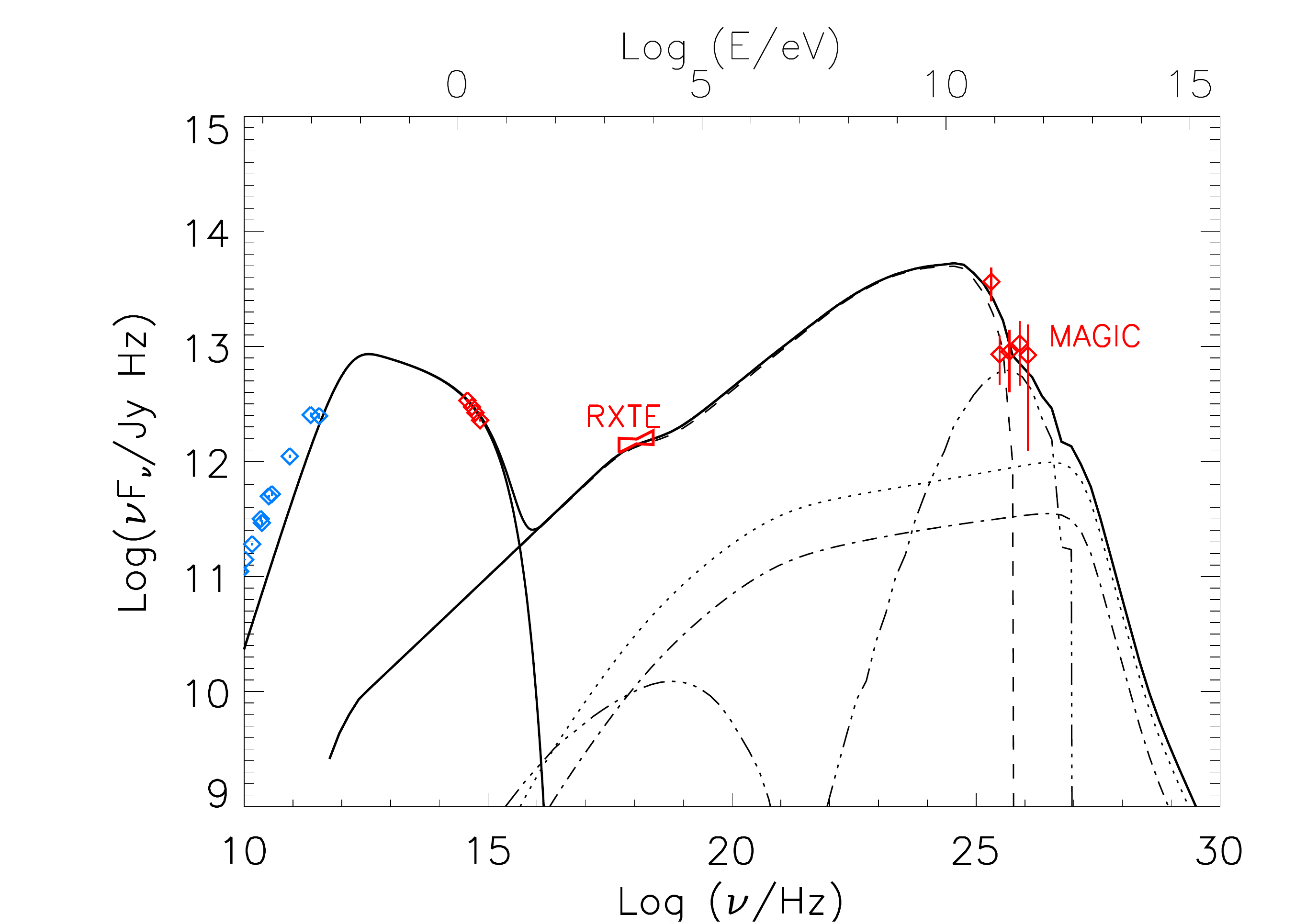}
  \includegraphics[width=0.5\textwidth]{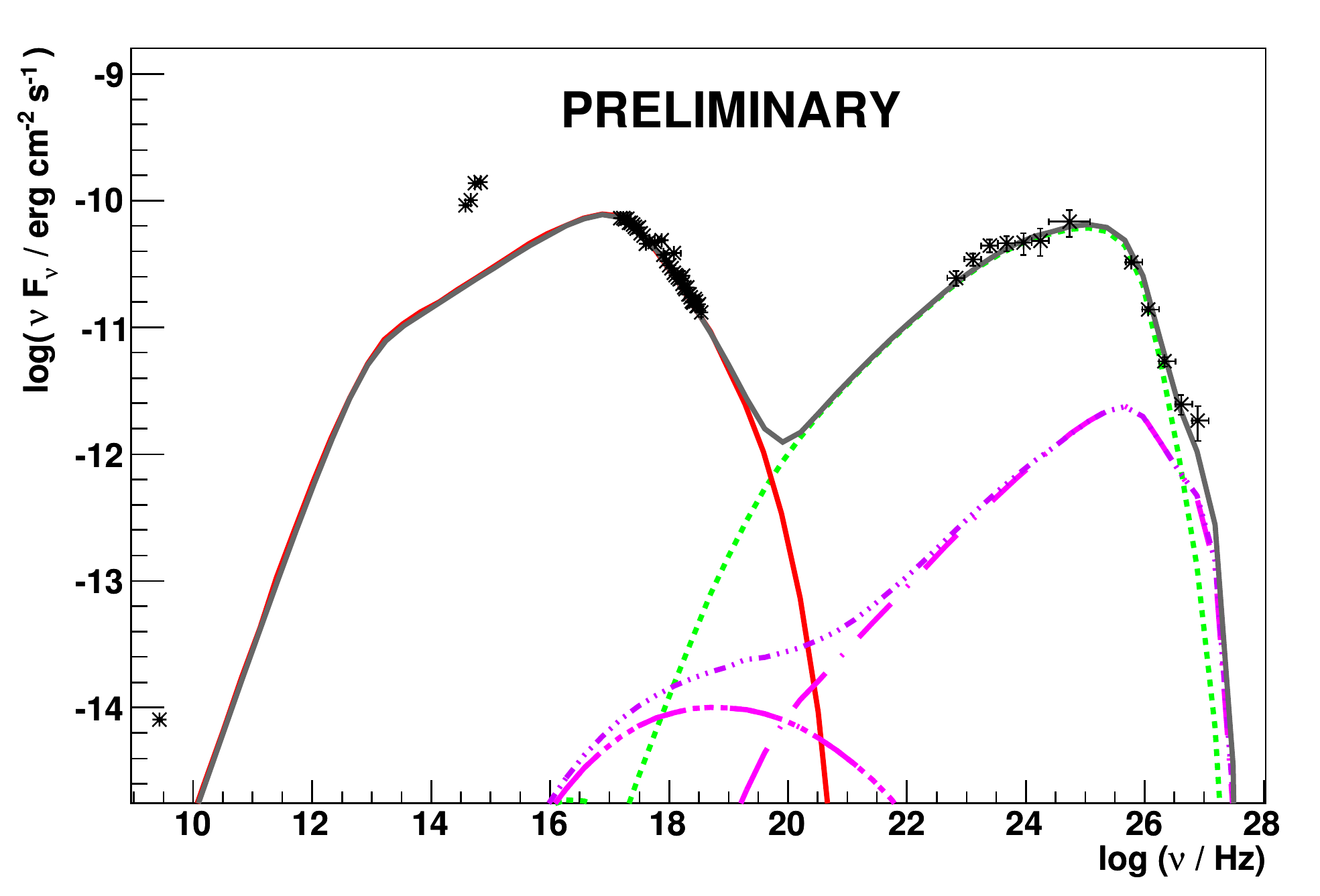}
  \caption{Reproducing SED with hadronic and lepto-hadronic models. Upper panel: The SED of the FSRQ 3C 279 compared to a hadronic model with several individual radiation components, namely proton synchrotron and cascade (dashed), muon-synchrotron and cascade (triple-dot-dashed), cascade from neutral pions (dotted), and cascade from charged pions (dot-dashed), from~\cite{Boe2008} (see also \cite{Boe2011}). Bottom panel: The SED of the HBL PKS~2155-304 as seen during the multiwavelength campaign of 2008 \cite{PKS2155mwl_09}, compared to a leptonic model (red line and dotted green lines for the synchrotron and SSC components). An additional component from proton-induced cascades is shown in magenta (dashed dotted lines) \cite{Cer2011}. High sensitivity and spectral resolution are needed to look for signatures of the various hadronic components with CTA. } 
  \label{fig:hadronic}
\end{figure}

Although it is not obvious to find a clear detectable signature that would allow the unambiguous distinction between leptonic and hadronic models, the highly improved information from CTA on the SED and its evolution, together with multiwavelength and multi-messenger data from other instruments, will help constrain model parameters and reject certain scenarios. An interesting target for such studies would be sources with significant internal or external photon fields, such as LBLs or FSRQs, in which photo-meson production and cascading play an important role in hadronic scenarios. Figure~\ref{fig:hadronic} (upper panel) shows one scenario, where cascades lead to a hardening of the SED at the highest energies, which will be clearly detectable with the better energy coverage and sensitivity of CTA. The search for spectral variability could provide another piece of information for the existence of cascading.

Photo-meson production in internal or external photon fields would lead to a characteristic cutoff energy of the proton spectrum, depending on the photon field density. This cutoff could be inferred from observations of the proton-synchrotron emission. Given the large number of sources accessible to CTA, it would be possible to study a relation between the cutoff energy and the blazar class, thus probing the blazar sequence within the hadronic framework. A refined understanding of the EBL, which could also induce features at the high energy end of the VHE spectrum, will be required and should be developed in parallel with the studies proposed here.

Apart from the two extreme cases, where the emission in the high energy bump stems either from the leptonic or from the hadronic population, one can also investigate mixed scenarios, in which both leptonic and hadronic processes contribute to the high energy bump. Such ``lepto-hadronic'' models can exhibit detectable spectral features in the high energy and VHE range as a result of the combination of different components. An illustrative example is shown in Figure~\ref{fig:hadronic} (bottom panel). These different components would also be expected to exhibit different temporal behaviour that one could hope to resolve with the high timing capabilities of CTA.

For all these studies, simultaneous multiwavelength coverage is needed to assess the electron population in the source from observations of the synchrotron bump. With this information, the existence of an additional hadronic component can be evaluated from an investigation of the VHE emission with CTA.

\subsection{Studying spectral features}

In a few cases, VHE spectra are already known to deviate from a pure power-law. This is  better seen for bright sources during active states. Spectral break and curvature have been detected in the SEDs of Mkn 421, Mkn 501 and PKS 2155-304 \cite{Felix_421,Albert_501,Abram2010}. The origin of such spectral features is not yet well identified and will require high quality spectra obtained with CTA. They could be the signature of different phenomena such as a cut-off due to the maximal Lorentz factor $\gamma_{\rm max}$ of the population of emitting particles, to the Klein-Nishina regime, or to some absorption effects. The detection of additional bumps in the SED could provide evidence for the presence of some hadronic component (see Fig.\ref{fig:hadronic}). 

\begin{figure} 
\includegraphics[width=0.4\textwidth]{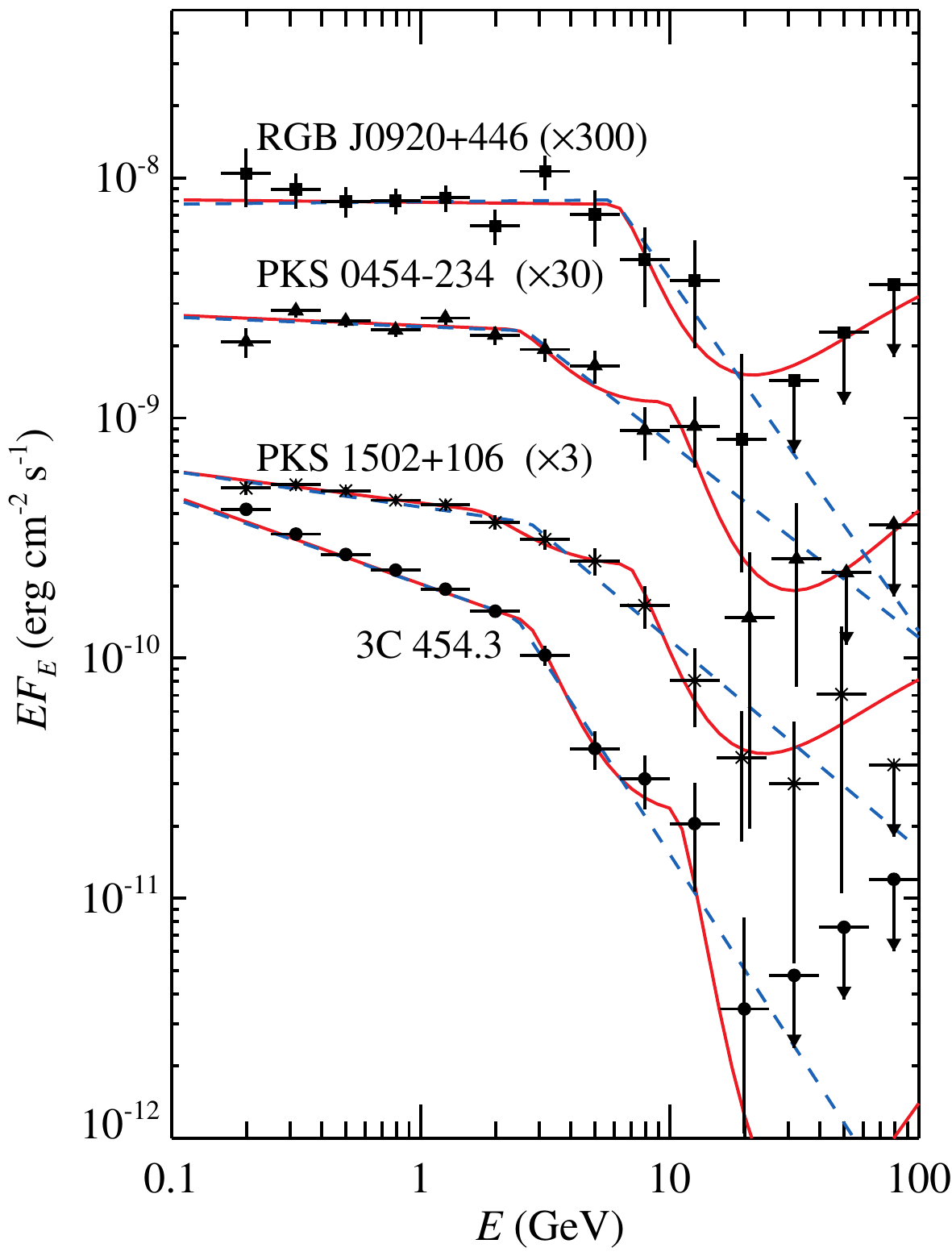}
\includegraphics[width=0.4\textwidth]{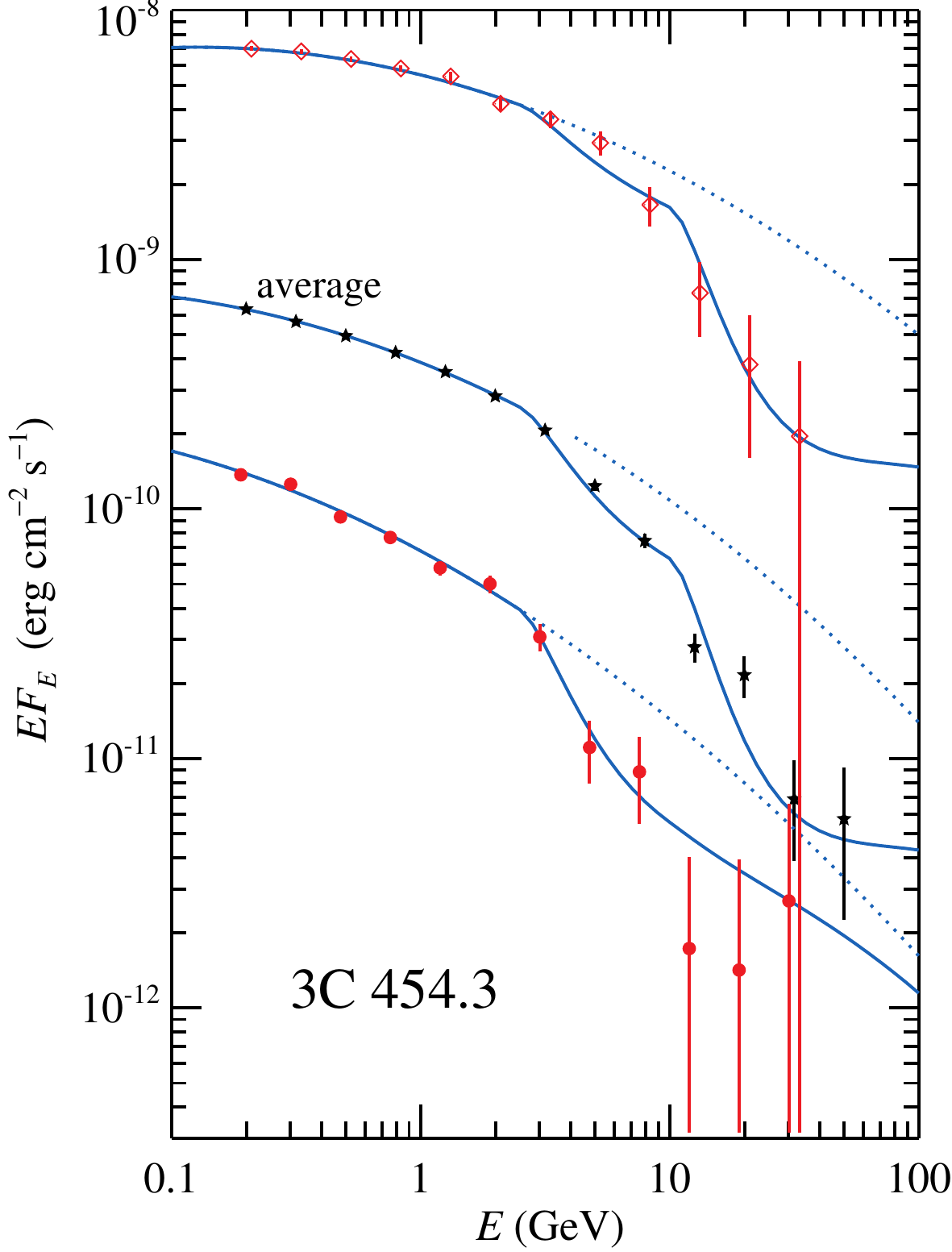}
\caption{Top: Spectral energy distribution of a few blazars as observed with {\em Fermi}/LAT 
during the first 6 months  of its operation.  
The  best-fit broken power law and  a power law with the double-absorber models are shown by  the
dashed and solid lines, respectively. 
Bottom: Spectral energy distribution of 3C~454.3 at low and high fluxes 
as well as averaged over the whole observation period of 2.5 years. 
The solid curves show the best-fit lognormal model with absorption by \mbox{He\,{\sc ii}} and H Ly recombination continua. 
The dotted curves show the unabsorbed lognormal distributions.  
} 	
\label{fig:all_spe}
\end{figure}

Indeed, recent observations by the {\em Fermi Gamma-ray Space Telescope} ({\em Fermi}) 
Large Area Telescope (LAT) revealed that the spectra of bright blazars 
cannot be described by a simple power law model  \cite{Abdo09_3C454.3,Abdo10_blazars}. 
A much better description is obtained with a broken power law, with break energies of a few GeV.   
In the brightest GeV blazars, the photon statistics allows spectral variations at different flux levels to be studied.
In spite of obvious spectral changes, the break energy in some sources like 3C454.3 turned out to be surprisingly stable \cite{Fermi10_3C454,Fermi11_3C454,SP11}. 
These breaks can be produced by  absorption of the GeV photons by the photon-photon ($\gamma\gamma$) 
pair production on the \mbox{He\,{\sc ii}} and \mbox{H\,{\sc i}} Lyman recombination continua (LyC) from  the BLR  
\cite{PS10}  (Fig. \ref{fig:all_spe}).  
The spectral breaks are expected at the energies $261\ \mbox{GeV}/(E_{\rm LyC} ({\rm eV}) [1+z])$, i.e. at about 
$5/(1+z)$ and $19/(1+z)$ GeV. 
It has been argued that a relatively high opacity in the \mbox{He\,{\sc ii}} LyC observed in a number of bright blazars
implies the location of the $\gamma$-ray emitting region within the highly ionized inner part of the BLR  \cite{PS10}. 
The absorption internal to the BLR thus complicates the determination of the intrinsic spectral shape of the bright blazars. 
The broad-band coverage  provided by  {\em Fermi} and CTA will be crucial for such a study. 

Photoionization models of the BLR predict also other strong lines, with Balmer lines (H$\alpha$, H$\beta$) and Paschen $\alpha$ being the strongest. These lines can produce absorption breaks at 100--150$/(1+z)$ and 400$/(1+z)$ GeV, respectively \cite{PS11}, i.e. within the CTA range. 
Thus with a spectral resolution of about 10\%,
the breaks can easily be  detected. 
The detailed fitting of the broad-band spectra including internal absorption within the BLR 
will give information on the column density of various soft photon sources that produce 
absorption in a given gamma-ray band. Thus, CTA will be able to map the BLR structure 
and to locate the gamma-ray emitting region with a higher precision than was possible up to now. 
Highly active states should provide the opportunity to obtain significant detections of such spectral features. Such type of studies will clearly be very sensitive to the low energy threshold of CTA.

\section{The origin of variability}

\subsection{Timing capabilities of CTA}

Blazars are usually detected in the gamma-ray band when in a high state, though the sensitivity of current IACT  has recently allowed the community to start probing the quiescent state of some of the strongest VHE sources \cite{Abram2010}. The high states of these objects are dominated by strong and burst-like flaring episodes that are characterized by  short, aperiodic variability events possibly dictated by the fast cooling times of the $\sim$ 10 TeV electrons thought responsible for the VHE emission. The detection and detailed studies of these short transient events is important to put constraints on the physics of the emission process and the astrophysics of AGN jets and is one of the key science cases for CTA.

Its huge effective area ($10^5$ to $10^7$~m$^2$, much larger than the ~m$^2$ effective areas reachable by space experiments) makes CTA an ideal instrument for the study of timing properties of blazars. Detections of fast variability of the VHE emission from some AGN indicate that their emission region is extremely compact. Such a compact emission region would most naturally occur close to the SMBH, in the blazar central engine.  Then timing properties of the AGN flares encode valuable information about the properties of the black holes and their immediate surroundings. However fast variability could also indicate that there are very compact substructures within the jet.  Current generation instruments are able to detect variability on the scales of several minutes in the case of the brightest blazar flares. Detecting characteristic temporal features with CTA, such as minimal variability times, or periodic or quasiperiodic events, could provide definite answers to the question of the size and possible location of the VHE emission region (active zone in extended or VLBI jets, base of the jet, or central engine magnetosphere) and stringent information on Doppler factors, physical emission mechanisms, particle acceleration, cooling and escape times, as well as on the intrinsic time scale of the central SMBH.

Variability time scales of AGN time series provide direct constraints on the $\gamma$-ray emitting zone. Short rise/decay times of an observed burst set an upper limit on the minimum variability time of the source, constraining by causality its size and Doppler factor. Complementary to variability studies in the temporal domain, one can derive the Fourier properties of a time series e.g., from Power Spectral Density (PSD) in a given frequency range, set by the observation live time and the sampling frequency of the lightcurve. Such analyses characterize the stochastic mechanism generating the flux modulation. These studies, inspired by the X-ray domain \cite{Marko2003}, are now feasible in the TeV range for a handful of dramatic outbursts, such as the exceptional flares of the blazar PKS 2155$-$304 seen by H.E.S.S. in July 2006 \cite{AhBigFlare} which produced the fastest and highest fluence flares ever observed in the VHE range from an AGN, and can thus serve as a proxy to study what are the limiting variability scales that CTA is likely to be able to probe from these kinds of objects.  Nonetheless, the lack of sensitivity and limited field of view of the current generation of IACT are responsible for their scarcity. The next generation, represented by CTA, will allow for faster sampling of the lightcurves and for the generalization of studies in both Fourier and temporal spaces. The expansion of this field will be enabled by the increased sensitivity and the lower energy threshold that CTA will afford (eg. $\sim$ 200 GeV for H.E.S.S. compared to 50 GeV for CTA). Hereafter we illustrate the consequences of these increased capabilities by simulating PKS 2155$-$304 flares as possibly monitored with CTA.

The lowering of the energy threshold will indeed allow one to monitor the source between 50 GeV and 200 GeV, whose flux is modeled here using the spectral energy density derived during the 2008 multiwavelength campaign on PKS 2155$-$304 \cite{PKS2155mwl_09,Sanch2009}. The increased energy coverage, combined with the large simulated effective area\footnote{The final results are barely dependent on the configuration used, yielding an uncertainty on the final temporal binning of around 5\%.} \cite{CTA-DS,Bern2008}, will allow CTA to significantly detect the source within a shorter duration than any current IACT, hence extending the high frequency part of the PSD. The low frequency part of the PKS 2155$-$304 spectrum during the high state is well described by a power law of index 2 \cite{Abram2010,Super2008}. Additional variance in the lightcurve beyond the reach of H.E.S.S. is derived from the Timmer and K$\ddot{\mathrm{o}}$nig method \cite{Timmer1995}, assuming the PSD extends to higher frequencies with the same index. Finally, the binning of the simulated CTA lightcurve is performed assuming an average significance that equals the one of the monitored H.E.S.S. lightcurve.\\
\par
\begin{figure}[h!]
\includegraphics[width=0.45\textwidth]{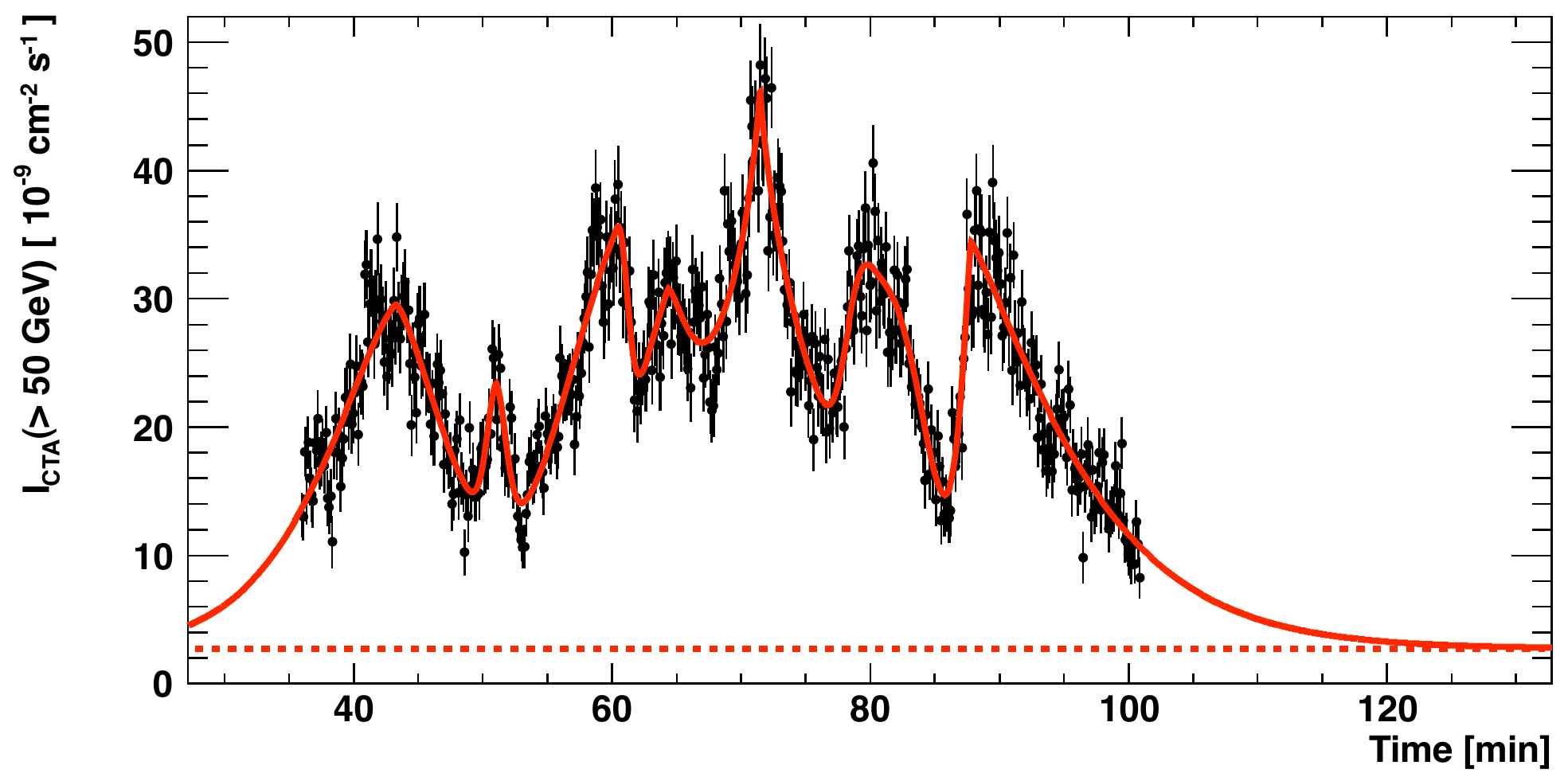}
\caption{Simulated integral flux of PKS 2155-304 above 50~GeV as CTA would monitor it. This simulation relies on an extension of the red noise behavior to high frequencies, generating the short time scale structures (second and fourth peaks). The data are binned in 7.5 seconds intervals.}
\label{fig:lcext}
\end{figure}

One specific CTA lightcurve realization is shown in Fig.~\ref{fig:lcext}. Provided the red noise behavior extends to higher frequencies, additional substructures which could not have been resolved by H.E.S.S. are revealed by the peak finding and fitting procedure. During such an event, the shortest significant rise time accessible to CTA would be $\tau_{\rm r} = 25 \pm 4$~s (second peak on Fig.~\ref{fig:lcext}), which is approximately seven times smaller than $\tau_{\rm r\ H.E.S.S.} = 173 \pm 28$~s. Such an upper limit on the smallest variability time scale constrains the ratio of the size of the emission region over the Doppler factor to $R\delta^{-1} \le c \tau_r/(1+Z) \sim 6.7 \times 10^{11}${\rm cm}, implying e.g. an emission region smaller than 1{\rm AU} for a Doppler factor of 20, an unusual value in the current acceleration models used for active galactic nuclei. The detection of such short time scale events would be a real challenge for jet formation models. \\

\par

This first example illustrates the great CTA timing capabilities which could detect events almost ten times shorter than the shortest ones detected by current IACT. This temporal resolution, below the minute time scale under exceptional circumstances, could impose  severe constraints on acceleration mechanisms and raise the question of the maximal Fourier frequency accessible to very high energy sources. However, such events are relatively rare and will require long term monitoring of large samples of AGN and the  triggering of observations under alarm to be efficiently detected. Below we present a more general statistical study which aims at assessing the typical gain in terms of temporal resolution of flaring events that CTA could monitor.  \\

To perform this statistical study, we chose to adopt unbinned, non-parametric methods, in order to take into account all the available timing information present in the light-curves and allow to probe for short variability events in a model-independent way. We generated a family of light-curves  
\cite{Timmer1995}, with an empirical PSD index $\beta \sim 2$, which are independent realizations of the process describing the H.E.S.S. observations of 2006. A second group of light-curves was also generated, extending the PSD to the higher frequencies which would be accessible by CTA, as described in our first example. These unbinned sequences of time-tagged events were further analyzed for variability using the Bayesian Blocks algorithm \cite{Ulisses_thesis,Scar98,Scar,Jack04},  which aim at finding the best partition of a sequence of events in a non-parametric fashion. The optimization of the Bayesian priors of the analysis is described in the Appendix B.

\begin{figure}[h!]
\begin{center}
\includegraphics[width=0.45\textwidth]{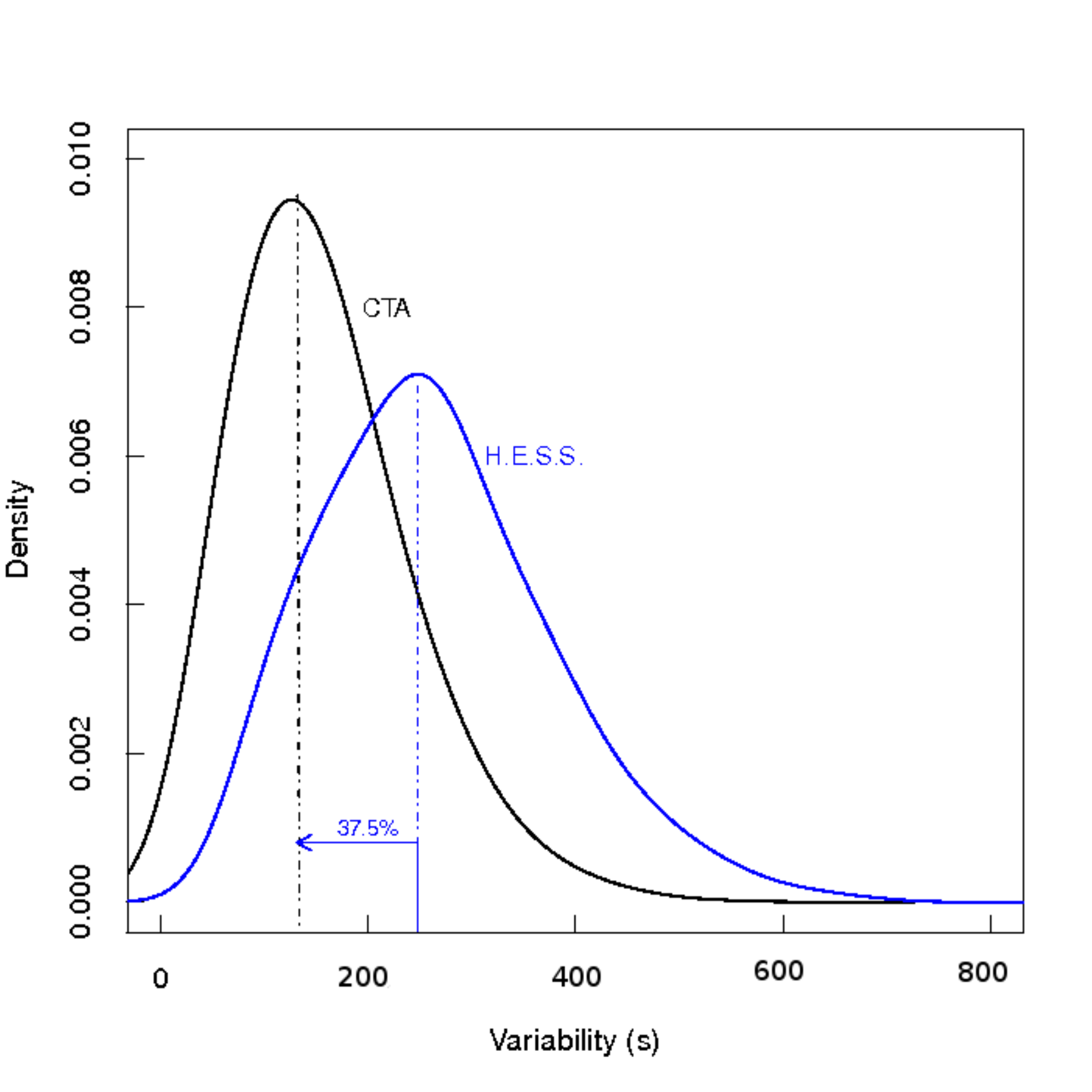}
\caption{Kernel density estimates for the distributions of the sizes of variability features detected from the H.E.S.S. and CTA simulated light-curve samples. The most probable typical flare size as seen with CTA, of the order of 100 seconds, is approximately $1\sigma$ from the typical values of 5 minutes  observed today.} 
\label{fig:var_cta_dist}
\end{center}
\end{figure}

Two groups of 1,000 light-curves were derived for H.E.S.S. and for the CTA extrapolation. Observe that the power law describing the PSD being essentially the same for both groups, the only difference between the two data sets is the sampling provided by the instrument. The distribution of flare doubling times $\tau_{\rm d}$ for both samples is shown in Fig.\ref{fig:var_cta_dist}. Many of the light-curve features not resolved in the H.E.S.S. sample can be separated in different events with the CTA resolution, shifting the peak of the distribution towards shorter duration events and thus revealing additional information about the timing structure of the source which was previously unnoticed. In particular, the most probable flare $\tau_{\rm d}$ derived for H.E.S.S. is of the order of 5 min, a typical value for the H.E.S.S. light-curve of PKS 2155-304 in 2006. With CTA, this value is expected to shift to 100 s, a factor of 3 shorter than H.E.S.S.

\begin{figure}[h!]
\begin{center}
\includegraphics[width=0.45\textwidth]{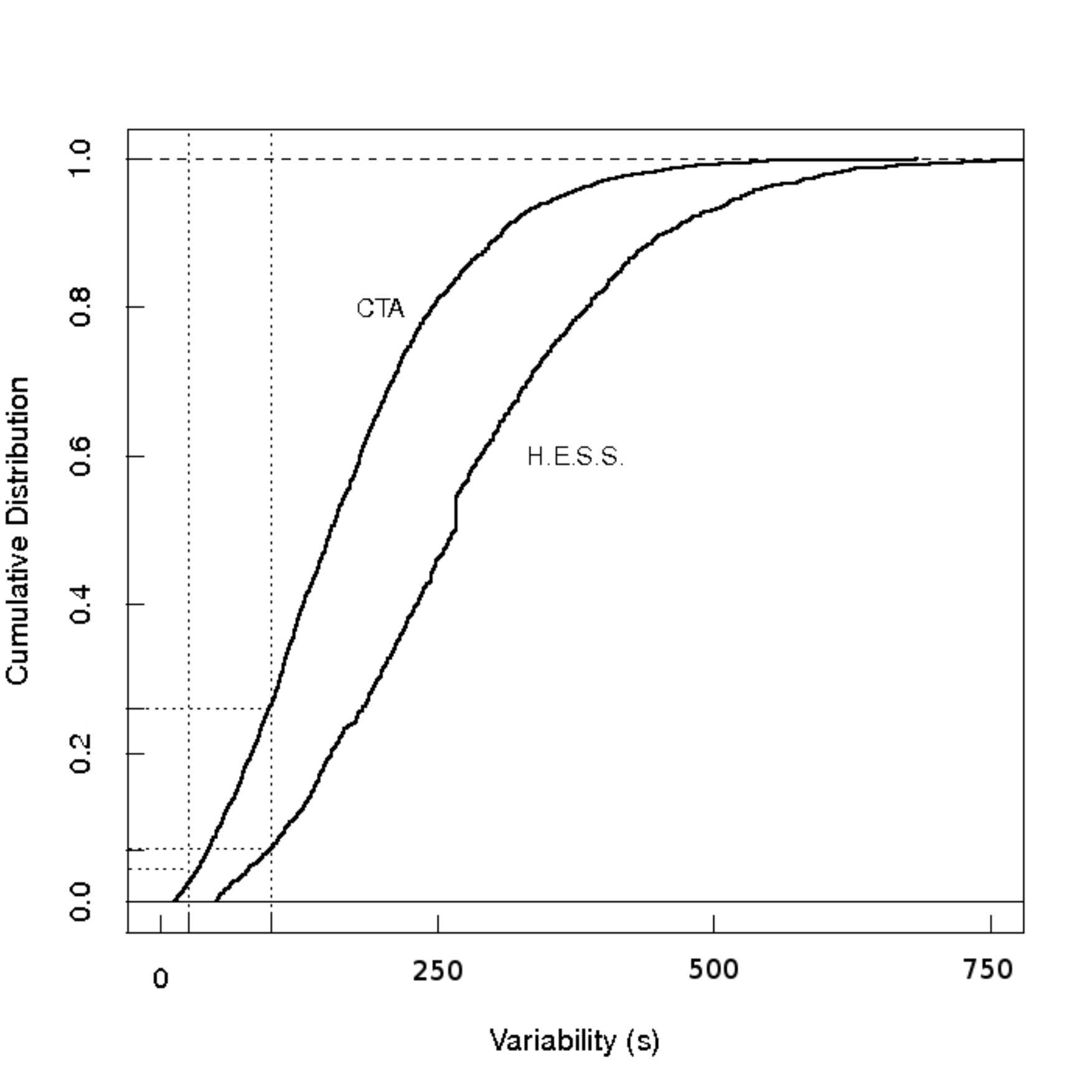}
\caption{Cumulative function of the same distributions shown in Fig.\ref{fig:var_cta_dist} (see text).
\label{fig:var_cta_cumdist}}
\end{center}
\end{figure}

The probability of occurrence and detection of a given ultra-fast flare as presented in the beginning of this section can be assessed by comparing the cumulative distribution of the flare durations as derived from our MC simulations (Fig.\ref{fig:var_cta_cumdist}). If short events are present in PKS 2155-like light-curves, these will be better sampled by CTA, up to the high resolution limit of the instrument. In the PKS 2155-304 light-curve as seen by H.E.S.S. in 2006 \cite{AhBigFlare}, the shortest variability event was of 173 s, which from simulations has an estimated probability of occurrence and detection by an instrument like H.E.S.S. of $\sim 13$\% (a rare event). With the improved capabilities of CTA, such an event will be rather usual, with a probability of $\sim 40$\%. The shortest burst shown in Fig.\ref{fig:lcext} as simulated for CTA, could not be detected by H.E.S.S. due to the limitations in sensitivity of the instrument. CTA, though, will be capable of seeing such a 25 s event if it is expected to be a rather rare occurrence, with probability below 10\%. 

To conclude, the advanced timing performance of CTA will allow to sample AGN light curves up to timescales significantly shorter than the ones achievable by the current generation of instruments, with the potential to probe variability well below the minute-scale for exceptional outbursts. The statistical study of typical simulated AGN light-curves shows that CTA will detect routinely many more variable events than current IACT, as a direct consequence of the better sensitivity, which will allow to resolve previously unnoticed short-time variability features. Typical flares duration are expected to be about 3 times shorter than those of the current generation, reflecting better the intrinsic structure of the sources. One can anticipate that such improvements in the timing analysis capabilities with the next generation of IACT observatories will be of great relevance for the study of AGN jet physics, bringing us closer to the fundamental limit of these variability timescales and to an understanding of the physical processes ultimately governing them.

\subsection{Multiwavelength and multi-messenger campaigns: monitoring and targets of opportunity (ToO)}

Given the complex variability properties of AGN, the flexibility of the CTA Observatory will be extremely useful. Long term correlated multiwavelength monitoring programs of a well-defined sample of AGN will be a necessity for several reasons, first to characterize quiescent and low-activity states and to estimate their VHE duty cycle, and also to send alarms for ToO during specific events (bright flares, rare events seen at lower energies or with other messengers, and also particularly low-activity states for instance in the optical range, which should be a good time to determine redshifts of remote BL Lac objects when spectral features from the host galaxy can be better detected above a low non-thermal optical continuum). The possibility to observe with CTA sub-arrays will be mandatory for such observing programs \cite{PoS}.   

Characterizing the VHE variability properties themselves should help identifying the mechanisms at the origin of the temporal evolution, and to  distinguish between features induced by specific bursts from those essentially due to a noise process. Multiwavelength monitoring should also contribute to identify the various emission zones,  to see whether they can be co-spatial or not, to locate them relatively to the others. This has been recently illustrated \cite{Marscher2010,Agudo2011}, combining the GeV gamma-ray lightcurves from Fermi satellite with X-ray, optical and VLBA monitoring data. With current generation instruments the VHE gamma-ray long term monitoring is possible only for handful of brightest AGN. With the CTA instead we should be able to clarify whether the VHE variability is dominated by the jet physics, with development of shocks and turbulence in the jet itself, or if the origin of the variability primarily lies in the accretion flow and the immediate vicinity of the central black hole, and to disentangle variability due to the global dynamics of the engine from variability induced by radiative processes (see section 4) studying the variation of the SED as a function of source activity. As an example, the binary black hole system OJ 287 could be a unique laboratory in this regard, thanks to the regular modulation of its parameters due to the orbital motion, which allows to observe the system, complex but rather well constrained, in different conditions over years \cite{OJ287}. Long timescale observations in X-rays and multiwavelength succeeded to answer such questions in the case of Seyfert galaxies \cite{McH}. Establishing for instance the relationship between the RMS variability and fluxes, they ruled out simple noise models and showed the presence of non-linear variability, possibly driven by fluctuation in the accretion flow  inducing a multiplicative process \cite{Lyub1997}. Multiplicative processes have recently be suspected in VHE blazars, as shown by temporal analyzes of the big flare of PKS 2155-304 \cite{Abram2010}. Variability studies of TeV AGN are still in their infancy, with variability characteristics still poorly determined and difficult to interpret \cite{Emma}. As shown above, CTA will open an avenue to deep time analysis at VHE. 

Generally speaking, AGN being very broad-band emitters over the whole electromagnetic spectrum, very promising synergies can be anticipated with most space and ground-based large infrastructures of the domain such as LOFAR, SKA, GAIA, JWST, and ELT. Based on the results obtained so far by the MOJAVE project and similar VLBI searches \cite{Lister1,Lister2,TANAMI}, one might expect that regular VLBI mapping at radio frequencies, coordinated with VHE campaigns, will significantly constrain the global picture of the AGN central core at the sub-{\rm pc} scale, linking the mechanism of jet formation, the ejection of radio knots, the growth of shocks and turbulence together with the localization, geometry and dynamics of the VHE emitting zones. It should also solve the persistent paradox of non-detection of superluminal velocities in the VLBI maps of VHE blazars while extremely large bulk Lorentz factors are needed for explaining the fluxes and variability \cite{Henri2011}. In the optical range, monitoring of the photometric and polarized continuum fluxes, and weak spectral lines from the broad line regions (if any), will allow to analyze  the unclear link between VHE and optical active states, to test the relative importance of EC contribution in the blazar sequence and especially FSRQ, and to make the connection between accretion events and VHE phenomena.  Simultaneous data in the X-ray and soft gamma-ray ranges will be essential to constrain the SED and the whole picture (see section 4). In particular, having an important time overlap between Fermi, X-ray satellites and CTA operations would be extremely rewarding. This should be considered when planning future high-energy satellites. 

The observational study of the compact and fast-varying AGN and possible cataclysmic events related to SMBH physics will clearly benefit from an efficient alert system within a global network. Operating CTA under alerts will be an important strategy to catch specific AGN behaviors. This procedure is already followed with success by present IACT. Alerts from optical monitoring, X-rays and gamma-rays have led to several discoveries and should be pursued and extended, especially to wide-field instruments like HAWC and LHAASO. Thanks to its long lifetime, CTA will also allow to search for remarkable electromagnetic counterparts of sources of non-photonic messengers detected by large experiments as ultra high energy cosmic rays with the PAO and the Telescope Array, neutrinos with IceCube and Km3Net and gravitational waves from SMBH with LISA \cite{Wang,Allard,Essey,Jeremy,Stone}. One can anticipate a great scientific return both in astrophysics and in fundamental physics from  any discovery of this kind. The common occurrence of VHE gamma-rays and high energy cosmic rays and neutrinos is already studied in the literature (see section 4.2), while the possible link between VHE gamma-rays and gravitational waves still deserves further analysis. Non-thermal electromagnetic counterparts are expected from black hole merger (see \cite{Jeremy} and references therein) and any coalescence event detected by LISA would be a compulsory ToO for CTA. Predictions based on the statistics of binary massive black holes estimate a possible rate of one major coalescence event every 5 years and hundreds of weaker events per year. However a better knowledge of the electromagnetic signal expected from such phenomena is still needed before quantifying the probability of serendipitous detection of coalescence events by CTA prior to LISA operations.

\section{Probing the intergalactic medium and backgrounds}

\subsection{AGN as beacons of gamma-rays}

 Beyond the studies of intrinsic AGN properties, monitoring of a large sample of AGN at different redshifts will provide constraints on cosmic background radiation, some cosmological matters, and fundamental physics of space-time. Indeed, many observing programs related to AGN will have to be thought of as large key projects with multiple aims, pursuing different scientific goals although requiring almost exactly the same type of data. 

Photons from AGN with sufficiently high energies can pair-produce with photons from the infrared, optical and/or ultraviolet EBL in processes of the type $\gamma_{\rm TeV}+\gamma_{\rm IR/O/UV}\rightarrow e^+ e^-$ \cite{Weekes,FAAharonian}. The IR/optical/UV EBL results from direct and reprocessed stellar (and AGN) emission at all redshifts. In the early Universe, decays of exotic particles or Population III stars may have made substantial contributions to the EBL. The fact that the EBL contains calorimetric information about the entire history of IR-UV emission in the Universe makes it an extremely interesting phenomenon for detailed studies. Although extragalactic pair-production processes reduce the visibility of high-redshift objects, studying the resultant attenuation features as a function of photon energy and redshift opens up exciting possibilities to determine the evolving intensity and spectrum of the IR-UV EBL, and in turn the cosmic history of star and galaxy formation in the Universe \cite{Mankuz2}. This alternative approach is complementary to direct measurements of the EBL, for which a major uncertainty is the subtraction of intense foreground components due to the zodiacal light. Measuring the EBL with gamma-ray absorption can circumvent this problem, and moreover can probe its time evolution through observations of sources at different redshifts. High quality spectra up to 100 TeV and analysis of spectral features such as cut-off and absorption features at extreme VHE will be mandatory in this regard. Identifying the stationary spectral pattern from variable features should help to disentangle EBL effects from intrinsic phenomena in the compact sources (see article on EBL, this issue). 

Monitoring of flaring AGN should also allow to check the Lorentz invariance and to put various limits on models of quantum gravity as VHE gamma-rays with very short wavelengths become sensitive to the microstructure of space-time. Tight constraints on relativity violations could be obtained while looking for a frequency-dependent velocity of gamma-rays due to vacuum dispersion (see the paper on Fundamental Physics, this issue). If present, such effects induce a possibly detectable time delay between arrival times of VHE flares seen at different energies \cite{AlbertQG,Amelino,Biller,AhQG,HESS2011}. They also modify the threshold condition for interaction of VHE particles with the EBL. Recent analysis of current IACT data did not allow any firm conclusion on this issue \cite{Albert07,Albert08,AhQG,HESS2011}. Such investigation would deeply benefit from the advent of VHE detectors with higher level capabilities. However any time lags possibly detected by temporal analysis of light curves will remain difficult to interpret, since it requires a good understanding of the source itself to disentangle any signature of vacuum dispersion along the line of sight from intrinsic time delays due to particle acceleration, emission and radiation transfer in the AGN \cite{Mastichiadis08,Bednarek}. Long term monitoring of a large sample of variable targets at various redshifts and collection of high quality light curves and spectra during ToO bright flares should be scheduled probably as a key-programme, to optimize the probability of such time delay detection. Such type of multipurpose programme will be very rewarding in term of scientific return versus observing time since the same set of data can address many different issues (AGN physics, EBL, fundamental physics ...).

\subsection{Pair halos, pair echos and intergalactic magnetic fields}

High-energy gamma-rays from AGN offer a unique potential
to probe weak intergalactic magnetic fields (IGMF), which represent a long standing question of astrophysics, astroparticle physics and cosmology.
There are various ways in which weak ubiquitous magnetic fields may have been generated
in the early Universe.
Such cosmological magnetic fields are potentially important for the description of the primordial universe,
and could be the ultimate origin of the magnetic fields seen today
in galaxies and clusters of galaxies
by serving as the seed fields for subsequent amplification by galactic dynamo mechanisms \cite{Asseo,Grasso:2000,Widrow:2002}.
In some regions such as intergalactic voids,
they may have survived to the present day as IGMF
without being affected by later magnetization from astrophysical sources \cite{Bertone:2006},
and therefore may provide us with valuable information
about physical processes in the early Universe.
So far, various mechanisms have been proposed for the generation of such cosmological magnetic fields,
including different types of cosmological phase transitions \cite{Grasso:2000, Widrow:2002}
or processes related to cosmic reionization \cite{Gnedin:2000, Langer:2005},
with predicted field amplitudes in the range $B \sim 10^{-25}-10^{-15}~{\rm G}$.
While they may suffice as seeds for galactic dynamos,
such tiny magnetic fields are extremely difficult to put in evidence 
through conventional methods such as Faraday rotation measurements
or by their effect on cosmic microwave background anisotropies.

A very powerful probe of such weak IGMF
may be offered by the secondary GeV-TeV components accompanying the primary TeV emission of blazars, which 
result from IC emission by $e^{-}e^{+}$ pairs
produced via intergalactic $\gamma\gamma$ interactions among primary TeV and EBL photons.
Depending on the IGMF value, such secondary components may be observable either as
``pair echos'' that arrive with a time delay relative to the primary emission
\cite{Plaga:1995, Dai:2002, Murase:2008, Takahashi:2011}, or as 
extended emission with a spatial extension around the primary source
\cite{Aharonian:1994, Neronov:2007, Neronov:2009}.  
The properties of the extended emission depend on the IGMF strength. Strong enough IGMF leads to full isotropization of the cascade emission and formation of a physical "pair halo" around the primary gamma-ray source \cite{Aharonian:1994}\footnote{For an isotropic pair halo to form, there must be a relatively high magnetic field ($B \ge 10^{-12}$~G) within $\sim 10$~{\rm Mpc} of the source to facilitate isotropic emission.}, while weak magnetic field leads to appearance of an extended emission with IGMF dependent size, which emission benefits from Doppler boosting. 

There have been no clear observational indications for either pair echo or halo emission so far, 
and deriving lower bounds to IGMF strengths 
from upper limits to such components obtained by GeV-TeV observations of selected blazars
appears very promising and still a matter of debate
\cite{Neronov:2010, Aleksic:2010, Tavecchio:2010, Dolag:2011, Dermer:2011, Taylor:2011, Takahashi:2011}.  
Existing limits rely on a combination of data obtained with different types of telescopes in the GeV and TeV bands. CTA will have the great advantage of being able to cover all the relevant energy range, from 10 GeV up to 10 TeV, within one observational facility. 
Improving the time coverage to better determine the GeV and TeV light-curves with Fermi and hopefully future GeV gamma-ray satellites, together with a global network of IACT including the two sites of CTA, would significantly benefit to this issue. 

The positive detection of the pair echo or extended emission components would provide highly valuable information on the IGMF, a goal that is expected to be achievable with CTA. Pair echos could be detected in the case of very tiny IGMF ($\le 10^{-16}$~G), and extended emission for higher IGMF values. 
CTA will conduct detailed measurements of spectral variability
with higher sensitivity and time resolution over a wider energy band
compared to current gamma-ray telescopes,
potentially allowing us to disentangle and positively identify the echo component 
from the primary emission,
and thereby probing very weak IGMF in the range $B \le 10^{-16}~{\rm G}$, totally out of reach with other means of investigation.

A fruitful synergy with wide-field instruments
such as HAWC or LHAASO can be expected in the CTA era,
since regular, long-term coverage of the multi-TeV emission of blazars by such facilities
should greatly improve our knowledge of their TeV activity on timescales of years to decades,
giving us a much better handle on the light curve of the primary emission
as input for more reliable predictions of the pair echo and extended emission properties. The prediction of the level of the cascade emission will be greatly improved also with the detailed measurements of the blazar spectra in the 10 TeV band with CTA itself, which will give the details of absorption of the primary gamma-rays in the intergalactic medium.  

\begin{figure}[h!]
\begin{center}
\includegraphics[width=0.35\textwidth]{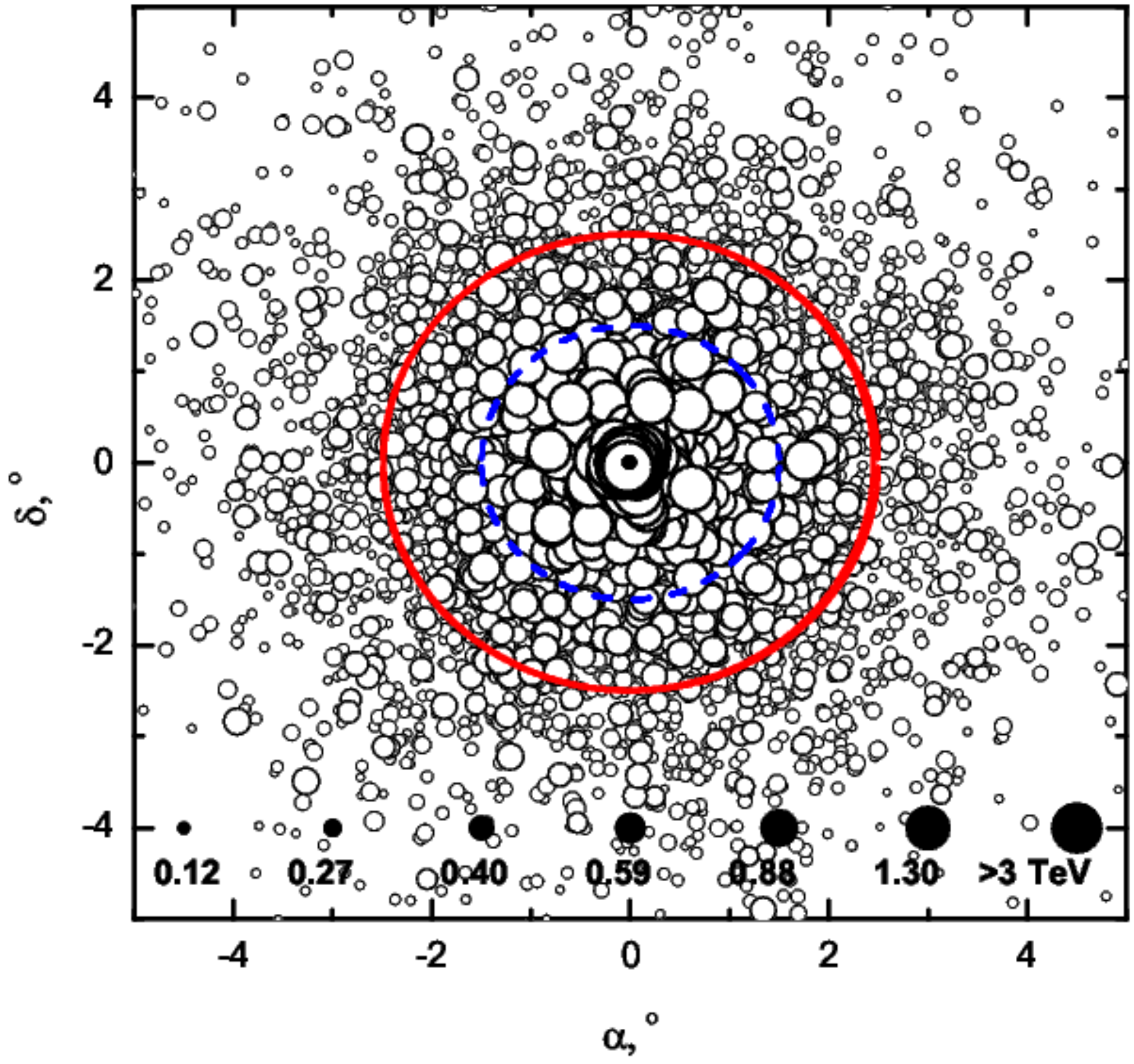}
\includegraphics[width=0.35\textwidth]{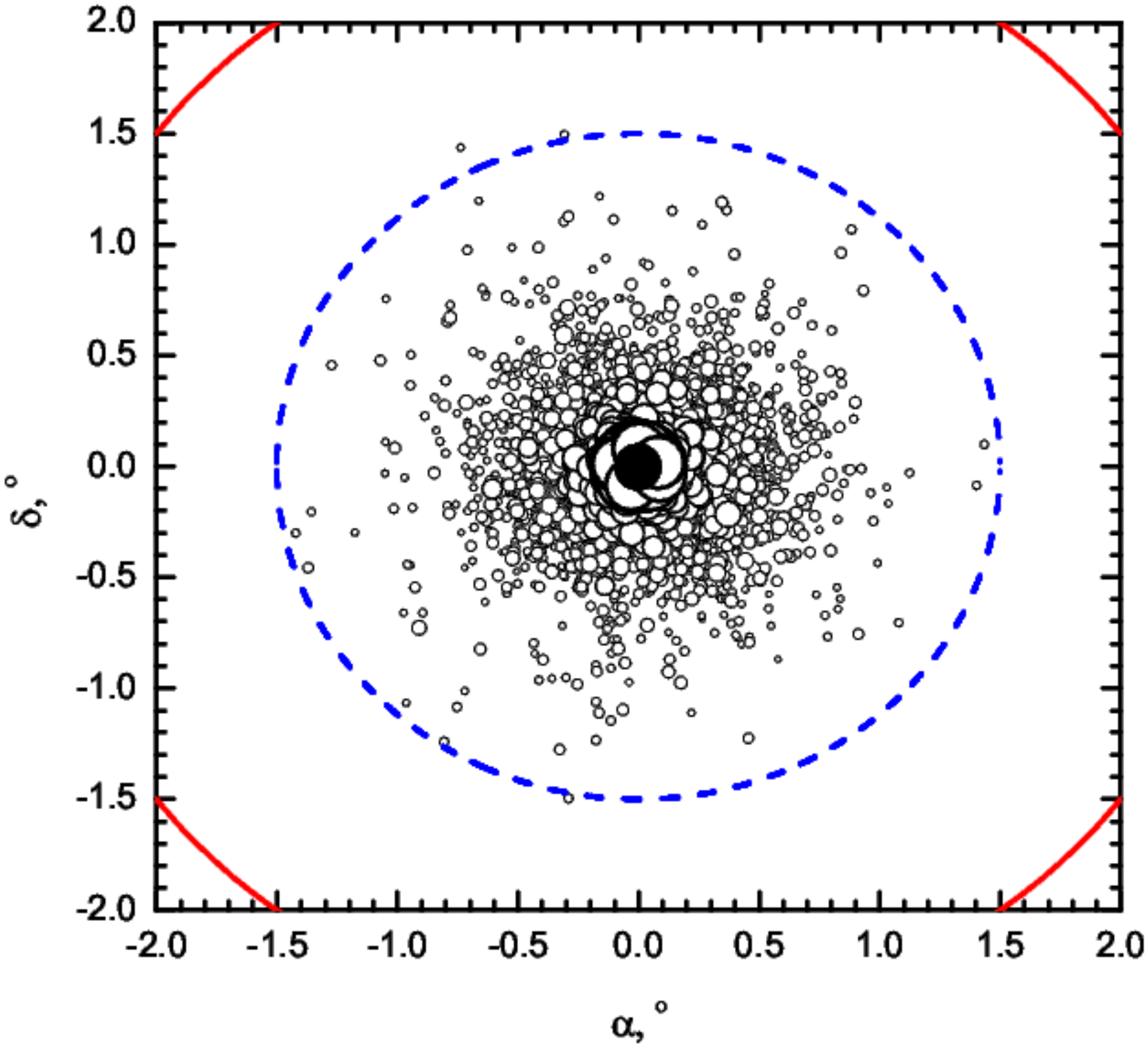}
  \caption{The arrival directions of the primary and secondary 
  gamma-rays (circles) from a source at a distance D = 120 {\rm Mpc}. 
  The IGMF strength is 10$^{-14}$~G (upper panel) and 10$^{-15}$~G (lower panel). 
  The sizes of the circles representing each photon are proportional to the photon 
  energies. The AGN intrinsic gamma-ray spectrum is typically described as $dN/dE \sim E^{-2} exp(-E/E_{cut})$ with the cutoff energy $E_{cut} \sim 300~TeV$. The blue dashed and red solid circles are radii of 1.5$^{\circ}$ and 2.5$^{\circ}$, respectively, and would perfectly fit into the FoV of CTA. 
  See \cite{Elyiv:2009} for details. \label{cascade}}
\end{center}
\end{figure}

If IGMF strengths are in a stronger range, $B \sim 10^{-16}-10^{-12}~{\rm G}$, the spatially-extended emission may be detectable and resolvable by CTA by virtue of its high sensitivity and angular resolution (see Fig. \ref{cascade}), in which case one would probe not only IGMF but also the lifetime of TeV activity for the source. High signal-to-noise detections of pair halos around nearby blazars ($z\sim 0.1-0.2$) would allow us to study the energy-dependent radial profile and thereby deduce unique information on the IGMF and on the EBL \cite{Neronov:2010b}. 

\begin{figure}[h!]
\begin{center}
  \includegraphics[width=0.45\textwidth]{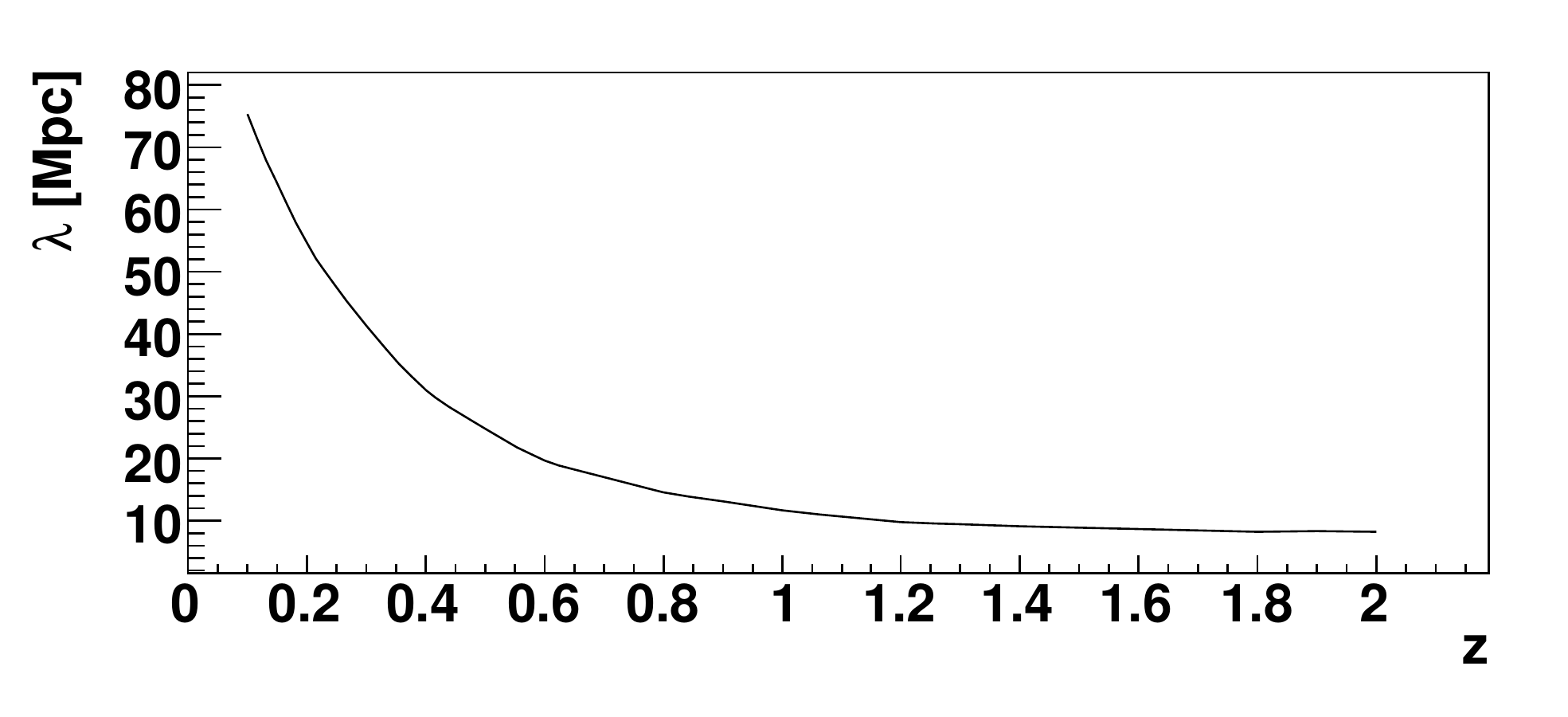}
  \includegraphics[width=0.45\textwidth]{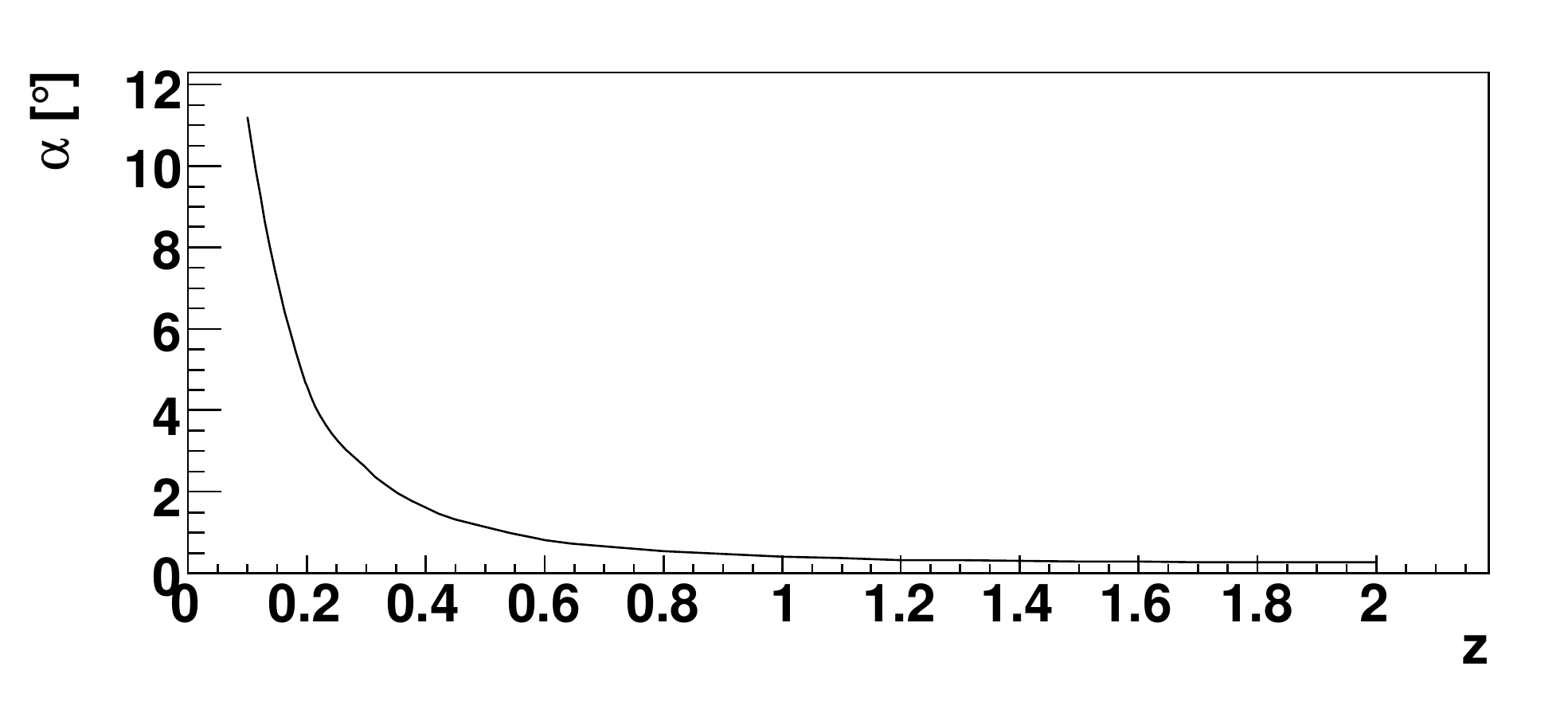}
  \caption{Radius in {\rm Mpc} (upper panel) and angular radius in  
degree (lower panel) at which the $\gamma$-ray emissivity of isotropic pair halos drops by a  
factor of $1/e$ as functions of redshift. The calculation assumes 10 TeV primary photons leading to 100 GeV  
secondary photons in the observer frame, an IGMF strength strong enough to isotropize the secondary pairs,  
and the EBL model of \cite{Franceschini:2008}.
   \label{halo}}
\end{center}
\end{figure}

If the IGMF is $B \ge 10^{-12}~{\rm G}$ and strong enough to effectively isotropize the secondary pairs,
the detection of isotropic pair halos from distant AGNs should be feasible \cite{Aharonian:1994,Coppi:1997}.
Figure \ref{halo} shows the expected sizes of such isotropic pair halos as a function of redshift,
generated by 10 TeV primary photons giving rise to $\sim$100 GeV secondary photons,
and using the EBL model of Franceschini et al. 2008 \cite{Franceschini:2008}.
Even though the properties of such isotropic halos no longer depend explicitly on the IGMF,
observation of their isotropic nature would still provide indicative lower bounds on its value, allow to check the self-consistency of our EBL description, 
and also give important information on the high-energy spectrum and beaming angle of the primary emission.

Although the halos are very large for redshifts $z\,<$ 0.5, 
a halo with a radius $<1^{\circ}$ at $z\,>$ 0.5 would fit into
the field of view of CTA telescopes and would allow for reliable
background subtraction. If one assumes that the source has a
constant $\nu F_{\rm \nu}$-flux from 100 GeV to 10 TeV, a simple estimate
shows that the halo fluxes are below a couple of percent compared 
to the primary 100 GeV emission. The halo fluxes would be much larger
if the photon index at the source was much harder than $\Gamma=2$. This is to be compared to the pair halo detection limits which should be actually reachable with CTA.

\begin{figure}[h!]
\begin{center}
\includegraphics[width=0.45\textwidth]{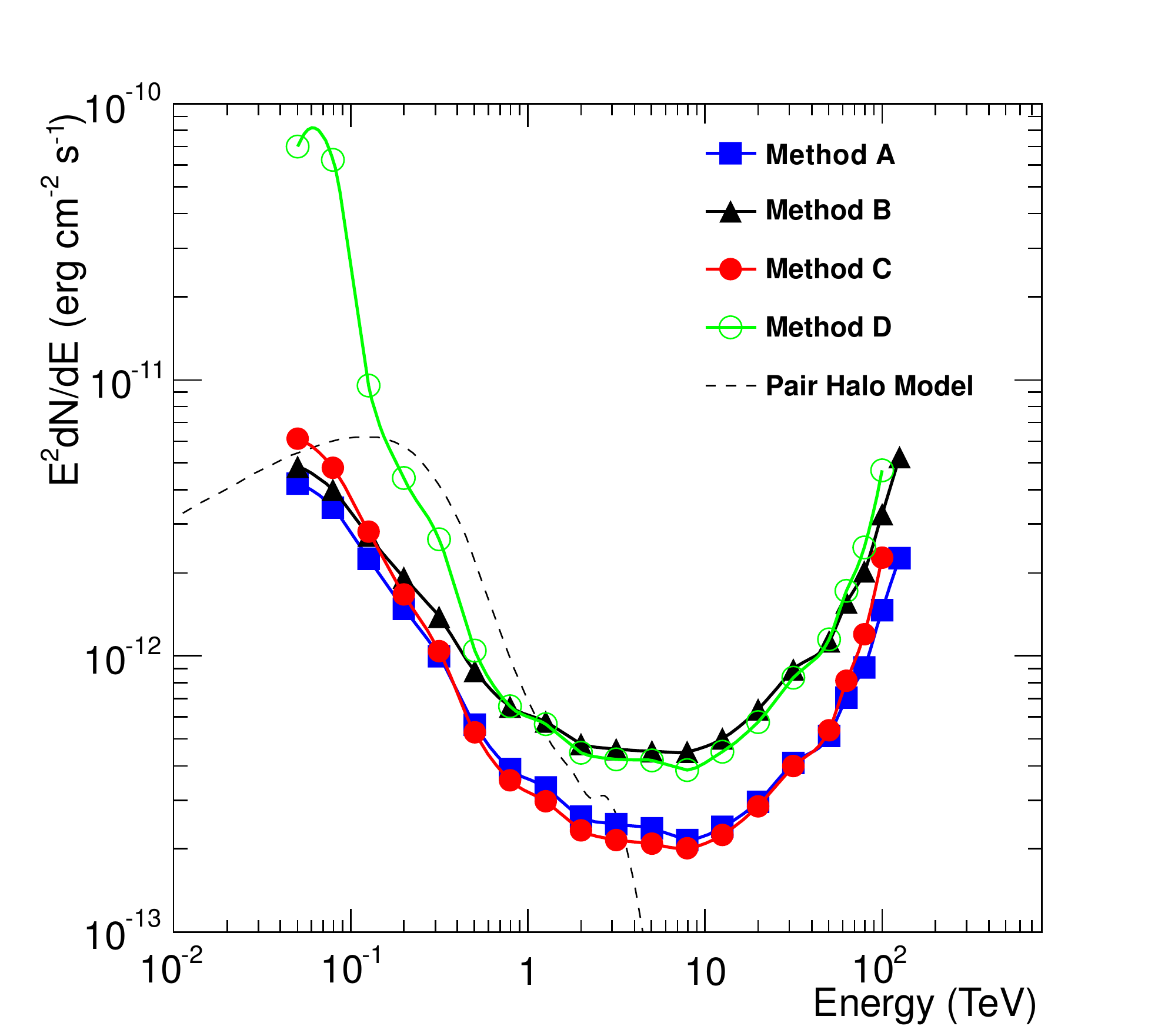}
\includegraphics[width=0.45\textwidth]{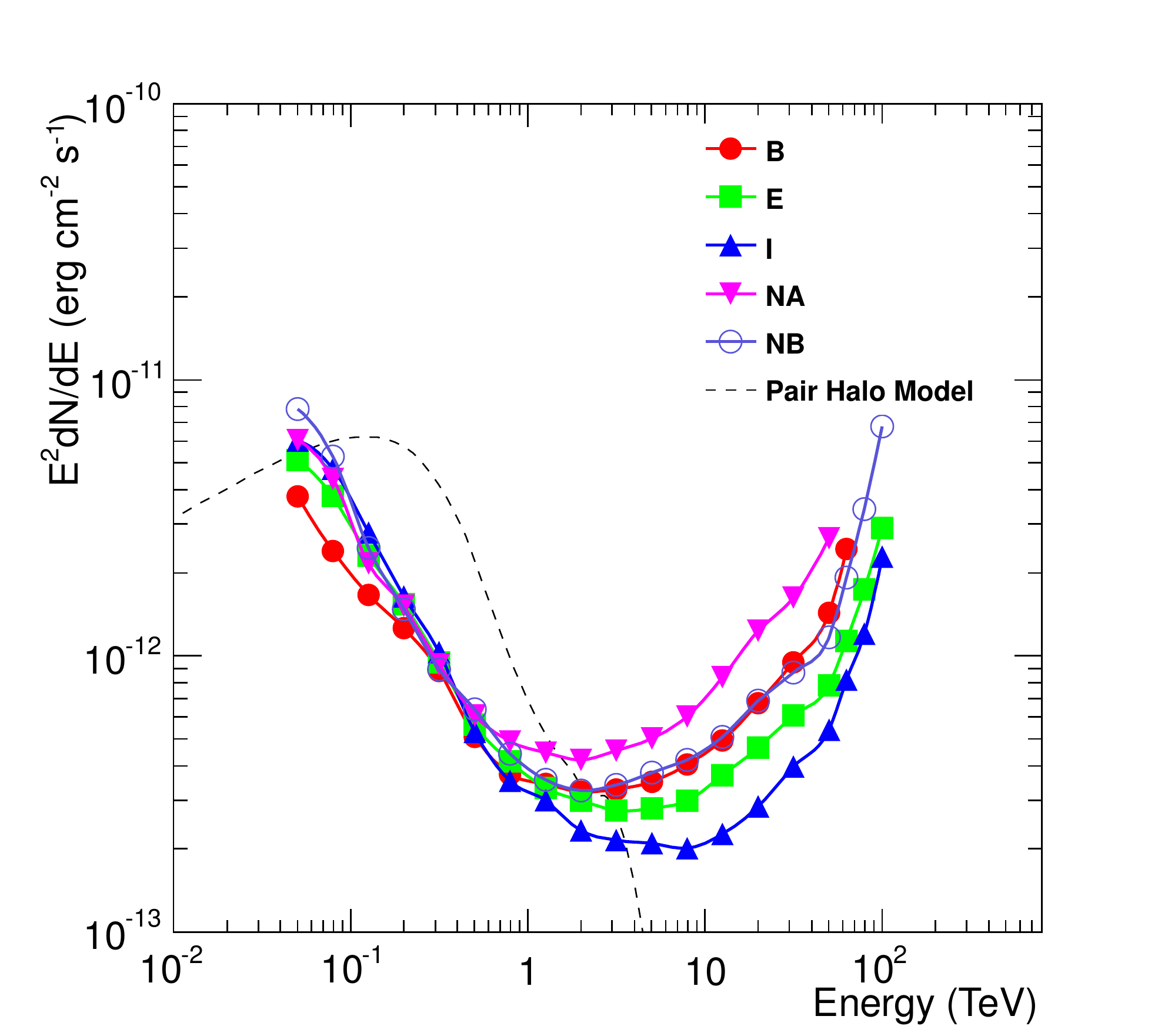}
\caption{\textbf{Upper panel :} A comparison of the sensitivity of CTA configuration I to pair halo emission for four different analysis methods. These methods are described in the main text.
\textbf{Lower panel :}
Flux estimates on the expected pair halo emission with CTA, for various CTA array configurations (50 hours observing time for each field, 20 degrees zenith angle observations), using analysis Method A.
A differential angular distribution of a pair halo at $z$ = 0.129 and E$_{\gamma}$ $>$ 100 GeV, taken from Fig. 6 in \cite{Eung} was used as our theoretical model (dashed line). 
\label{fig:Lisa}}
\end{center}
\end{figure}

The expected flux sensitivity for pair halo detection with CTA has been
calculated for four separate approaches. In all cases a differential
angular distribution of a pair halo at $z$ = 0.129 and E$_{\gamma}$
$>$ 100 GeV, taken from Fig. 6 of \cite{Eung} was used as our model halo profile, with a functional form: dN / d$\theta^{2}$ $\propto$ $\theta^{−5/3}$. The differential energy flux
sensitivity for the four different approaches is shown in Fig.\ref{fig:Lisa}. Methods A and B rely on a 5 $\sigma$ excess above the expected
background, calculated using Equation 17 in \cite{Li&Ma}. Method A searches for some overall extended emission with a
halo profile for $\theta<0.32^{\circ}$. Method B probes a region in which a point-like central source would no
longer be dominant ($0.11^{\circ}$ $<$ $\theta$ $<$ $0.32^{\circ}$) and therefore provides the most basic method of establishing the detection of extended emission. For methods C and D the
``goodness of fit'' for a halo profile (convolved with the CTA PSF) fitted to simulated CTA data is used to determine the expected flux sensitivity.
Method C compares a halo fit to the null hypothesis of background
fluctuations only, and the optimal $\theta$ limit for the
fitting was found to be $\approx$~0.2$^{\circ}$.
Method D however, tests how well a point-like source and a pair halo can
be distinguished. This is done by assessing the difference between the
likelihood obtained for a PSF hypothesis and that obtained for
a pair halo function, with the limiting flux defined as that at which
the point-source hypothesis can be rejected at the 5
$\sigma$ level. The rather high flux sensitivities for Methods B and D
in Fig.\ref{fig:Lisa} indicate the difficulty of
identifying halo-like emission at this confidence level (CL). If a 95 $\%$ CL is taken instead
of 5 $\sigma$, then the sensitivity of Method D is comparable to
Method A and C. Therefore a potentially extended source detected at
the sensitivity limit with a significance of 5 $\sigma$ could be
distinguished from a point source with about 95$\%$ confidence.  
Note that in Fig.\ref{fig:Lisa} the flux quoted is that
within a 1$^{\circ}$ region of the source.  The lower panel of Fig.\ref{fig:Lisa} shows the
pair halo sensitivity using Method A derived for a range of different
CTA candidate configurations. Config. I appears to be the most
suitable setup for pair halo studies.

To summarize, any detection of pair halos, extended emission, or echos with CTA would be a great step forwards in the characterization of cosmic magnetic fields and backgrounds. It could also open new paths to cosmologists and astroparticle physicists, if constraints on the primordial magnetic field could be deduced from the IGMF properties \cite{Grasso:2000}. 

\subsection{Observing the diffuse VHE background with CTA}

\begin{figure}[h!]
\begin{center}
\includegraphics[width=0.45\textwidth]{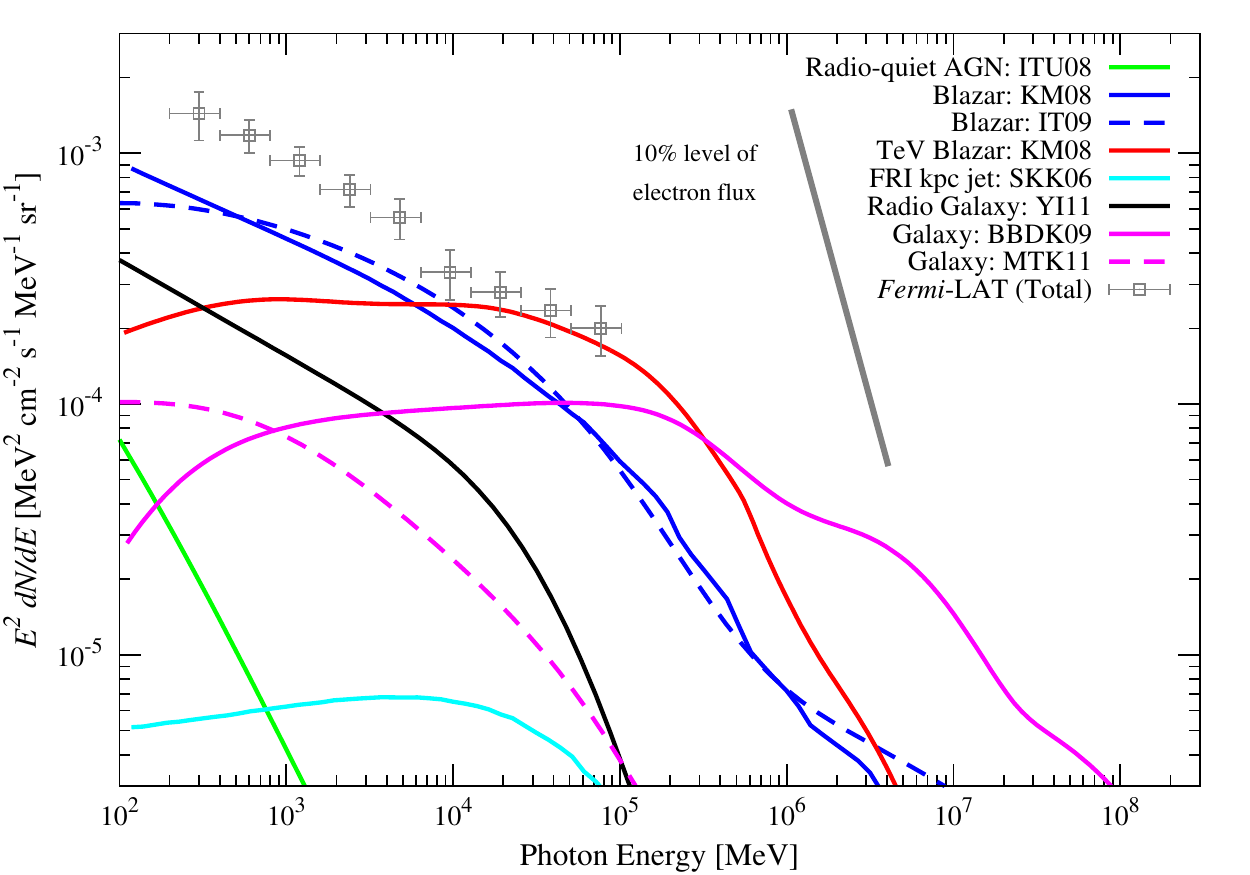}
\caption{The different components of the VHE extragalactic diffuse background, from ITU08 \cite{ITU08}, KM08 \cite{KM08}, IT09 \cite{IT09}, SKK06 \cite{SKK06}, YI11 \cite{YI11}, BBDK09 \cite{BBDK09}, MTK11 \cite{MTK11} and Fermi-LAT \cite{Fermi-LAT}. 
\label{fig:VHEdiffuse}}
\end{center}
\end{figure}

Extragalactic diffuse background radiation has been observed from
radio to GeV energies.  Evidence for diffuse extragalactic gamma-ray background (EGRB)
has been reported first by the SAS 2 satellite \cite{Fichtel78}, and more firmly by the EGRET detector \cite{Sree98,Strong04}.  
The separation of an extragalactic component from foreground and galactic
emission is generally challenging and model-dependent, and the
detection of EGRB was a matter of debate \cite{Keshet04}.
The origin of EGRB is still unsolved, but there must be
some contribution from unresolved or unidentified astrophysical
sources, especially blazars, active nuclei and galaxies (see Fig
\ref{fig:VHEdiffuse}). More exotic processes like $e^+ e^-$ pair halos or
even self-annihilating dark matter could also contribute.

Recently Fermi reported their new results about EGRB \cite{Abdo10a,Abdo10b}.
The reported flux is consistent with the EGRET data at 100 MeV, but the Fermi spectrum
is softer than EGRET and the Fermi flux is lower than EGRET by a
factor of 2--3 at 10 GeV.  The Fermi team suggested that unresolved
blazars can explain at most $\sim$23\% of the Fermi EGRB flux at $>$
100 MeV, indicating a significant contribution from other populations.
Indeed, the Fermi EGRB spectrum is well described by a
power-law with photon index of $\gamma = 2.4$, which is in good
agreement with the prediction based on theoretical models of EGRB from
astrophysical sources like blazars, radio galaxies, and star-forming
galaxies \cite{Yoshi1,Inoue11,Makiya11}. Against the total EGRB flux
(unresolved diffuse plus bright resolved sources, $\sim 1.4 \times
10^{-5} \ \rm cm^{-2} s^{-1}$ at $>$ 100 MeV), the predictions by
these models are 44\% by blazars, 18\% by radio galaxies, and 7\% by
star-forming galaxies.  Therefore, $\gtrsim$70\% of the total EGRB
flux can be explained by the known population of gamma-ray
emitters. The remaining 30\% may come from new, still undetected
populations, or from systematic uncertainties in the EGRB measurement,
e.g., foreground subtraction. Fermi EGRB measurement below 100 GeV has already set an upper-limit on the EGRB
itself above 100 GeV \cite{Inoue-Ioka}. The limit is conservative for the cascade
emission from EGRB interacting with the cosmic microwave-to-optical background
radiation not to exceed the current EGRB measurement. The cascade component fits the
measured VHE EGRB spectrum. However, taking into account known components, the
upper limit can be approximated as $E^2dN/dE<4.5\times10^{-5}(E/100\,{\rm GeV})^{-0.7}\
{\rm MeV/cm^2/s/sr}$  \cite{Inoue-Ioka} which is below the Fermi EGRB measurement
above 100 GeV. This suggests the existence of new physics in this energy range and shows
the need to study the VHE EGRB for a better understanding of the VHE gamma-ray
sky.

Detection of the VHE gamma-ray background would provide important
information to solve the EGRB problem.  Especially, we expect a
turnover above 100 GeV due to intergalactic absorption by the EBL, and
the break energy depends on the redshift distribution of the sources
contributing to EGRB. Moreover, HBL population like 1ES 0229+200 are 
not detected by Fermi but by current IACT. However, it is challenging to determine the
diffuse background level by an observation using a ground-based
Cherenkov telescope, because the photon events must be discriminated
from the much more frequent cosmic-ray electron events.  In the
measurement of cosmic-ray flux, the H.E.S.S. team found that the
distribution of the depth of shower maximum ($X_{\rm \max}$) indicates less
than 10\% of photon contribution in the electron flux, though they
also mentioned that $\sim$50\% photon contribution cannot be excluded
when the systematic uncertainty in $X_{\rm \max}$ determination is
conservatively taken into account. The 10\% level of electron flux
is indicated in Fig.\ref{fig:VHEdiffuse}. The expected EGRB signal in the TeV 
range is about 1\% of the electron flux. 
Detecting it with CTA will be challenging, but if it is achieved, 
we will get an important insight into the extragalactic VHE populations.

\section{Conclusion}

The discovery of AGN at extreme energies was a real breakthrough performed by IACT of the 90�s decade.  The current IACT increase the VHE extragalactic sample, and provide basic spectral and variability characteristics. They start constraining the physical processes at work and to catch a glimpse of the extragalactic VHE populations. Together with EGRET and later on Fermi, they show that the emitted gamma-ray power can represent a significant part of the non-thermal activity energy budget, and they raise a lot of intriguing questions. 

CTA will have the unique potential to open new science directions beyond the conventional AGN paradigm, and to shed new light on AGN and SMBH physics, populations, and evolution. Especially the jump achieved in sensitivity and spectral resolution, and the large spectral coverage from a few tens of GeV to several tens of TeV will be mandatory in this regard. Improvement of the angular resolution and of astrometric accuracy will facilitate source identification and may lead to the discovery of diffuse or multiple VHE components associated to AGN. The numerous modes of observation foreseen for the CTA array are perfectly appropriate to AGN observing programs and should definitely be implemented. Needless to say that AGN studies will strongly benefit from having access to the whole sky. Developing the two sites of the CTA array, one in the southern and one in the northern hemisphere, will be mandatory in this regard. Having them at different longitudes would allow to improve the time coverage for variable sources reachable from both sites. As far as the current CTA candidate arrays are concerned, emphasis on LSTs could provide a crucial boost in the effective area in the low-energy region where high-redshift sources are expected to be most prominent. Solutions without LSTs would have a negative impact on AGN population studies and every effort should be made to smoothly connect the low-energy regime of CTA ($\leq 30$ GeV) with {\it Fermi}. Conversely, the VHE fluxes of most of the bright Fermi blazars are largely above the CTA sensitivity limit and interesting high-statistics spectral information on these blazars will be available. In particular, observing over a large spectral range up to several tens of TeV with a good spectral resolution will make it possible to find out whether the observed cut-offs in the blazar spectra are intrinsic to the source or are induced by the effect of EBL absorption, and to analyze the maximal energies of particles responsible for the VHE emission and extreme acceleration processes. 

Hundreds of VHE AGN will be detected over the years of CTA operations. It will be the role of CTA to produce refined light curves to probe the shortest time scales, high quality spectra to constrain acceleration, emission and absorption effects, and maps to locate the VHE emission zones of nearby radiogalaxies. AGN monitoring, serendipitous discoveries and deep surveys will gather meaningful samples of VHE sources to contribute to the elaboration of a convincing unifying scheme for AGN. AGN physics appear as a science driver for CTA, exploring present mysteries such as mechanisms to accelerate VHE particles in extremely compact zones. Generally speaking, VHE instruments offer a completely new and independent way to explore AGN and their most efficient and energetic regions. They should provide a decisive key for the understanding of the origin of several non-thermal phenomena at work in such sources, in particular the missing link between accretion physics, SMBH magnetospheres and jet formation and propagation, which have been studied for several decades at lower energies and still remain fundamental open questions of contemporaneous astrophysics. 

As a caveat, it should be noted that all quantitative estimates presented in this article are preliminary, as further refinements to the CTA array configurations and improvements to the analysis tools are being worked on. Uncertainties in the spectral parameters and the EBL model could introduce biases. However, it is clear that we can look forward to a densely populated sky in the TeV range with CTA. This will open new opportunities to extragalactic studies, including dedicated multiwavelength campaigns and studies of very rapid flares. Furthermore, CTA will matter for research on blazar evolution, with the detection of high-redshift blazars.

CTA will be a crucial instrument for multi-messenger astronomy. Indeed, hadronic scenarios for TeV emission predict the simultaneous production of neutrinos and cosmic rays. Bright objects detected by CTA will be prime candidates for non-photonic sources potentially reachable by large cosmic ray and neutrinos detectors such as the PAO and the KM3Net. Conversely, any future detection of cosmic non-photonic sources of neutrinos, cosmic rays or even gravitational waves will benefit from follow-up observations with CTA to identify their nature, further explore their properties and constrain the parameters. 

While observing AGN with CTA, many other questions less directly related to AGN topics, can be simultaneously explored, using AGN as bright remote beacons to analyse various properties of the intervening media, such as the extragalactic background light, and the incredible perspective to measure extremely tiny values of the intergalactic magnetic field, or even the possible quantum properties of space-time.  Unexpectedly, CTA turns out to be an ideal instrument to probe the nature of the seed IGMF and could solve the problem of the origin of magnetic fields in galaxies and galaxy clusters which is one of the long-standing unsolved problems of astrophysics and cosmology. 

An important synergy between the analysis of the future VHE data and further developments of theoretical related fields is anticipated and will be mandatory to handle the flow of new information expected from the CTA observatory.

\section*{Acknowledgments}
The authors thank the CTA collaboration and many scientists for discussion and comments on the paper, with special thanks to C. Dermer, T. Kneiske, E. Lindfors and the CTA internal referees M. Persic and G. Romero.
The authors acknowledge the support of the CNRS and of the visitor programme of the Paris Observatory through the LEA ELGA. They thank M. Daniel, D. Emmanoulopoulos, and B. Thiam for their help. N.M. gratefully acknowledges support from the Spanish MICINN through a Ram\'on y Cajal fellowship under project code FPA2010-22056-C06-06.

\appendix

\section{Appendix on extrapolation of Fermi spectra}
\label{sec:app_fermi}

For the conservative estimate, based only on AGN with confirmed redshifts in the Bzcat or V\'{e}ron (13th edition) catalog and without any flags in the 2FGL catalog, 508 sources from the cross-correlated 2FGL catalog were selected for further processing, including 288 flat-spectrum radio quasars (FSRQs), 156 BL Lacs, 7 radio galaxies (RGs), 5 Seyfert galaxies, 2 starburst galaxies (SBGs) and 50 AGN of other classes. In the second study, for which additional sources of redshift estimates where used and the 2FGL analysis flags where ignored, a subset of 561 {\it Fermi} sources was selected, including 340 FSRQs, 171 BL Lacs, 10 RGs, 6 Seyfert galaxies, 3 SBGs and 31 AGN of other classes.

To extrapolate the spectra to the highest energies, we use the integral flux from Fermi measurements between 1 and 100 GeV (``F1000'') or the differential flux at the pivot energy (``Flux\_Density''), together with the spectral index $(\Gamma)$ given in the 2FGL catalog. For each individual source, we adopt the corresponding power-law or LogParabola parametrization as described in \cite{fermi}. For very hard sources following a power-law with $\Gamma<2$ in the Fermi range, a straight extrapolation could create runaway integrations, therefore we apply an artificial break in the power law with a photon index of  $\Gamma_2$ = 2.5 above 100 GeV to soften such spectra. The latter is in broad agreement with observed AGN features. 

Using the extrapolated and absorbed AGN spectra, we integrate the flux of particles per energy bin weighted with the effective areas for the different candidate arrays. The resulting expression is subsequently multiplied by the observation time, yielding the total number of detected photons. For each source, the significance was calculated using Equation 17 in \cite{Li&Ma} assuming $N_{\rm on}$ (on region) to be the number of source photons plus the number of photons derived from the simulated background rates,  and $N_{\rm off}$ fixed at the background rate (off-region). The variable $\alpha$ is given by the ratio of the sizes of the two regions times the ratio of the exposure times and the respective acceptances. 

For the first study, $\alpha$ is assumed to be 0.1. The detection of a source requires a significance of the signal over noise ratio of at least 5$\sigma$, a minimum of 7 excess events and a signal exceeding 3\% of the background. \footnote{The {\tt CTAmacrosv5} and {\tt CTAmacrosv6} tools for the estimation of the instrumental response were used for this study.} For the second study, an $\alpha$ of 0.2 is used and a signal over 5\% of the background is required. To maximise the number of detections in this study the energy threshold for the detection is varied such as to optimize the signal from each source.

\section{Appendix on variability statistical analysis}
\label{sec:variability}

The unbinned light-curves were generated using the algorithm of \cite{Timmer1995}, as a sequence of time-tagged events, which can then be directly inspected for variability with no recourse to binning. To analyze the variability properties of the unbinned datasets in a non-parametric fashion, we adopted a Bayesian algorithm called "Bayesian blocks". The method uses a dynamic algorithm to analyze the entire light-curve and find the optimal way of dividing it into a series of contiguous blocks, k, each consistently describable as an homogeneous Poisson process of rates $\lambda_{\rm k}$. The change-point between each block marks the position where a significant change in the rate of arrival of events  occurred, the entire event-sequence being thus treated as an inhomogeneous Poisson process made up from the juxtaposition of constant-rate processes. The approach is therefore very convenient as a model-independent analysis of the time-series, in the sense that the partitioning of the light-curve is based only on the (trivial) hypothesis that the event sequence follows a Poisson distribution. At every time in which the hypothesis of constant rate for the distribution is no longer valid for a subsection of the events, a new partition (or block) is created, within which the approximation of homogeneity is again valid. In this way the partitioning naturally selects the flares in the time series and the variability timescale of a given feature can be directly identified with the size of the corresponding block. If one wants to quote the variability in function of the more common quantity of the "doubling time" $\tau_{\rm d}$ of the flux, a simple correspondence can be made, by which the size of the block $\Delta t_{\rm k}$ can be taken as approximately $2\sigma$ of a Gaussian fit to the flare events. In the case of generalized Gaussian profiles, such as presented in the text, the correspondence can be made in terms of $\Delta t_{\rm k} \sim \sigma_{\rm r}+\sigma_{\rm d}$, where the different rise and decay times of the profile are taken into account. For a Gaussian profile, the conversion factor is: $\Delta t_{\rm k} = 2/ln(2)\sigma \approx 3 \tau_{\rm d}$.

\begin{figure}[h!]
\begin{center}
\includegraphics[width=0.45\textwidth]{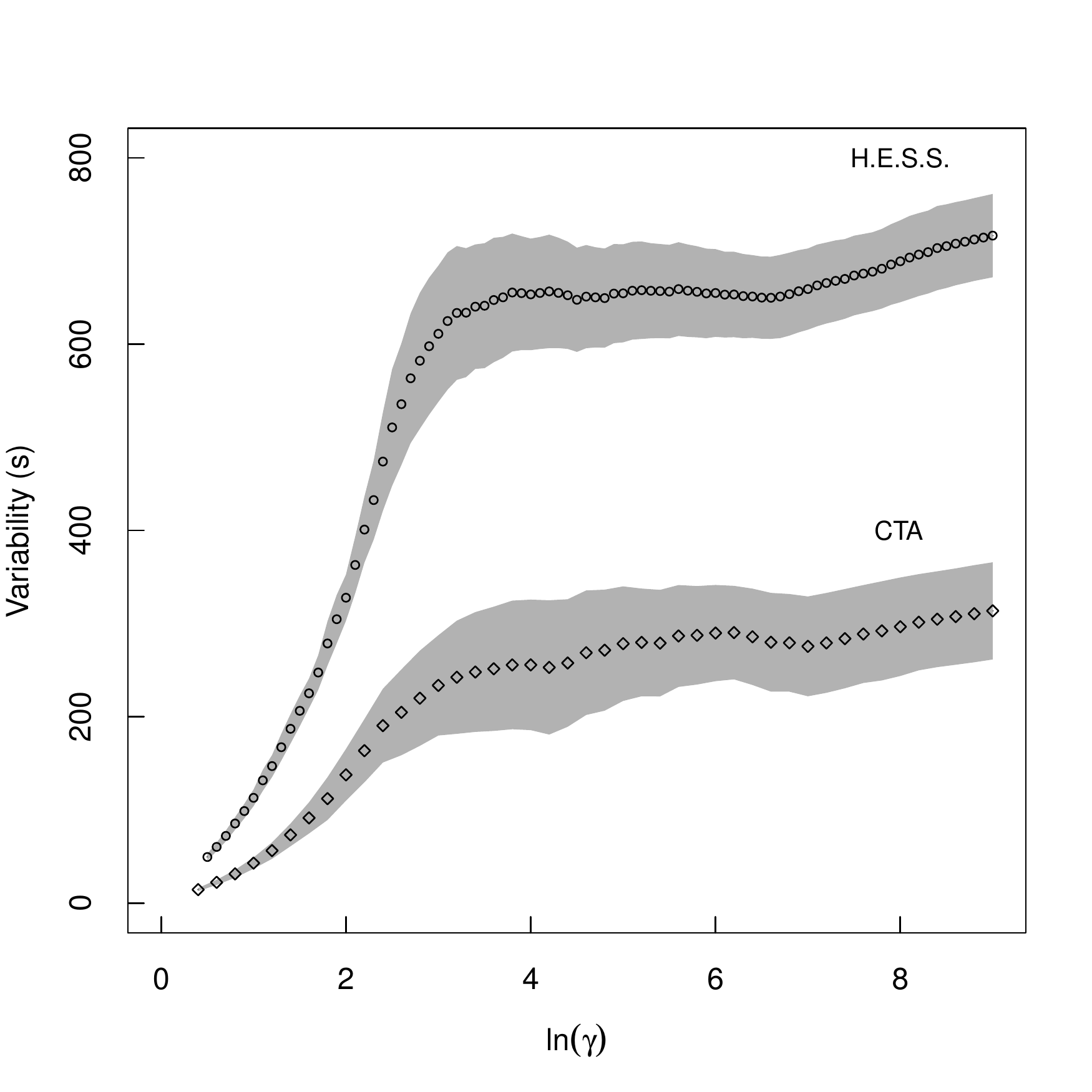}
\caption{Analysis of the stability of the prior to the partition model, $ln(\gamma)$. The plot shows the distribution of the sizes of the recovered blocks versus the prior values. As can be seen, for too low priors, the light-curve is over-divided, as there is little penalization for the modeling of the series with too many blocks. Conversely, for prior values greater than $\ln(gamma) \sim 8$, the number of blocks starts to slowly decrease (sizes of blocks increase), indicating that the penalization  introduced by the prior is now strongly influencing the final partition. Between 3-6 the prior shows an optimal, stable behaviour.
\label{fig:var_cta}}
\end{center}
\end{figure}

As with any Bayesian method, an a priori hypothesis on the time series model must be tested against the data. In the case of the Bayesian blocks, the prior is on the number of blocks which will compose the partition model of the event sequence. Without entering into much detail, the algorithm uses a geometric prior which penalizes for excessive division of the light-curve into too many parts, as it is likely that the number of blocks (flares), k, is $k \ll N$, the number of events in the sequence. For each new block added in the light-curve, a penalizing term $ln(\gamma)$ is added to the log-likelihood of the partition model. The value of the prior $\gamma$ here is a free parameter to be determined by simulation, but a strength of the model is that for a large range of values of $\gamma$, the model finds a stable, optimal solution, as can be seen in Fig.\ref{fig:var_cta}. For the analysis shown in the text, a prior of $ln(\gamma)=6$ was used, corresponding to a probability odds for block division of $\sim 5 \times 10^2$.

The results presented in Fig.\ref{fig:var_cta} were derived by simulating two groups of 1000 light-curves of about $10^3$ and $10^4$ events each, for H.E.S.S. and the CTA extrapolation. The two curves shown present the mean block size $\Delta t_{\rm k}$ and the variance of the histogrammed distributions for both sets of light-curves: H.E.S.S.-like (upper-curve) and CTA-like (lower-curve). The plateau part of the curve indicates the regions where the algorithm gives stable solutions to the partitioning of the light-curve, chosen for the analysis. the gray-bands on the figure indicate the $1\sigma$ levels of the histogrammed distributions of the block sizes.

\end{document}